\documentclass[12pt]{article}

\usepackage{scicite}

\usepackage{color}
\usepackage[dvipsnames]{xcolor}
\usepackage{graphicx}	
\usepackage{amssymb}
\usepackage{enumitem}
\usepackage{bm}
\usepackage{caption}
\usepackage{subcaption}
\usepackage{mathtools}
\usepackage{titlesec}
\usepackage{bbm}

\usepackage{colortbl}
\definecolor{Gray}{gray}{0.9}

\usepackage{hyperref}

\makeatletter
\renewcommand*{\@biblabel}[1]{\hfill#1.}
\makeatother

\newcounter{myequation}
\makeatletter
\@addtoreset{equation}{myequation}
\makeatother

\makeatletter
\renewcommand*{\@textcolor}[3]{%
  \protect\leavevmode
  \begingroup
    \color#1{#2}#3%
  \endgroup
}
\makeatother

\newcommand{\norm}[1]{\lVert#1\rVert}
\newcommand{\abs}[1]{\lvert#1\rvert}

\usepackage{times}

\newenvironment{sciabstract}{%
\begin{quote} \bf}
{\end{quote}}

\renewcommand{\figurename}{Fig.}

\title{Network structural origin of instabilities \\in large complex systems}

\author{
\hspace{-6mm}Chao Duan,$^{1,2,*}$ Takashi Nishikawa,$^{2,3,*}$ Deniz Eroglu,$^{2,4}$ Adilson E. Motter$^{2,3}$\\
\\
\hspace{-6mm}\normalsize{$^{1}$School of Electrical Engineering, Xi'an Jiaotong University, Xi'an, 710049, China}\\
\hspace{-6mm}\normalsize{$^{2}$Department of Physics and Astronomy, Northwestern University, Evanston, IL 60208, USA}\\
\hspace{-6mm}\normalsize{$^{3}$Northwestern Institute on Complex Systems, Northwestern University, Evanston, IL 60208, USA}\\
\hspace{-6mm}\normalsize{$^{4}$Department of Molecular Biology and Genetics, Kadir Has University, 34083 Istanbul, Turkey}\\
\hspace{-6mm}{$^{*}$\small These authors contributed equally to this work.}\\
}

\date{}

\begin{document} 


\maketitle 

\baselineskip18pt

\vspace{-2mm}\noindent{\bf One Sentence Summary:}
Structural imbalances underlie nonnormality-induced amplification of perturbations in network systems.

\phantomsection
\addcontentsline{toc}{section}{Abstract}
\begin{sciabstract}
A central issue in the study of large complex network systems, such as power grids, financial networks, and ecological systems, is to understand their response to dynamical perturbations.
Recent studies recognize that many real networks show \emph{nonnormality} and that nonnormality can give rise to \emph{reactivity}---the capacity of a linearly stable system to amplify its response to perturbations, oftentimes exciting nonlinear instabilities.
Here, we identify network structural properties underlying the pervasiveness of nonnormality and reactivity in real directed networks, which we establish using the most extensive data set of such networks studied in this context to date.
The identified properties are imbalances between incoming and outgoing network links and paths at each node.
Based on this characterization, we develop a theory that quantitatively predicts nonnormality and reactivity and explains the observed pervasiveness.
We suggest that these results can be used to design, upgrade, control, and manage networks to avoid or promote network instabilities.
\end{sciabstract}

\vspace{10pt}
\begin{center}
\small Published in \href{https://doi.org/10.1126/sciadv.abm8310}{\textit{Science Advances} \textbf{8}, eabm8310} (2022)
\end{center}

\baselineskip18pt

\clearpage

\phantomsection
\addcontentsline{toc}{section}{Introduction}
\section*{Introduction}

The dynamical stability of large complex network systems is an intriguing problem.
The basic question of what properties of such systems govern their stability has attracted much interest, which was initially sparked by a 1972 article by Robert May predicting that sufficiently large systems should be linearly unstable despite the observed stability of large real ecological systems \cite{May:1972}.
While the literature on this problem and discrepancies between theory and observation has focused mostly on May's original context (ecological networks \cite{May:1973,McCann:2000,Allesina:2012,Tang:2014,Gravel:2016,Jacquet:2016}, including microbiome communities \cite{Coyte:2015,Murall:2017,Butler:2018}), the problem is relevant for large network systems in general, including financial networks \cite{Haldane:2011,Moran:2019}, power networks \cite{Hastings:1984}, and immune system networks \cite{Hastings:1984}.
The problem acquires a new dimension when the Jacobian matrix $M$ determining the linear stability is \textit{nonnormal} \cite{Trefethen:2005} (i.e., $MM^T \neq M^TM$, where $M^T$ denotes the transpose of $M$). 
This is because a small perturbation in such a system can cause the resulting state deviation to initially grow and become large enough to excite nonlinear instabilities, even when the system is linearly stable \cite{Neubert:1997,Trefethen:2005,Asllani:2018a,Nicolaou:2020}.
The system's capacity to exhibit initial growth of deviations in the linear regime is termed \textit{reactivity} \cite{Neubert:1997}, which is known to relate to a spectral property of the matrix $M$.
Even though the initial interest in nonnormality and reactivity emerged in hydrodynamics \cite{Trefethen:1993,Farrell:1996,Schmid:2007}, these properties have recently gained attention in the study of network systems, including
ecological \cite{Tang:2014,Asllani:2018b}, neuronal \cite{Hennequin:2012,Gudowska-Nowak:2020}, chemical reaction \cite{Nicolaou:2020}, and communication networks \cite{Baggio:2020}, 
as well as in the study of pattern formation in networks \cite{Biancalani:2017,Muolo:2019} and control of networks \cite{Lindmark:2020}.
The literature has begun to reveal how prevalent nonnormality and reactivity are in real-world networks \cite{Asllani:2018a}, but the fundamental question 
of which network structural mechanisms underlie the apparent prevalence of nonnormality and reactivity has not yet been addressed.

In this Article, we address this question by deriving rigorous conditions for nonnormality and reactivity that can be applied to any directed network (rather than to the average over an ensemble of networks \cite{Tang:2014}) and interpreted in terms of the structure of the given network.
For nonnormality, the condition is that there is an imbalance between incoming and outgoing links at a node or a pair of nodes in terms of their numbers and/or weights.  For reactivity, the condition is that there is an imbalance between the eigenvector centrality of a node associated with incoming network paths (including their weights) and the eigenvector centrality associated with outgoing paths.
We use these conditions to show that, in a broad class of directed networks, the probability that the coupling matrices are both nonnormal and reactive approaches $1$ quickly as the network size increases.
We prove our results for large networks using a general network model that permits arbitrary distributions of possibly correlated in- and out-degrees (the number of incoming and outgoing links at a node, respectively) and arbitrary distributions of link weights. 
We also validate the prevalence of nonnormality and reactivity using a data set of $251$ real networks.  This set is the largest and most diverse collection of directed networks---which also includes the largest networks (with up to nearly $8$ million nodes)---ever considered in this context.

These findings indicate that no additional global organization of connectivity is necessary to generically observe nonnormality and reactivity.
This is important given that large-scale structures can have dynamical consequences, such as the stability promoted by trophic coherence in food-web networks \cite{Johnson:2014} and the non-monotonicity supported by linear chain structures in chemical reaction networks \cite{Nicolaou:2020}.
In addition to establishing the prevalence of nonnormality and reactivity, we develop a quantitative theory that directly relates the extent of the degree and centrality imbalances in a given network to the extent of nonnormality and reactivity, respectively.
Thus, our results reveal the network structural features responsible for nonnormality and reactivity, contributing to the much needed fundamental understanding of the relationship between dynamical and structural properties of directed networks \cite{Dorogovtsev:2001,Schwartz:2002,Garlaschelli:2004,Bianconi:2008,Leicht:2008,Masuda:2009,Malliaros:2013,Ermann:2015,Liu:2016,Timar:2017,Fruchart:2021}.

\phantomsection
\addcontentsline{toc}{section}{Results}
\section*{Results}

\phantomsection
\addcontentsline{toc}{subsection}{Nonnormality and reactivity of network systems}
\subsection*{Nonnormality and reactivity of network systems}

Given that many real network systems operate near an equilibrium, here we consider the class of (nonlinear) systems whose linearization around a given reference equilibrium state is described by
\begin{equation}\label{eqn:main-system}
\dot{x_i} = -\alpha_i x_i + \sum_{j=1}^n A_{ij} x_j, \quad i=1,\ldots,n,
\end{equation}
where $x_i$ is the (scalar) deviation from the reference state for the $i$th node.  The matrix $A = (A_{ij})$ can be regarded as the weighted adjacency matrix of the system's directed interaction network: $A_{ij} \neq 0$ if node $j$ is connected to node $i$, and $A_{ij} = 0$ otherwise.  The parameter $\alpha_i$ and the diagonal element $A_{ii}$ represent the node dynamics and any self-link at node $i$, respectively.  We assume that the time scales of the dynamics in Eq.~\eqref{eqn:main-system} are much shorter than those of the evolution of the interaction network structure, so that $A_{ij}$ can be regarded as constant.
For concreteness, we also assume $\alpha_i = \alpha$ for all $i$ in the following unless otherwise indicated.  We note, however, that our results on random networks are valid for heterogeneous $\alpha_i$ and that the presence of the heterogeneity is generally expected to increase both the nonnormality and the reactivity of the Jacobian matrix of the system (see materials and methods for details).
For $\alpha_i = \alpha$, the Jacobian matrix $M$ and the adjacency matrix $A$ are related as $M = A - \alpha I_n$, with $I_n$ denoting the $n \times n$ identity matrix.  Thus, the reference state is asymptotically stable if and only if $\text{Re}\,\lambda_1(M) < 0$, or $\text{Re}\,\lambda_1(A) < \alpha$, where $\lambda_1(X)$ denotes the eigenvalue with the largest real part for any matrix $X$.
Assuming $A_{ij} \ge 0$ for $i \neq j$, as observed in many real networks, 
$\lambda_1(A)$ is guaranteed to be real by the Perron-Frobenius Theorem for non-negative matrices \cite{Horn:2012}, regardless of whether the network is strongly connected (i.e., any two nodes are connected by directed paths in both directions).
While we have implicitly assumed one-dimensional node dynamics in Eq.~\eqref{eqn:main-system} for clarity, we also establish an exact relation between the Jacobian and adjacency matrices for a broader class of systems.  In particular, this relation shows how the nonnormality and reactivity of the adjacency matrix generically imply the same properties for the Jacobian matrix (see supplementary text, Sec.~\ref{sec-si-contri-Jacobian} for details).

We first show how the nonnormality and reactivity of system~\eqref{eqn:main-system} can be expressed as properties of the network structure described by $A$ under the uniform $\alpha_i$ assumption.
Noting that matrix $D := MM^T - M^TM$ represents the deviation of the Jacobian matrix $M$ from being normal, the nonnormality of the system can be quantified by the Frobenius norm $\lVert D \rVert_\text{F} := \sqrt{\sum_i \sum_j \lvert D_{ij} \rvert^2}$ \cite{Elsner:1987}.
Since $D = MM^T - M^TM = A A^T - A^T A$ (and hence does not depend on $\alpha$), the nonnormality of $M$ is reduced to the nonnormality of $A$.
To characterize the reactivity of system~\eqref{eqn:main-system} mathematically, we consider the maximum exponential rate of initial growth of the state deviation vector that can result from a perturbation of the initial state.
This rate is given by $\lambda_1\bigl((M+M^T)/2\bigr) = \lambda_1(H) - \alpha$, where $H := (A+A^T)/2$ is the symmetric part of $A$.  Thus, the state deviation can grow if $\lambda_1(H) > \alpha$.
Combining this with the stability requirement $\lambda_1(A) < \alpha$, we define reactivity as a property of the interaction network structure $A$ (including its dependence on the reference state): $A$ is said to be reactive if there are $\alpha$ values for which $\lambda_1(H) > \alpha > \lambda_1(A)$.
That is, $A$ is reactive if, within the linear regime, the system can be both stable and capable of exhibiting initial growth of state deviations.  Such growth can push the system state out of the region in which the linearization in Eq.~\eqref{eqn:main-system} is valid, potentially inducing nonlinear instabilities.
The reactivity condition can be expressed as
\begin{equation}\label{eqn:reactivity-condition}
\lambda_\Delta(A) := \lambda_1(H) - \lambda_1(A) > 0,
\end{equation}
and thus we use $\lambda_\Delta(A)$ as a measure of the reactivity of the interaction network structure $A$ (this definition can be extended to allow for negative weights by replacing $\lambda_1(A)$ with $\text{Re}\,\lambda_1(A)$).
This measure is independent of $\alpha$, in contrast to the reactivity for a specific $\alpha$, which could be defined as $\lambda_1(H) - \alpha$ following previous studies \cite{Neubert:1997,Farrell:1996,Trefethen:2005,Schmid:2007}.
In the general case of heterogeneous $\alpha_i$, both nonnormality and reactivity can be defined in the same way after absorbing the heterogeneity into the diagonal elements of the matrix $A$ (see materials and methods for details).

We note that $\lambda_\Delta(A) \ge 0$ always holds (see Eq.~\eqref{eqn:squeeze} in materials and methods) and $\lambda_\Delta(A) > 0$ implies that the initial growth rate of state vector deviation $\frac{d\norm{x}}{dt}\bigr\rvert_{t=0} / \norm{x_0}$ in Eq.~\eqref{eqn:main-system} for $\alpha = 0$ can be strictly larger than $\lambda_1(A)$ when the initial state vector $x_0$ is chosen to be the eigenvector corresponding to $\lambda_1(H)$, where we use $\lVert \cdot \rVert$ to denote the $2$-norm and $x := (x_1, \ldots, x_n)^T$ to denote the state vector. This is the case because $\frac{d\norm{x}}{dt}\bigr\rvert_{t=0} / \norm{x_0} = x_0^T H x_0 / (x_0^T x_0) = \lambda_1(H) > \lambda_1(A)$.
We also note that $\lambda_\Delta(A)$ is invariant under any coordinate transformation if we concurrently apply the transformation to the observable for the system~\eqref{eqn:main-system}, since $\lambda_\Delta(A)$ is based on the value of the observable rather than its coordinate-specific representation.  For the reactivity measure defined in Eq.~\eqref{eqn:reactivity-condition}, the observable is the $2$-norm of the system state vector.
This choice is standard in the literature (see, e.g., Refs.~\cite{Tang:2014,Trefethen:2005,Neubert:1997,Farrell:1996,Asllani:2018b,Biancalani:2017,Muolo:2019,Asllani:2018a}), as it permits a convenient characterization through eigenvalues, but for certain network processes different measures of deviations may be more natural and would lead to different definitions of reactivity (see, e.g., the $1$-norm used in Refs.~\cite{Townley:2007,Stott:2011,Huang:2015} and the absolute value of a one-dimensional projection used in Ref.~\cite{Nicolaou:2020}).

We used the conditions and measures just defined to study nonnormality and reactivity in a large data set of $251$ real directed networks, which consists of $63$ biological, $51$ informational, $79$ social, $39$ technological, and $19$ economic/game networks and avoids repetition of similar networks.
The data set used here substantially expands on an earlier study \cite{Asllani:2018a} with respect to the number of networks, the largest network size, and the diversity of network types.
We verified that $A$ is nonnormal (i.e., $\lVert D \rVert_\text{F} > 0$) for all $251$ networks and is reactive (i.e., $\lambda_\Delta(A) > 0$) for all but one network (see materials and methods for details on the network data and our findings).
\begin{figure}
\phantomsection
\addcontentsline{toc}{subsection}{Fig 1: Nonnormality and reactivity of real networks}
\begin{center}
\includegraphics{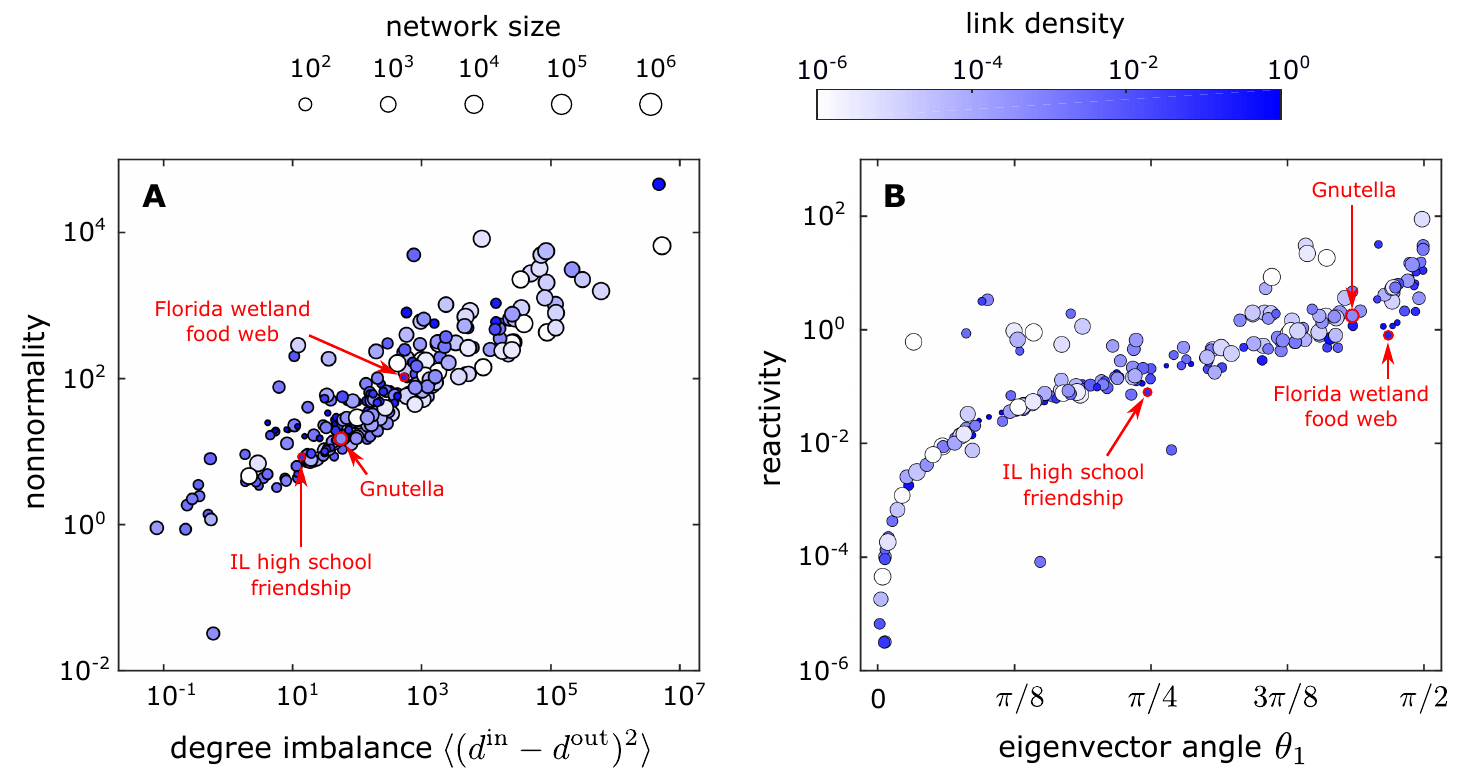}
\caption{\textbf{Nonnormality and reactivity of real networks.}
(\textbf{A})~Nonnormality vs.\ the in- and out-degree imbalance quantified by the average $\langle (d^\text{in} - d^\text{out})^2 \rangle$ of the squared difference between $d_i^\text{in}$ and $d_i^\text{out}$ over all nodes $i$ for the $251$ real networks considered (see materials and methods for details on the data).
The measure of nonnormality $\norm{D}_\text{F}$ is normalized by $\sqrt{n} \langle w^2 \rangle$ based on the scaling predicted by our theory for random networks, where $\langle w^2 \rangle := \sum_i \sum_j A_{ij}^2 / (n\bar{d})$ is the mean squared weight over all links in the network and $\bar{d}$ is the average degree; see Eq.~\eqref{eqn:D_F_theory_weighted}.
The size and color of a circle indicate the network size and link density, respectively.
(\textbf{B})~Reactivity vs.\ the in- and out-centrality imbalance quantified by the angle $\theta_1$ (in radian) between the left and right eigenvectors associated with the largest eigenvalue of $A$.  The measure of reactivity $\lambda_{\Delta}(A)$ is normalized by $\lambda_1(A)$ based on our theory in Eq.~\eqref{eqn:lambda_delta_theory}. The plot shows only the $212$ non-degenerate reactive networks (i.e., those having non-degenerate $\lambda_1(A)$ and satisfying $\lambda_{\Delta}(A) > 0$; see materials and methods for details). 
We observe significant levels of nonnormality and reactivity in a vast majority of the networks: $\norm{D}_\text{F} / (\sqrt{n}\langle w^2 \rangle) > 1$ for $248$ networks ($\approx 99$\%) and $\lambda_{\Delta}(A) / \lambda_1(A) > 0.01$ for $188$ non-degenerate networks ($\approx 88$\%).
}
\label{fig:real_networks}
\end{center}
\end{figure}
Quantitatively, Fig.~\ref{fig:real_networks}A reveals a strong positive correlation between the level of nonnormality $\norm{D}_\text{F}$ and the overall level of imbalance between the nodes' in- and out-degrees, measured by the average $\langle (d^\text{in} - d^\text{out})^2 \rangle := \sum_i (d_i^\text{in} - d_i^\text{out})^2/n$, where $d_i^\text{in}$ and $d_i^\text{out}$ are respectively the in- and out-degrees of node $i$ (i.e., the numbers of incoming and outgoing links from/to other nodes).
Similarly, Fig.~\ref{fig:real_networks}B shows a monotonic relation between the reactivity $\lambda_{\Delta}(A)$ and a measure of imbalance, in this case between the eigenvector centrality associated with the incoming and outgoing paths, quantified by the angle $\theta_1$ between the left and right eigenvectors corresponding to $\lambda_1(A)$ (to be precisely defined below).
Both relations are observed for each type of networks (fig.~\ref{fig:real_networks_SI}) and will be theoretically derived and computationally validated in a later section.

\phantomsection
\addcontentsline{toc}{subsection}{Imbalance conditions for nonnormality and reactivity}
\subsection*{Imbalance conditions for nonnormality and reactivity}

\begin{figure}
\phantomsection
\addcontentsline{toc}{subsection}{Fig 2: Topological and spectral features inducing nonnormality and reactivity}
\begin{center}
\includegraphics[width=1.0\textwidth]{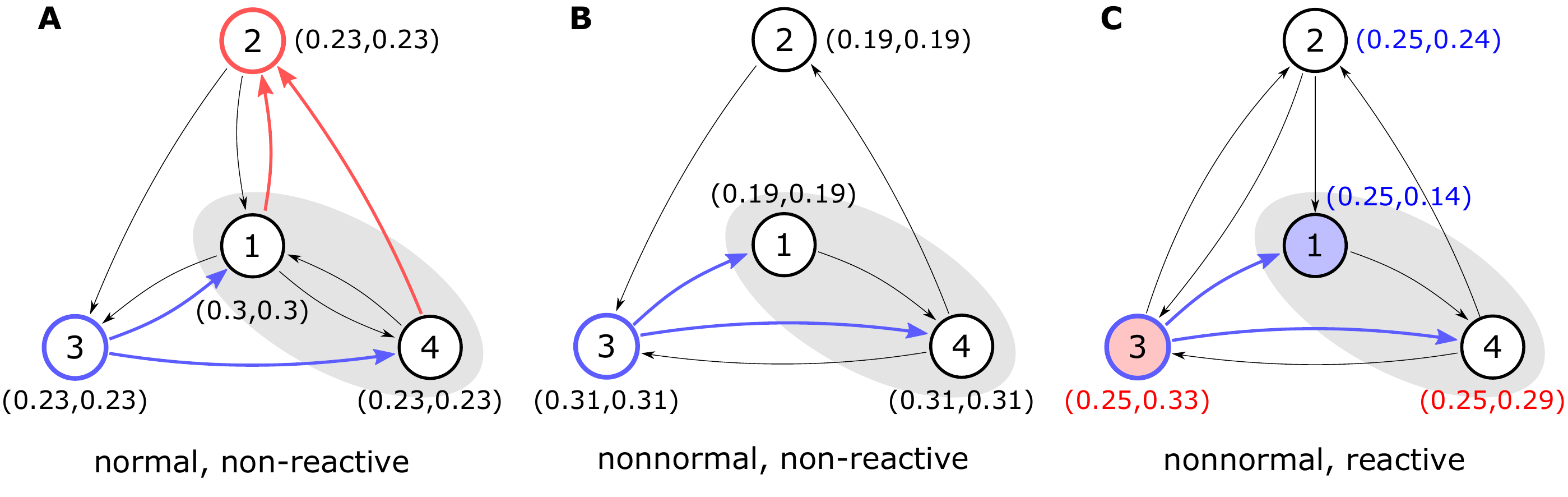}
\caption{\textbf{Topological and spectral features inducing nonnormality and reactivity.}
The four-node unweighted example networks shown illustrate all three possible cases (the normal-reactive case is not possible).
The eigenvector in-centrality $v_{1i}$ and the eigenvector out-centrality $u_{1i}$, normalized so that $\norm{v_1}=\norm{u_1}=1$, are indicated near each node $i$ as $(v_{1i}, u_{1i})$.
Note that we have $\delta_i^\text{in} = d_i^\text{in}$ and $\delta_i^\text{out} = d_i^\text{out}$ as the networks are unweighted.
(\textbf{A})~Network for which $A$ is normal (and thus non-reactive) because the in- and out-degrees match ($d_i^\text{in} = d_i^\text{out} = 2$) for all nodes $i$ and the number of shared in- and out-neighbors match ($d_{ij}^\text{in} = d_{ij}^\text{out} = 1$) for all pairs of nodes $i$ and $j \neq i$, as illustrated for $i=1$, $j=4$ (blue: shared in-neighbor; red: shared out-neighbor).
(\textbf{B})~Network for which $A$ is nonnormal because the second sum in Eq.~\eqref{eqn:D_F_def} is nonzero due to $d_{ij}^\text{in} \neq d_{ij}^\text{out}$ for some $i \neq j$, e.g., $d_{14}^\text{in} = 1 \neq d_{14}^\text{out} = 0$.
However, $A$ is non-reactive because the in- and out-centralities are balanced at each node, i.e., $v_{1i} = u_{1i}$ for all $i$.
(\textbf{C})~Network for which $A$ is nonnormal and reactive.
Nonnormality in this case is guaranteed not only by $d_{ij}^\text{in} \neq d_{ij}^\text{out}$ for some $i \neq j$ (e.g., for $i=1$, $j=4$), but also by $d_i^\text{in} \neq d_i^\text{out}$ for some $i$ (blue node: $d_i^\text{in} > d_i^\text{out}$; red node:  $d_i^\text{in} < d_i^\text{out}$). 
Reactivity is guaranteed by $v_{1i} \neq u_{1i}$ for all nodes (blue numbers: $v_{1i} > u_{1i}$; red numbers: $v_{1i} < u_{1i}$).
}
\label{fig:illust_examples}
\end{center}
\end{figure}

To understand the mechanisms underlying the correlations observed above, we first examine the network structural features that are responsible for the nonnormality and reactivity of $A$.
The measure of nonnormality introduced above can be expressed as 
\begin{equation}\label{eqn:D_F_def}
\lVert D \rVert_\text{F}^2 = \sum_{i=1}^n (\delta_i^\text{in} - \delta_i^\text{out})^2 + 2 \mathop{\sum\sum}_{i<j} (d_{ij}^\text{in} - d_{ij}^\text{out})^2,
\end{equation}
where $\delta_i^\text{in} := \sum_{k \neq i} A_{ik}^2$, $\delta_i^\text{out} := \sum_{k \neq i} A_{ki}^2$, $d_{ij}^\text{in} := \sum_k A_{ik} A_{jk}$, and $d_{ij}^\text{out} := \sum_k A_{ki} A_{kj}$.
The variables $\delta_i^\text{in}$ and $\delta_i^\text{out}$ can be regarded as generalized in- and out-degrees, respectively, since they reduce to $d_{i}^\text{in}$ and $d_{i}^\text{out}$ for unweighted networks (i.e., if $A_{ij} \in \{0,1\}$).
For distinct nodes $i$ and $j$, the variable $d_{ij}^\text{in}$ ($d_{ij}^\text{out}$) can be interpreted as a further generalization of the in-degree (out-degree) to a pair of nodes, and it reduces to the number of common in-neighbors (out-neighbors) shared by the two nodes in the case of unweighted networks.
Equation~\eqref{eqn:D_F_def}, despite being immediate from the definition of $D$ and its Frobenius norm, provides an insightful decomposition of nonnormality into two types of imbalances between incoming and outgoing links: the first term is the square sum of the generalized in- and out-degree differences at individual nodes, while the second is an analogous square sum for node pairs.
From Eq.~\eqref{eqn:D_F_def}, we see that $A$ is normal if and only if the generalized in- and out-degrees are equal for each node and for each pair of nodes (see Fig.~\ref{fig:illust_examples}A for an illustrative example).
It is thus sufficient to have just a single node whose in- and out-degrees differ in order to make $A$ nonnormal.
However, even without such a node, $A$ can still be nonnormal if there is a node pair with an imbalance between their common weighted in-neighbors and the common weighted out-neighbors (making the second sum nonzero in Eq.~\eqref{eqn:D_F_def}), as illustrated in Fig.~\ref{fig:illust_examples}B.
Furthermore, Eq.~\eqref{eqn:D_F_def} shows that a larger total imbalance between incoming and outgoing links implies a more nonnormal $A$.

For reactivity, a stronger condition is needed, since nonnormal $A$ does not have to be reactive, as illustrated by the network in Fig.~\ref{fig:illust_examples}B.  
In particular, for a $d$-regular network, defined as an unweighted network in which the in- and out-degrees are all equal to a constant integer $d\ge0$, the adjacency matrix $A$ can be nonnormal but can never be reactive.
To see this, we first note that $\lambda_1(A)$ is bounded between the minimum and maximum row sum of $A$ because this eigenvalue is the spectral radius of $A$ (see, e.g., Theorem 8.1.22 in Ref.~\citen{Horn:2012}) and thus is equal to $d$ for any $d$-regular network.  
Since the same argument applies to the symmetric part of $A$, we have $\lambda_1(H)=d$, implying that $\lambda_{\Delta}(A) = d - d = 0$, i.e., $A$ is non-reactive. 
Nonetheless, for many $d$-regular networks, the second term in Eq.~\eqref{eqn:D_F_def} is strictly positive, rendering $A$ nonnormal.

Given that nonnormality is only a necessary condition for reactivity, we now present a condition guaranteeing reactivity: $A$ is reactive if the left and right eigenspaces associated with its largest eigenvalue $\lambda_1(A)$ are distinct (see materials and methods for a proof).
In the generic case in which these eigenspaces are one-dimensional, the reactivity condition is equivalent to having a strictly positive angle $\theta_1$ between the right eigenvector $v_1$ and left eigenvector $u_1$.  We always choose the acute angle so that $\theta_1 \le \pi/2$, where the quantity $1/\abs{\cos\theta_1}$ is known as the eigenvalue condition number in the literature \cite{Trefethen:2005}.
This condition is illustrated by the example network in Fig.~\ref{fig:illust_examples}C.
While the non-orthogonality of right eigenvectors defining nonnormality has been shown to often lead to reactivity (e.g., in Refs.~\cite{Murphy:2009,Chalker:1998}), our condition $\theta_1>0$ captures reactivity more precisely, as it is equivalent to the specific non-orthogonality between the right eigenvector $v_1$ and some other right eigenvector.
The condition $\theta_1>0$ also translates to the existence of at least one node $i$ for which $v_{1i} \neq u_{1i}$, where $v_{1i}$ is the \textit{eigenvector in-centrality}, defined to be the $i$th component of the right eigenvector $v_1$ (and hence associated with incoming paths to the node), and $u_{1i}$ is the \textit{eigenvector out-centrality}, defined similarly through the left eigenvector $u_1$ (and the outgoing paths).
The inequality $v_{1i} \neq u_{1i}$ can thus be interpreted as an imbalance between the incoming and outgoing ``flow'' of centrality.
In the example of $d$-regular networks above, if $\lambda_1(A)$ is non-degenerate (so that $v_1$ and $u_1$ are unique and $\theta_1$ is well defined), we find that $v_{1i} = u_{1i}$ for all $i$ and thus $\theta_1 = 0$, as expected from the result above that $A$ is non-reactive for all $d$-regular networks.
In general, the imbalances $\delta_i^\text{in} - \delta_i^\text{out}$ and $v_{1i} - u_{1i}$ for individual nodes and $d_{ij}^\text{in} - d_{ij}^\text{out}$ for pairs of nodes can be either positive or negative and tend to be distributed heterogeneously across a given network, as illustrated by the three example real networks in fig.~\ref{fig:real_networks_examples}.

While we focus mainly on the adjacency matrix $A$, different interaction
matrices can also be considered in Eq.~\eqref{eqn:main-system}. This includes the Laplacian matrix, which we show can be nonnormal even if $A$ is not, and vice versa (a fact that has been generally overlooked; see fig.~\ref{fig:A_L_normality} for illustrative examples).

\phantomsection
\addcontentsline{toc}{subsection}{Prevalence of nonnormality and reactivity}
\subsection*{Prevalence of nonnormality and reactivity}

We now address the question of how often nonnormality and reactivity are expected to be observed by considering a model of random directed weighted networks with a given number of nodes $n$, a given joint probability distribution for the in- and out-degrees, and a given distribution of link weights.
By allowing for arbitrary distributions for the in- and out-degrees, an arbitrary correlation between them, and an arbitrary distribution of weights, the model is capable of capturing the essential elements of real networks involved in the conditions established above for nonnormality and reactivity.
This model is a generalization of the one studied by Chung, Lu, and Vu in Ref.~\citen{Chung:2003} and allows us to specify a general joint distribution of in- and out-degrees (rather than the expected node degrees) and a general weight distribution.  A network realization under this model is generated as follows.
First, the (possibly non-integer) expected in- and out-degrees of each node are randomly drawn from the given joint distribution.
Then, for each $i$ and $j$, a directed link is created from node $j$ to node $i$ with a probability proportional to the product of the expected in-degree of node $i$ and the expected out-degree of node $j$.
Finally, a random weight $A_{ij}$ is drawn from the given distribution for each link $j \rightarrow i$.
We note that a special case of this model produces unweighted networks, and self-links can be excluded if desired.
For further details on this generalized Chung--Lu--Vu (GCLV) model, see materials and methods.

For weighted networks generated by the GCLV model, we show that the adjacency matrix $A$ is almost always nonnormal in the limit of large network size $n$ (see supplementary text, Sec.~\ref{sec-si-proofs}, for a proof for unweighted networks and its extensions to weighted and Laplacian-coupled networks).
More precisely, we show that the probability of having at least one node whose generalized in-degree and out-degree are different (implying nonnormality of $A$, as discussed above) tends to one as $n\to\infty$.
In practice, the probability that $A$ is nonnormal grows quickly and is very close to one even for networks with less than $10$ nodes, as shown numerically for both weighted and unweighted networks in fig.~\ref{fig:prob-norm-nonreact}A and table~\ref{table:prob-norm-nonreact} for two classes of in-/out-degree distributions: \textit{i}) the gamma distribution, whose probability density function is $p(x) \sim x^{a-1}e^{-bx}$, $x>0$, where $a$ and $b$ are parameters; and \textit{ii}) the Dirac delta distribution centered at $d$, which renders the model equivalent to the Erd\H{o}s-R\'{e}yi (ER) networks with fixed mean degree $d$ (and thus with $n$-dependent connection probability $p=d/n$).
The same appears to hold true for other random network models, as verified in fig.~\ref{fig:prob-norm-nonreact}A for the unweighted ER networks with fixed $p$ and random $d$-regular networks and also verified in table~\ref{table:prob-norm-nonreact} for the weighted versions of these random networks.

Nonnormality does not necessarily imply reactivity.
For large networks, however, we can show that, if $A$ is nonnormal, then it is also reactive in almost all cases.
This is because the conditional probability $\mathbb{P}(\text{$A$ is reactive} \,\vert\, \text{$A$ is nonnormal}) \ge \mathbb{P}(\text{$A$ is reactive})$, as reactivity implies nonnormality, and because we show that $\mathbb{P}(\text{$A$ is reactive})$ approaches one as $n\to\infty$ for the GCLV model (see supplementary text, Sec.~\ref{sec-si-proofs}, for a proof, including the case of Laplacian-coupled networks).
More precisely, we prove that the probability of having distinct left and right eigenspaces associated with $\lambda_1(A)$ (sufficient for reactivity, as noted earlier) converges to one.
When estimated numerically for finite $n$, the actual probability that $A$ is reactive and the conditional probability that $A$ is reactive given that it is nonnormal are again very close to one even for small $n$, as shown in fig.~\ref{fig:prob-norm-nonreact}, B and C, respectively (which also shows similar results for two other random network models).
All these results support the observation from Fig.~\ref{fig:real_networks} that the reactivity of $A$ is prevalent among real networks.

\phantomsection
\addcontentsline{toc}{subsection}{Quantitative characterization of nonnormality and reactivity}
\subsection*{Quantitative characterization of nonnormality and reactivity}

\begin{figure}
\phantomsection
\addcontentsline{toc}{subsection}{Fig 3: Validating theoretical predictions for nonnormality and reactivity}
\begin{center}
\includegraphics[width=0.97\textwidth]{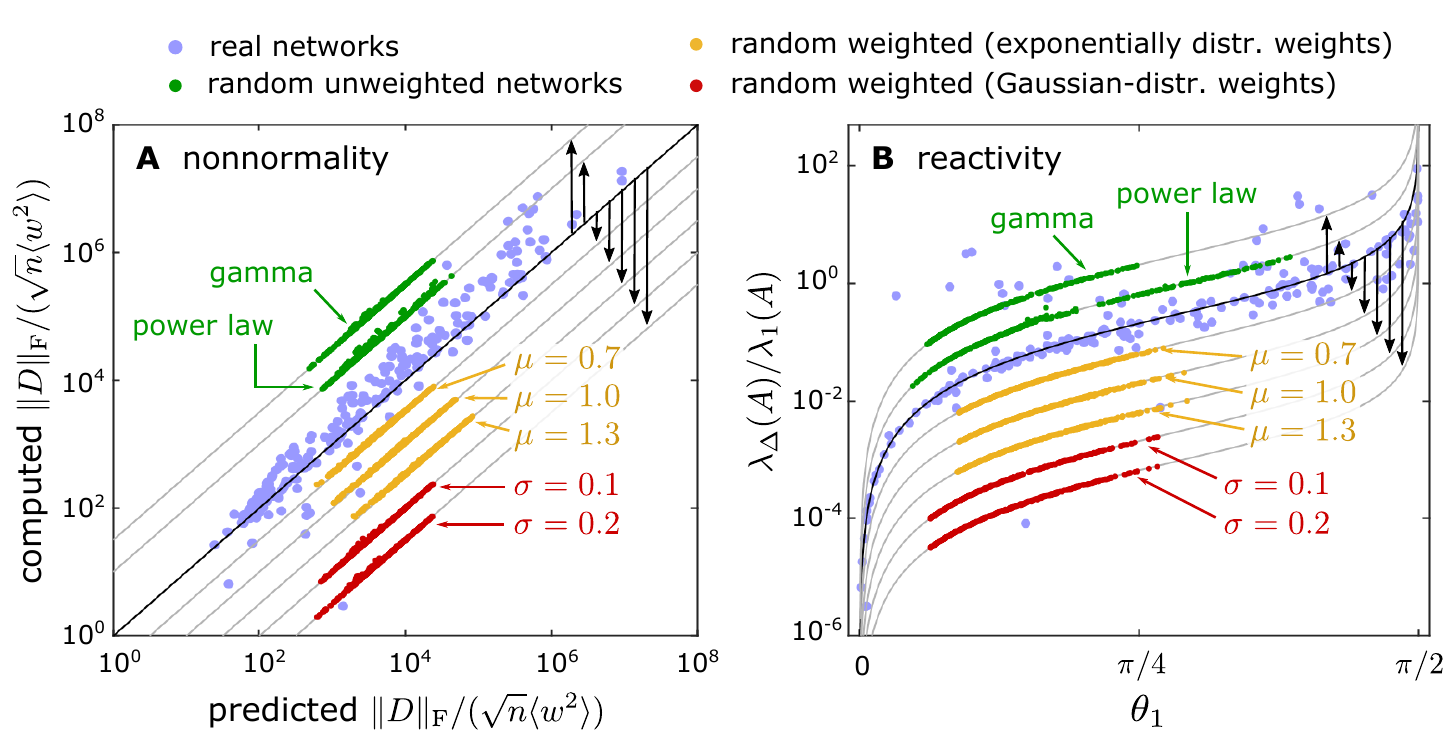}
\caption{
\textbf{Validating theoretical predictions for nonnormality and reactivity.}
(\textbf{A})~Computed nonnormality vs.\ the corresponding prediction from Eq.~\eqref{eqn:D_F_theory} for the real networks in Fig.~\ref{fig:real_networks}A (blue).
Plots are also shown for random networks generated by the GCLV model: unweighted networks with the gamma or power-law distribution for the in- and out-degrees (green) and weighted networks with gamma-distributed degree distributions and exponentially distributed weights (orange; mean $\mu = 0.7$, $1.0$, or $1.3$) or Gaussian-distributed weights (red; mean one and standard deviation $\sigma=0.1$ or $0.2$), keeping only positive weights.
For each class of networks, we used $20$ equally spaced values of $n$ between $10^3$ and $10^5$ on the logarithmic scale.
For each $n$, we show $20$ realizations of the model, with the parameters of the distribution drawn randomly for each network realization (see materials and methods for details).
The continuous lines correspond to Eq.~\eqref{eqn:D_F_theory_weighted}.
The plots for the random networks and the corresponding lines are shifted vertically to avoid overlapping.
(\textbf{B})~Reactivity vs.\ the eigenvector angle $\theta_1$.
The continuous curves correspond to Eq.~\eqref{eqn:lambda_delta_theory}.
The plot shows the same random networks as in (A) and the real networks in Fig.~\ref{fig:real_networks}B, to which Eq.~\eqref{eqn:lambda_delta_theory} is applicable.
}
\label{fig:theory_validation}
\end{center}
\end{figure}

To understand the correlations observed for the real networks in Fig.~\ref{fig:real_networks}, we now derive theoretical estimates of the nonnormality $\norm{D}_\text{F}$ and the reactivity $\lambda_{\Delta}(A)$ for random networks.
For the nonnormality, we first consider networks generated by the unweighted GCLV model with no self-links (i.e., $A_{ij}\in\{0,1\}$ and $A_{ii}=0$ for all $i$ and $j$) given a fixed set of in-degrees $d_i^\text{in}$ and out-degrees $d_i^\text{out}$ for all nodes.
In this case, the first sum in Eq.~\eqref{eqn:D_F_def} is fixed and proportional to the (constant) average single-node degree imbalance $\langle (d^\text{in} - d^\text{out})^2 \rangle$, while the second sum representing the imbalances at the node pair level is a random variable.
Assuming $n \gg 1$ and approximating the second sum with its expected value (see materials and methods for details), we have 
\begin{equation}\label{eqn:D_F_theory}
\begin{split}
\lVert D \rVert_\text{F}^2 
&\approx n \langle (d^\text{in} - d^\text{out})^2 \rangle + n \bigl[ \langle (d^\text{in} - d^\text{out})^2 \rangle + 2\langle d^\text{in} d^\text{out} \rangle - 2 \overline{d} \,\bigr]\\
&= 2n \bigl[ \langle (d^\text{in} - d^\text{out})^2 \rangle + \langle d^\text{in} d^\text{out} \rangle  - \overline{d} \,\bigr],
\end{split}
\end{equation}
where $\langle z \rangle := \sum_i z_i/n$ denotes the average of $z_i$ over nodes $i$ and $\overline{d} := \langle d^\text{in} \rangle = \langle d^\text{out} \rangle$ is the average degree.
Thus, in the limit of large networks, nonnormality increases with the network size $n$ as $\norm{D}_\text{F} \sim n^{1/2}$, with an $n$-independent prefactor expressed as a simple function of the in- and out-degree imbalance $\langle (d^\text{in} - d^\text{out})^2 \rangle$, the in- and out-degree correlation $\langle d^\text{in} d^\text{out} \rangle$, and the average degree $\overline{d}$.
We also derive extensions of this formula to the weighted GCLV model (allowing self-links) and to Laplacian-coupled networks (see materials and methods), which have additional terms and factors involving the diagonal elements $A_{ii}$ and the statistics of link weights (see Eqs.~\eqref{eqn:D_F_theory_weighted}, \eqref{eqn:DA_weighted}, and \eqref{eqn:D_F_theory_laplacian}).
For the reactivity $\lambda_{\Delta}(A)$, we make use of the observation that the spectral gap between the leading eigenvalue $\lambda_1(A)$ and the remaining eigenvalues $\lambda_2(A),\ldots,\lambda_n(A)$ is often large for the adjacency matrix $A$ of large random networks (which is the case, e.g., for the ER networks \cite{Juhasz:1982} and for the GCLV model when the mean degree is large \cite{Neri:2020}).
For such networks, we expect $\lambda_1(H)$ to be well-approximated by $\lambda_1(H_1)$, where $H_1 := (A_1+A_1^T)/2$ is the symmetric part of the leading component $A_1 := \lambda_1(A) v_1 u_1^T$ and $u_{1}$ and $v_{1}$ are respectively the left and right (column) eigenvectors associated with the leading eigenvalue $\lambda_1(A)$, normalized so that $\norm{v_1} = 1$ and $u_1^T v_1 = 1$.
Among the $212$ real networks in Fig.~\ref{fig:real_networks}B (which have non-degenerate $\lambda_1(A)$, ensuring that $H_1$ is well defined), we indeed observe $\lambda_1(H) \approx \lambda_1(H_1)$ for most, as shown in fig.~\ref{fig_A1_approx}A.  For any (possibly weighted) network satisfying $\lambda_1(H) \approx \lambda_1(H_1)$ and having non-degenerate $\lambda_1(A)$ (even when it is not strongly connected), we derive a simple expression for $\lambda_{\Delta}(A)$ with a prefactor that depends monotonically on the angle $\theta_1$ between the leading left and right eigenvectors $u_1$ and $v_1$ (see materials and methods for a derivation):
\begin{equation}\label{eqn:lambda_delta_theory}
\lambda_{\Delta}(A) \approx \frac{1-\cos\theta_1}{2\cos\theta_1} \cdot \lambda_1(A).
\end{equation}
We find that Eqs.~\eqref{eqn:D_F_theory} and \eqref{eqn:lambda_delta_theory} provide good approximations, as validated in Fig.~\ref{fig:theory_validation} for several classes of random networks generated by the GCLV model (green, orange, and red dots) as well as for the real networks used in Fig.~\ref{fig:real_networks} (blue dots).
In addition, they capture the general tendency observed in Fig.~\ref{fig:real_networks} for nonnormality and reactivity to increase with the degree imbalance and the eigenvector angle in real networks.

Our results show that $A$ is more nonnormal and reactive when the network is  weighted.
In particular, Eqs.~\eqref{eqn:D_F_def} and \eqref{eqn:D_F_theory_weighted} indicate that allowing for weights in random networks can only increase the probability of observing imbalances that induce nonnormality (i.e., the probability that $\norm{D}_\text{F} \neq 0$) and that a larger variance for the weights leads to a larger extent of nonnormality.
Likewise, the eigenvector condition and Eq.~\eqref{eqn:lambda_delta_theory} indicate that the presence of weights, whose randomness would be reflected in the eigenvectors, is expected to increase in-/out-centrality imbalances, leading to a higher probability that $A$ is reactive and to a larger extent of reactivity.

The tendency for $A$ to be dominated by the leading eigenvalue $\lambda_1(A)$ and the corresponding eigen-component $A_1$ also explains why large directed networks are almost always both nonnormal and reactive.
To see this, we first note that $A_1$ satisfies the following exact relation linking its nonnormality and reactivity to each other and to the eigenvector angle:
\begin{equation}\label{eqn:relation-dominant}
\lambda_1^2 + \sqrt{\lambda_1^4 + 2\lVert D_1 \rVert_\text{F}^2}
= 2\bigl( \lambda_1 + 2\lambda_{\Delta}(A_1) \bigr)^2
= \frac{2\lambda_1^2}{\cos^2\theta_1},
\end{equation}
where we denote $\lambda_1 = \lambda_1(A_1) = \lambda_1(A)$ for brevity.  Here, we note that $A_1$ and $A$ have the same eigenvector angle $\theta_1$ associated with $\lambda_1$, and we introduce $D_1 := A_1 A_1^T - A_1^T A_1$ to quantify the nonnormality of $A_1$.
From relation~\eqref{eqn:relation-dominant}, it immediately follows that, for the leading component $A_1$, nonnormality ($\lVert D_1 \rVert_\text{F} > 0$), reactivity ($\lambda_{\Delta}(A_1) > 0$), and having a strictly positive eigenvector angle ($\theta_1 > 0$) are all mathematically equivalent to each other.
Moreover, relation~\eqref{eqn:relation-dominant} shows that an increase in any one of these three measures implies an increase in all the other measures.
Building on these observations, we find that random networks whose $A$ is dominated by $A_1$ generally satisfy relation~\eqref{eqn:relation-dominant} approximately with $\lVert D_1 \rVert_\text{F}$ and $\lambda_{\Delta}(A_1)$ replaced by $\lVert D \rVert_\text{F}$ and $\lambda_{\Delta}(A)$, respectively, which implies that nonnormality and reactivity are approximately equivalent for such networks.
This finite-$n$ approximate equivalence complements the rigorous results we established above in the limit of large network size.
For the real networks in Fig.~\ref{fig:theory_validation}B, relation~\eqref{eqn:relation-dominant} approximately holds true when $\lVert D_1 \rVert_\text{F}$ and $\lambda_{\Delta}(A_1)$ are replaced by $\lVert D \rVert_\text{F}$ and $\lambda_{\Delta}(A)$, respectively (fig.~\ref{fig_A1_approx}, B to D).

\begin{figure}
\begin{center}
\includegraphics[width=0.91\textwidth]{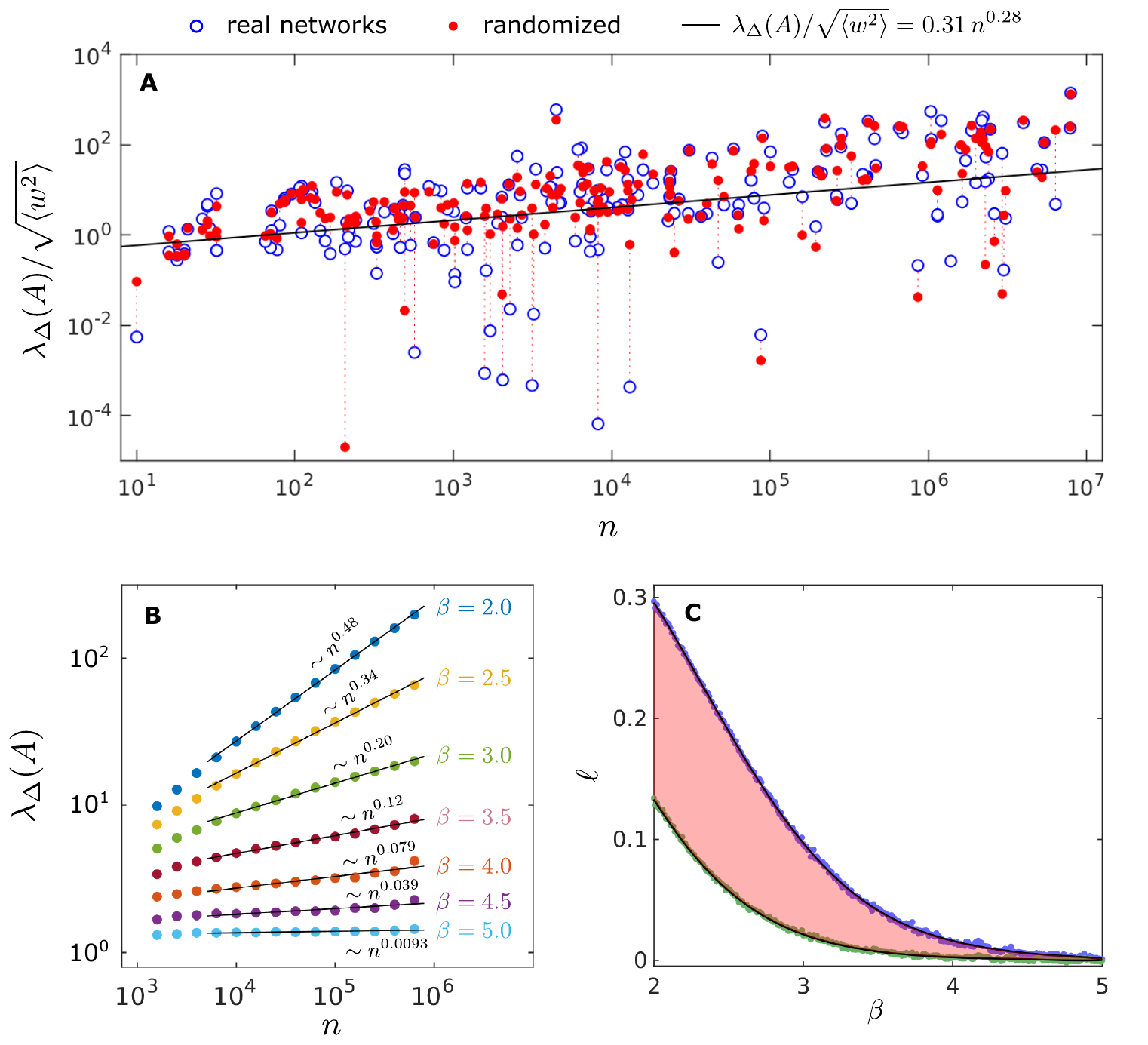}
\vspace{-7pt}
\caption{
\textbf{Increase of reactivity with network size.}
(\textbf{A})~Reactivity vs.\ network size $n$ for the real networks in Fig.~\ref{fig:theory_validation}B (blue circles).
We normalize $\lambda_{\Delta}(A)$ by $\sqrt{\langle w^2 \rangle}$ to facilitate comparison between networks with different scales for link weights.
The black line indicates the least-squares fit (on the logarithmic scale), reflecting an increasing trend for the reactivity.
Red dots indicate randomized versions of the real networks (see materials and methods for details), which more closely follow the trend line.
(\textbf{B})~Scaling of $\lambda_{\Delta}(A)$ with $n$ for the GCLV model with the same discrete power-law degree distribution $p(x) \sim x^{-\beta}$ used in Fig.~\ref{fig:theory_validation}, but for a given value of $\beta$ and uncorrelated $\tilde{d}_i^{\text{in}}$ and $\tilde{d}_i^{\text{out}}$, with link weights drawn from the positive domain of the Gaussian distribution with mean one and standard deviation $0.1$.
For each $\beta$, the plot shows the average over $200$ network realizations.
We observe that the scaling exponent for $\lambda_{\Delta}(A)$ depends on the parameter $\beta$ of the degree distribution.
(\textbf{C})~Reactivity--stability phase diagram for system~\eqref{eqn:main-system} with an $n$-dependent parameter $\alpha = \alpha(n) \sim n^{\ell}$, where $\ell$ is a constant, and for the same power-law distribution with parameter $\beta$ as in (B).
Shaded in red is the region in which the maximum growth rate of state deviations $\lambda_1(H)-\alpha(n)$ increases with $n$ despite the concurrent increase of linear stability [i.e., decrease of $\lambda_1(A)-\alpha(n)$]x.
See materials and methods for details on the procedure to identify the boundaries of this region.}
\label{fig_scaling_reactivity_stability}
\end{center}
\end{figure}

For networks satisfying Eqs.~\eqref{eqn:D_F_theory} and \eqref{eqn:relation-dominant} approximately, we see that, if $\lambda_1=\lambda_1(A)$ is bounded as $n$ increases, nonnormality and reactivity would scale with $n$ as $\lVert D \rVert_\text{F} \sim n^{1/2}$ and $\lambda_{\Delta}(A) \sim n^{1/4}$, respectively.
For the real networks, $\lambda_{\Delta}(A)$ tends to increase with $n$ (even when normalized by $\langle w^2 \rangle$, the r.m.s.\ of the link weights), which is a trend observed even more strongly when the networks are randomized while holding the in- and out-degrees fixed (Fig.~\ref{fig_scaling_reactivity_stability}A).
For random networks with a power-law degree distribution $p(x) \sim x^{-\beta}$ and Gaussian-distributed link weights, the reactivity $\lambda_{\Delta}(A)$ scales with $n$ with an exponent that depends on the power-law parameter $\beta$ (Fig.~\ref{fig_scaling_reactivity_stability}B).
The increase of $\lambda_{\Delta}(A)$ with $n$ indicates that, as the system becomes larger, there will be a wider range of $\alpha$ for which the system can simultaneously exhibit a more pronounced transient response to a small perturbation and stronger linear stability.
Specifically, Fig.~\ref{fig_scaling_reactivity_stability}C establishes that, if the node stability parameter $\alpha$ in Eq.~\eqref{eqn:main-system} has an $n$-dependence $\alpha(n) \sim n^{\ell}$ with a constant $\ell$, then there is a region in the $\beta$ vs.\ $\ell$ parameter space (shaded red) for which the maximum initial growth rate of perturbations given by $\lambda_1(H) - \alpha(n)$ increases with $n$ even though the first-order dynamics become more stable (i.e., $\lambda_1(A) - \alpha(n)$ decreases).
Below this region the growth rate increases while linear stability decreases with $n$, which is doubly destabilizing, and above this region the opposite is observed.

\phantomsection
\addcontentsline{toc}{subsection}{Fig 4: Increase of reactivity with network size}

\phantomsection
\addcontentsline{toc}{section}{Discussion}
\section*{Discussion}

Our demonstration that nonnormality and reactivity are stronger and more prevalent for larger networks implies that conventional modal stability analysis alone is not sufficiently informative: the system can be less stable against perturbations than the modal analysis indicates, and this gap can grow with the system size.
Even if such a system is linearly stable and the perturbation is well within the linear regime, the transient response may bring the system sufficiently far from the equilibrium that nonlinear instabilities can be induced, possibly in the form of a cascade \cite{Motter:2017aa}.
Our findings thus suggest that, following a perturbation, there is a parameter region for which larger networks can exhibit larger oscillations in the linearized system and thus the possibility of a rare but substantial instability in the nonlinear system.
Does such an instability cause a permanent transition to a different state? If so, to which state does the system transition?  
The answers to these questions depend on the size and direction of the perturbation, the extent of reactivity in the system, and the global structure of the state space.
The latter, in particular, requires specific knowledge of the nonlinearity of the system under consideration.

Given our validation with extensive network data and general mathematical analysis, these conclusions are applicable to a wide range of real network systems, and are thus not limited to the networks previously considered in addressing the complexity--stability problem.
In particular, knowing that nonnormality and reactivity can be caused by the imbalance of incoming/outgoing connections or of the node's in-/out-centrality will likely be useful in designing large complex technological (and possibly synthetic biological) systems.
Importantly, our results show that non-uniform link weights and self-links, which are aspects of complexity often neglected in the systematic study of networks, tend to increase nonnormality.
While here we focused on reactivity as a widely used measure of transient response, 
we anticipate similar results for other measures, such as pseudospectra \cite{Trefethen:2005}, singular values \cite{Farrell:1996}, and the perturbation-averaged evolution of the state vector \cite{Tarnowski:2020}.
We also suggest that the structural characterization of nonnormality and reactivity can be relevant for the study of generalized networks of current interest, such as those accounting for temporal, multi-layer, and higher-order (non-pairwise) interactions.
Ultimately, the demonstration that nonnormality and reactivity tend to be more prevalent and more consequential for stability as the system grows in size opens up new vistas of network dynamics in complex systems.

\phantomsection
\addcontentsline{toc}{section}{Materials and Methods}
\section*{Materials and Methods}

\phantomsection
\addcontentsline{toc}{subsection}{Impact of node parameter heterogeneity on nonnormality and reactivity}
\subsection*{Impact of $\boldsymbol{\alpha_i}$ heterogeneity on nonnormality and reactivity}
Any heterogeneity in $\alpha_i$ can be subtracted from the first term and absorbed into the second in Eq.~\eqref{eqn:main-system} by replacing $\alpha_i$ and $A_{ii}$ with $\alpha = \max_i \alpha_i$ and $A_{ii} + (\alpha - \alpha_i)$, respectively.
In Eq.~\eqref{eqn:EX_ij2}, this would contribute to the heterogeneity of $\beta_i$, extending the validity of the nonnormality approximation in that equation to the general case of heterogeneous $\alpha_i$.
It then follows from Eqs.~\eqref{eqn:DF_deriv_weighted} and \eqref{eqn:EX_ij2} that the expected nonnormality $\lVert D \rVert_\text{F}$ for the GCLV model generally increases with the heterogeneity of $\alpha_i$. 
Combining this with the approximate relation between nonnormality and reactivity in Eq.~\eqref{eqn:relation-dominant}, we see that the reactivity also tend to increase with the heterogeneity of $\alpha_i$ in random networks.
Moreover, the modification of $\alpha_i$ and $A_{ii}$ in Eq.~\eqref{eqn:main-system} used here can also be applied to Eq.~\eqref{eqn:lambda_delta_theory}, which would extend the reactivity approximation formula to heterogeneous $\alpha_i$.
The conclusion above on the impact of $\alpha_i$ heterogeneity in random networks holds true in particular when $\alpha_i$ and $A_{ii}$ are uncorrelated, while specific correlations can in principle lead to a reduction in the heterogeneity of $\beta_i$ and thus in $\mathbb{E}\bigl(\lVert D \rVert_\text{F}^2\bigr)$.

After absorbing $\alpha_i$ heterogeneity into $A_{ii}$ in Eq.~\eqref{eqn:main-system} as in the previous paragraph, the nonnormality and reactivity of the system can be defined in the same fashion, i.e., as $\lVert D \rVert_\text{F} = \lVert AA^T - A^TA \rVert_\text{F}$ and $\lambda_\Delta(A) = \lambda_1(H) - \lambda_1(A)$, respectively, but using the modified $A$.
While this definition allows us to derive results for random networks (as described in the previous paragraph), the real networks were analyzed under the uniform $\alpha_i$ assumption since the values of $\alpha_i$ were not available in the data set.

\phantomsection
\addcontentsline{toc}{subsection}{Sufficient condition for reactivity}
\subsection*{Sufficient condition for reactivity}
To establish that having distinct (real) left and right eigenspaces associated with $\lambda_1(A)$ implies reactivity (i.e., $\lambda_\Delta(A)>0$), we will prove the contrapositive: $\lambda_\Delta(A)=0$ implies that these eigenspaces coincide.  Suppose that $\lambda_\Delta(A)=0$, which implies $\lambda_1(H) = \lambda_1(A)$.  We will seek to show that any right eigenvector is also a left eigenvector, and vice versa.  Let $v$ be a (real) right eigenvector of $A$ associated with the eigenvalue $\lambda_1(A)$.  Without loss of generality, we can assume $v$ to be normalized so that $v^T v=1$.  Then,
\begin{equation}\label{eqn:squeeze}
\begin{split}
\lambda_1(H) = \max_{x \neq 0} \frac{x^T H x}{x^Tx} \ge v^T H v =  \lambda_1(A),
\end{split}
\end{equation}
noting that $v^T H v = v^T (A + A^T) v/2 = \lambda_1(A) v^T v = \lambda_1(A)$.
Since $\lambda_1(H) = \lambda_1(A)$, the inequality in Eq.~\eqref{eqn:squeeze} becomes an equality, implying that $v$ is a solution of the maximization problem and thus a (right) eigenvector of $H$ associated with $\lambda_1(H)$, i.e., $H v = \lambda_1(H) v$.
This, together with $2H=A+A^{T}$ and $Av=\lambda_1(A)v$, yields $v^T A=(2\lambda_1(H) - \lambda_1(A))v^T = \lambda_1(A)v^T$.
We thus conclude that $v$ is a left eigenvector of $A$ associated with $\lambda_1(A)$, in addition to being a right eigenvector.
To show the opposite direction, we now let $v$ be any left eigenvector of $A$ associated with the eigenvalue $\lambda_1(A)$.
This again implies $v^T H v = \lambda_1(A)$ and turns Eq.~\eqref{eqn:squeeze} into an equality, implying that $v$ is a (left) eigenvector of $H$ associated with $\lambda_1(H)$, i.e., $v^T H = v^T \lambda_1(H)$.
From this, an argument similar to the one above shows that $v$ is also a right eigenvector of $A$ associated with the eigenvalue $\lambda_1(A)$.
Therefore, we conclude that the left and right eigenspaces of $A$ corresponding to $\lambda_1(A)$ must coincide, which completes the proof.

\phantomsection
\addcontentsline{toc}{subsection}{Real network data}
\subsection*{Real network data}
We retrieved raw data for $534$ distinct directed networks available from Koblenz Network Collection (KONECT) \cite{konect,konect_doc}, the Netzschleuder Network Catalogue and Repository \cite{Netzschleuder}, the Colorado Index of Complex Networks (ICON) \cite{icon}, and the Matrix Market Repository \cite{matrixmart}.
From KONECT, Netzschleuder, and ICON, we obtained all distinct directed non-bipartite networks that were available in the form of adjacency lists.
From the Matrix Market Repository, we retrieved a total of $17$ directed non-bipartite networks among the largest available in each technological or economic application domain to compensate for the relatively few networks of those types available from the other three data sources.

To eliminate redundancy among these networks, we identified and excluded duplicates by comparing the description, number of nodes, number of links, and other network statistics.
Since the resulting set still contained a disproportionately large number of wiki-talk and wiki-link networks ($28$ and $167$, respectively), we removed this bias by excluding all but the largest network from each of these two groups.
This led to our final selection of $251$ networks, consisting of $62$ biological networks, $51$ informational networks, $79$ social networks, $40$ technological networks, and $19$ economic/game networks.
The network size ranges from $n=10$ to $n=7.9\times 10^6$, and the link density ranges from $4.0 \times 10^{-7}$ to $1.0$.
The adjacency matrix $A$ of each network was constructed as follows.  For each link, the corresponding $A_{ij}$ was set to the weight from the data if available and set to one otherwise.  Multiple links between the same pair of nodes were combined into a single link whose weight equals the sum of the weights of the original links.  For a self-link, the corresponding $A_{ii}$ was set based on the information from the data if available and $A_{ii} = 0$ otherwise.
The constructed adjacency matrices were then used to compute the nonnormality measure $\lVert D \rVert_\text{F}$, reactivity measure $\lambda_{\Delta}(A)$, degree imbalance $\langle (d^\text{in} - d^\text{out})^2 \rangle$, eigenvector angle $\theta_1$, and other properties of the networks.
Of these $251$ networks, $38$ had degenerate or nearly degenerate $\lambda_1(A)$, which we numerically identified using the criterion $\theta_1>1.57$ (recalling that $\theta_1 \le \pi/2$ by definition and that $\theta_1 = \pi/2$ when $\lambda_1(A)$ is degenerate).

The minimum level of nonnormality observed among the $251$ networks in Fig.~\ref{fig:real_networks}A was for a $8{,}192$-node network of interactions in a square dielectric waveguide, with $\norm{D}_\text{F} / (\sqrt{n}\langle w^2 \rangle) \approx 0.032$.
The only non-reactive network in the data set was one representing child-parent relationships from an online genealogical website called WikiTree, for which $\lambda_\Delta(A)$ and $\theta_1$ are both estimated to be zero to machine precision. Among the reactive networks that are non-degenerate, shown in Fig.~\ref{fig:real_networks}B, the minimum value of $\lambda_\Delta(A)/\lambda_1(A) \approx 3.1 \times 10^{-6}$ was observed for the $3{,}140$-node network of inter-county migration in the U.S.

\phantomsection
\addcontentsline{toc}{subsection}{Directed random network model}
\subsection*{Directed random network model}
To sample a network realization of size $n$ from the GCLV model, we first draw the expected in-degree $\tilde{d}_i^{\text{in}} \ge 1$ and out-degree $\tilde{d}_i^{\text{out}} \ge 1$ of different nodes $i=1,\ldots,n$ independently from a common joint probability distribution that does not depend on the network size $n$ and has finite second moments.
For each $i$, the random variables $\tilde{d}_i^{\text{in}}$ and $\tilde{d}_i^{\text{out}}$ are not restricted to be integers and need not to be independent (and thus can be correlated).
Following Ref.~\citen{Chung:2003}, the actual in- and out-degrees $d_i^{\text{in}}$ and $d_i^{\text{out}}$ of the network are then determined as a result of randomly creating a directed link from node $j$ to node $i$ with the probability
\begin{equation}\label{eqn:def_rho}
\mathbb{P}(\text{link $j \rightarrow i$ exists} \,\vert\, \tilde{d}^\text{in}_1, \ldots, \tilde{d}^\text{in}_n, \tilde{d}^\text{out}_1, \ldots, \tilde{d}^\text{out}_n)
= \rho_{ij} := \frac{\tilde{d}_i^{\text{in}}\tilde{d}_j^{\text{out}}}{\sum_{k=1}^{n}\tilde{d}_k^{\text{in}}}
\end{equation}
for each $i$ and $j$.
We then independently draw random weights $A_{ij}>0$ for the resulting links from a given distribution, while we set $A_{ij}=0$ if node $j$ is not connected to node $i$.
Note that this process can create self-links with random weights, but we could choose to prohibit them by setting $A_{ii} = 0$ for all $i$.
The unweighted version of this model is obtained if we instead set $A_{ij}=1$ for all links.

For the (marginal) distributions of $\tilde{d}_i^{\text{in}}$ and $\tilde{d}_i^{\text{out}}$, we assume that $\rho_{ij} \le c$ for some constant $\frac{1}{2} \le c < 1$ and that the extreme values $\tilde{d}_{\text{max}}^{\text{in}} := \underset{1\leq i \leq n}{\text{max}} \tilde{d}_i^{\text{in}}$ and $\tilde{d}_{\text{max}}^{\text{out}} := \underset{1\leq i \leq n}{\text{max}} \tilde{d}_i^{\text{out}}$ asymptotically follow the so-called generalized extreme value (GEV) distributions \cite{de2007extreme} after appropriate normalization (see below for more details).
We note that the first assumption is only a slight addition to the second, as the second implies that $\rho_{ij} \to 0$ in probability as $n \to \infty$ (see supplementary text, Sec.~\ref{sec-si-conv}, for a proof).
We also assume that $\mathbb{E}(\tilde{d}_i^{\text{in}}) = \mathbb{E}(\tilde{d}_i^{\text{out}})$, so that we have $\underset{n\to\infty}{\lim} \bigl( \sum_{i=1}^{n}\tilde{d}_i^{\text{in}} - \sum_{i=1}^{n}\tilde{d}_i^{\text{out}} \bigr)/n = 0$ almost surely (which can be shown using the strong law of large numbers).
Denoting the mean of the expected in- or out-degree by $\tilde{d} := \mathbb{E}(\tilde{d}_i^{\text{in}}) = \mathbb{E}(\tilde{d}_i^{\text{out}})$ (which does not depend on $i$), we assume $\tilde{d}>1$.
This holds true because $\tilde{d}_i^{\text{in}}, \tilde{d}_i^{\text{out}} \ge 1$ unless the joint distribution of $\tilde{d}_i^{\text{in}}$ and $\tilde{d}_i^{\text{out}}$ is singular with all probability density concentrated at $\tilde{d}_i^{\text{in}}=\tilde{d}_i^{\text{out}}=1$.
These assumptions guarantee that, with probability one, the means of the actual in-degrees $d_i^{\text{in}}$ and out-degrees $d_i^{\text{out}}$ over the Bernoulli distribution of $A_{ij}$ indeed match $\tilde{d}_i^{\text{in}}$ and $\tilde{d}_i^{\text{out}}$, respectively, in the limit of $n\to\infty$.
Thus, the given joint distribution of $\tilde{d}_i^{\text{in}}$ and $\tilde{d}_i^{\text{out}}$ can be interpreted as the expected degree distribution for this random network model.

For $\tilde{d}_{\text{max}}^{\text{in}}$, the GEV assumption mentioned above is more precisely described as follows: 
there exist sequences of normalization constants $a_n>0$ and $b_n$ such that 
\begin{equation}\label{eqn:profkn}
\lim_{n\to\infty}\mathbb{P}\biggl( \frac{\tilde{d}_{\max}^{\text{in}} - b_n}{a_n} \le x \biggr) =  G_{\gamma}(x)
\end{equation}
and $\underset{n\to\infty}{\lim}a_n/n = \underset{n\to\infty}{\lim}b_n/n = 0$.
Here, the GEV cumulative distribution function $G_{\gamma}(x)$ is given as follows:
\begin{equation}
\begin{split}
\text{For $\gamma<0$:} \quad &G_{\gamma}(x) = \begin{cases}
\exp\bigl( -(1+\gamma x)^{-1/\gamma} \bigr), & x \le -1/\gamma, \\
1, & x > -1/\gamma,
\end{cases} \\
\text{For $\gamma=0$:} \quad &G_\gamma(x) = \exp(-e^{-x}) \text{ for all $x$}, \\
\text{For $\gamma>0$:} \quad & G_{\gamma}(x) = \begin{cases}
0, & x \le -1/\gamma, \\
\exp\bigl( -(1+\gamma x)^{-1/\gamma} \bigr), & x > -1/\gamma.
\end{cases} \\
\end{split}
\end{equation}
The parameter $\gamma$ is called the extreme value index and is determined solely by the distribution of $\tilde{d}_i^{\text{in}}$.
For $\tilde{d}_{\text{max}}^{\text{out}}$, we assume the same, but possibly with different $a_n$, $b_n$, and $\gamma$.
The GEV distributions are the only ones that can arise as the limit distribution of the maximum of independent and identically distributed random variables if the limit in Eq.~\eqref{eqn:profkn} exists, and they include the well-known Gumbel, Fr\'echet, and Weibull distribution as special cases.

The GEV assumption is mild and is satisfied for almost all commonly encountered distributions.  It can be verified using the explicit conditions for Eq.~\eqref{eqn:profkn} given in Ref.~\citen{de2007extreme}.
For example, one of these conditions (Theorem 1.1.8) shows that it is satisfied by any power-law distribution with exponent $\beta>2$, minimum expected degree $\tilde{d}_{\text{min}} \ge 1$, and the density function given by $p(x) = C x^{-\beta}$ if $x \ge \tilde{d}_{\text{min}}$, and $p(x) = 0$ otherwise, where $C = (\beta-1)\tilde{d}_{\text{min}}^{\beta-1}$ is the normalization constant.
The extreme value index in this case is $\gamma=1/(\beta-1)$, and the sequences of constants in Eq.~\eqref{eqn:profkn} are $a_n = n^\gamma \gamma \tilde{d}_{\text{min}}$ and $b_n = n^\gamma \tilde{d}_{\text{min}}$.
Since $\beta>2$ (which is also the condition for this distribution to have a finite mean), we have $\gamma<1$, implying $\underset{n\to\infty}{\lim}a_n/n = \underset{n\to\infty}{\lim}\tilde{d}_{\min}\gamma/n^{1-\gamma} = 0$ and $\underset{n\to\infty}{\lim}b_n/n = \underset{n\to\infty}{\lim}\tilde{d}_{\min}/n^{1-\gamma} = 0$.

\phantomsection
\addcontentsline{toc}{subsection}{Nonnormality approximation for unweighted networks}
\subsection*{Nonnormality approximation for unweighted networks}
Our derivation of Eq.~\eqref{eqn:D_F_theory} is for large random networks generated by the unweighted version of the GCLV model.
Since Eq.~\eqref{eqn:D_F_theory} expresses $\lVert D \rVert_\text{F}^2$ as a function of the in-degrees $d^\text{in}_i$ and out-degrees $d^\text{out}_i$, the derivation is based on calculating the conditional expected value
$\mathbb{E}\bigl( \lVert D \rVert_\text{F}^2 \,\big\vert\, d^\text{in}_1, \ldots, d^\text{in}_n, d^\text{out}_1, \ldots, d^\text{out}_n \bigr)$, which we denote simply by $\mathbb{E}(\lVert D \rVert_\text{F}^2)$ in this section (in which we also use similar simplified notations for other expected values, variances, and probabilities).
Taking the (conditional) expected value of Eq.~\eqref{eqn:D_F_def}, we obtain
\begin{equation}\label{eqn:DF_deriv_1}
\begin{aligned}
\mathbb{E}\bigl(\lVert D \rVert_\text{F}^2\bigr) 
&= \sum_{i=1}^n (d_i^\text{in} - d_i^\text{out})^2 + \mathbb{E}\Bigl( \mathop{\sum\sum}_{i \neq j} (d_{ij}^\text{in} - d_{ij}^\text{out})^2 \Bigr)\\
& = n \langle (d^\text{in} - d^\text{out})^2 \rangle + \mathop{\sum\sum}_{i \neq j} \mathbb{E}(Y_{ij}^2),
\end{aligned}
\end{equation}
where we define $Y_{ij} := d_{ij}^\text{in} - d_{ij}^\text{out} = \sum_{k\neq i,j}(A_{ik}A_{jk}-A_{ki}A_{kj})$ (recalling the assumption $A_{ii}=0$, which will be relaxed in the next section).
We note that, while $d_i^\text{in}$ and $d_i^\text{out}$ are held fixed, $d_{ij}^\text{in}$ and $d_{ij}^\text{out}$ are random.

We now seek to estimate $\mathbb{E}(Y_{ij}^2)$ in Eq.~\eqref{eqn:DF_deriv_1} by calculating $\mathbb{E}(Y_{ij})$ and $\text{Var}(Y_{ij})$, which can be broken down to individual terms $\mathbb{E}(A_{ik}A_{jk})$ and $\mathbb{E}(A_{ki}A_{kj})$.
We first note that $\mathbb{E}(A_{ik}A_{jk})
= \mathbb{P}(A_{ik}A_{jk}=1) = \mathbb{P}(A_{ik}=1 \text{ and } A_{jk}=1)$.
To compute this probability, we define
\begin{equation}
Q_{i,n}^\text{in}(d) := \mathbb{P} \biggl( \sum_{k=1}^n A_{ik} = d \biggr)
\quad\text{and}\quad
Q_{j,n}^\text{out}(d) := \mathbb{P} \biggl( \sum_{k=1}^n A_{kj} = d \biggr),
\end{equation}
which represents the probability that the in-degree of node $i$ is $d$ and the probability that the out-degree of node $j$ is $d$, respectively.
Assuming the random variable $\tilde{d}_i^\text{in}$ fixed for the moment, we have
\begin{equation}\label{eqn:13}
\begin{aligned}
Q_{i,n}^\text{in}(d)
&= \sum_\alpha \prod_{k=1}^n  \rho_{ik}^{\alpha_k} (1-\rho_{ik})^{1-\alpha_k}\\
&= \sum_\alpha \biggl(\,\prod_{k:\,\alpha_k =1} \rho_{ik} \biggr)\biggl(\,\prod_{k:\,\alpha_k =0}  (1-\rho_{ik}) \biggr)\\
&\approx \sum_\alpha \biggl(\,\prod_{k:\,\alpha_k =1} \frac{\tilde{d}_i^{\text{in}}\tilde{d}_k^{\text{out}}}{n\overline{d}} \biggr) \exp \biggl(-\sum_{k:\,\alpha_k =0} \frac{\tilde{d}_i^{\text{in}}\tilde{d}_k^{\text{out}}}{n\overline{d}} \biggr)\\
&\approx \binom{n}{d} \frac{(\tilde{d}_i^{\text{in}})^d}{n^d\overline{d}^d} \cdot \frac{1}{\binom{n}{d}}\sum_\alpha \biggl(\,\prod_{k:\,\alpha_k =1} \tilde{d}_k^{\text{out}} \biggr) \exp \biggl(- \frac{\tilde{d}_i^{\text{in}}}{\overline{d}} \cdot \frac{1}{n}\sum_{k:\,\alpha_k = 0} \tilde{d}_k^{\text{out}} \biggr)\\
&\approx \binom{n}{d} \cdot \frac{(\tilde{d}_i^\text{in})^d}{n^d} \cdot \exp(-\tilde{d}_i^\text{in}) ,
\end{aligned}
\end{equation}
where $\sum_\alpha$ denotes the summation over all $\alpha = \{\alpha_1, \ldots, \alpha_n\} \in \{0,1\}^n$ in which exactly $d$ elements are equal to one, and $k:\alpha_k = 0, 1$ denotes all $k$ such that $\alpha_k = 0, 1$, respectively. Equation~\eqref{eqn:13} also used the following approximations for large network size $n$ (with $d$ fixed): $1-\rho_{ik} \approx e^{-\rho_{ik}}$ valid when $\rho_{ik}$ is small (which is the case because $\rho_{ik} \to 0$ in probability as $n\rightarrow \infty$; see supplementary text, Sec.~\ref{sec-si-conv}); $\frac{1}{n}\sum_k \tilde{d}_k^\text{in} \approx \overline{d}$ and $\frac{1}{n}\sum_{k:\,\alpha_k = 0} \tilde{d}_k^\text{out} \approx \frac{1}{n}\sum_k \tilde{d}_k^\text{out} \approx \overline{d}$ (based on the strong law of large numbers); and $\sum_\alpha \prod_{k:\,\alpha_k =1} \tilde{d}_k^{\text{out}} / \binom{n}{d} \approx \overline{d}^d$ (a special case of the theorem proved in Ref.~\citen{halasz1976elementary}).
Analogously, holding $\tilde{d}_j^\text{out}$ fixed temporarily, we have
\begin{equation}
Q_{j,n}^\text{out}(d)
\approx \binom{n}{d} \cdot \frac{(\tilde{d}_j^\text{out})^d}{n^d} \cdot \exp(-\tilde{d}_j^\text{out}) ,
\end{equation}
and similar arguments also show that
\begin{equation}
\begin{split}
\mathbb{P} \biggl( \sum_{k \neq j} A_{ik} = d \biggr) = Q_{i,n-1}^\text{in}(d),\quad
\mathbb{P} \biggl(\, \sum_{k \neq j_1,j_2} A_{ik} = d \,\biggr) = Q_{i,n-2}^\text{in}(d),\\
\mathbb{P} \biggl( \sum_{k \neq i} A_{kj} = d \biggr) = Q_{j,n-1}^\text{out}(d),\quad
\mathbb{P} \biggl(\, \sum_{k \neq i_1,i_2} A_{kj} = d \,\biggr) = Q_{j,n-2}^\text{out}(d).
\end{split}
\end{equation}

We can now calculate $\mathbb{E}(A_{ik}A_{jk})$ and $\mathbb{E}(A_{ki}A_{kj})$ from the definition of the conditional expected values:
\begin{equation}\label{eqn:EAikpro}
\begin{aligned}
\mathbb{E}(A_{ik}A_{jk})
&= \mathbb{P}(A_{ik}=1 \text{ and } A_{jk}=1 \,\big|\, d_i^\text{in}, d_j^\text{in}, d_k^\text{out}) \\
&= \frac{\rho_{ik}\rho_{jk} Q_{i,n-1}^\text{in}(d_i^\text{in}-1) \, Q_{j,n-1}^\text{in}(d_j^\text{in}-1) \, Q_{k,n-2}^\text{out}(d_k^\text{out}-2)}{Q_{i,n}^\text{in}(d_i^\text{in}) \, Q_{j,n}^\text{in}(d_j^\text{in}) \, Q_{k,n}^\text{out}(d_k^\text{out})} \\
&\approx \frac{d_{i}^\text{in}d_{j}^\text{in}d_k^\text{out}(d_k^\text{out}-1)}{(\overline{d}n)^2}, \\
\mathbb{E}(A_{ki}A_{kj})
&\approx \frac{d_{i}^\text{out}d_{j}^\text{out}d_k^\text{in}(d_k^\text{in}-1)}{(\overline{d}n)^2}.
\end{aligned}
\end{equation}
We note that these expressions do not depend on $\tilde{d}_i^\text{in}$ nor $\tilde{d}_i^\text{out}$, and hence the estimates are valid even when these variables are allowed to be random.
We thus have
\begin{equation}\label{eqn:EvarAij}
\begin{aligned}
\mathbb{E}(A_{ik}A_{jk}-A_{ki}A_{kj}) &= \mathbb{E}(A_{ik}A_{jk}) - \mathbb{E}(A_{ki}A_{kj}) \\
&\approx \frac{d_{i}^\text{in}d_{j}^\text{in}d_k^\text{out}(d_k^\text{out}-1)}{(\overline{d}n)^2} - \frac{d_{i}^\text{out}d_{j}^\text{out}d_k^\text{in}(d_k^\text{in}-1)}{(\overline{d}n)^2}, \\
\text{Var}(A_{ik}A_{jk}-A_{ki}A_{kj}) &=\text{Var} (A_{ik}A_{jk}) + \text{Var}(A_{ki}A_{kj}) \\
&= \mathbb{E}(A_{ik}A_{jk})-\mathbb{E}(A_{ik}A_{jk})^2 + \mathbb{E}(A_{ki}A_{kj})-\mathbb{E}(A_{ki}A_{kj})^2\\
&\approx  \mathbb{E}(A_{ik}A_{jk}) + \mathbb{E}(A_{ki}A_{kj}) \\
&\approx \frac{d_{i}^\text{in}d_{j}^\text{in}d_k^\text{out}(d_k^\text{out}-1)}{(\overline{d}n)^2} + \frac{d_{i}^\text{out}d_{j}^\text{out}d_k^\text{in}(d_k^\text{in}-1)}{(\overline{d}n)^2},
\end{aligned}
\end{equation}
where we used that the high-order terms $\mathbb{E}(A_{ik}A_{jk})^2$ and $\mathbb{E}(A_{ki}A_{kj})^2$ are negligible for large $n$.
Summing these over $k$ (including the terms $k=i,j$, which are small compared to other terms and hence do not affect the sums for large $n$), we have
\begin{align}
\mathbb{E}(Y_{ij}) &\approx \frac{1}{n} \biggl[ \frac{d_{i}^\text{in}d_{j}^\text{in} \bigl( \langle ( d^\text{out})^2 \rangle - \overline{d} \,\bigr) }{\overline{d}^2} - \frac{d_{i}^\text{out}d_{j}^\text{out} \bigl( \langle ( d^\text{in})^2 \rangle - \overline{d} \,\bigr) }{\overline{d}^2} \biggr], \\
\text{Var}(Y_{ij}) &\approx \frac{1}{n} \biggl[ \frac{d_{i}^\text{in}d_{j}^\text{in} \bigl( \langle ( d^\text{out})^2 \rangle - \overline{d} \,\bigr) }{\overline{d}^2} + \frac{d_{i}^\text{out}d_{j}^\text{out} \bigl( \langle ( d^\text{in})^2 \rangle - \overline{d} \,\bigr) }{\overline{d}^2} \biggr].
\end{align}

The term $\mathbb{E}(Y_{ij}^2)$ in Eq.~\eqref{eqn:DF_deriv_1} can now be estimated for large $n$ as
\begin{equation}\label{eqn:EXij}
\mathbb{E}(Y_{ij}^2) = \text{Var}(Y_{ij}) + \mathbb{E}(Y_{ij})^2  \approx \text{Var}(Y_{ij}),
\end{equation}
since $\mathbb{E}(Y_{ij})^2$ is negligible compared to $\text{Var}(Y_{ij})$, and we have
\begin{equation}\label{eqn:sumEXij}
\begin{aligned}
\mathop{\sum\sum}_{i \neq j} \mathbb{E}(Y_{ij}^2) 
&\approx \frac{1}{n} \sum_{i=1}^n \sum_{j=1}^n \biggl[ \frac{d_{i}^\text{in}d_{j}^\text{in} \bigl( \langle ( d^\text{out})^2 \rangle - \overline{d} \,\bigr) }{\overline{d}^2} + \frac{d_{i}^\text{out}d_{j}^\text{out} \bigl( \langle ( d^\text{in})^2 \rangle - \overline{d} \,\bigr) }{\overline{d}^2} \biggr] \\
&= n[ \langle ( d^\text{in})^2 \rangle +\langle ( d^\text{out})^2 \rangle - 2\overline{d}],
 \end{aligned}
\end{equation}
(retaining the $i=j$ term, which is small compared to the sum of the other terms).
Substituting this into Eq.~\eqref{eqn:DF_deriv_1} yields Eq.~\eqref{eqn:D_F_theory}.

\phantomsection
\addcontentsline{toc}{subsection}{Nonnormality approximation for weighted networks}
\subsection*{Nonnormality approximation for weighted networks}
Here, we consider the weighted version of the GCLV model, assuming that the distribution of link weights squared has a finite mean $\mathbb{E}(w^2)$ and a finite variance $\text{Var}(w^2)$.  Assuming first that there are no self-links, i.e., $A_{ii} = 0$, we will show that the nonnormality $\lVert D \rVert_\text{F}$ can be approximated for large $n$ as
\begin{equation}\label{eqn:D_F_theory_weighted}
\lVert D \rVert_\text{F}^2 
\approx 2n \bigl[ \langle (d^\text{in} - d^\text{out})^2 \rangle + \langle d^\text{in} d^\text{out} \rangle  - \overline{d} \,\bigr]\mathbb{E}^2(w^2)+2n\bar{d}\text{Var}(w^2),
\end{equation}
which extends Eq.~\eqref{eqn:D_F_theory} to weighted networks.
When the link weight variance $\text{Var}(w^2)$ is small, this equation yields the scaling $\lVert D \rVert_\text{F} \sim \sqrt{n}\,\mathbb{E}(w^2) \approx \sqrt{n} \langle w^2 \rangle$.
As in the case of unweighted networks, the derivation is based on taking the conditional expectation given fixed values of $d^\text{in}_i$ and $d^\text{out}_i$.  The weighted version of Eq.~\eqref{eqn:DF_deriv_1} reads:
\begin{equation}\label{eqn:DF_deriv_weighted}
\begin{aligned}
\mathbb{E}\bigl(\lVert D \rVert_\text{F}^2\bigr) 
= \sum_{i=1}^n \mathbb{E}(Y_{ii}^2) + \mathop{\sum\sum}_{i \neq j} \mathbb{E}(Y_{ij}^2),
\end{aligned}
\end{equation}
where $Y_{ij} = \sum_{k\neq i,j}(A_{ik}A_{jk}-A_{ki}A_{kj})$ and thus $Y_{ii} = \sum_{k\neq i}(A_{ik}^2-A_{ki}^2)$. 
Following the derivation of Eq.~\eqref{eqn:EvarAij} while accounting for the random weights, we have
\begin{equation}\label{eqn:EAikpro2}
\mathbb{E}(A_{ik}A_{jk}-A_{ki}A_{kj})
\approx \biggl( \frac{d_{i}^\text{in}d_{j}^\text{in}d_k^\text{out}(d_k^\text{out}-1)}{(\overline{d}n)^2} - \frac{d_{i}^\text{out}d_{j}^\text{out}d_k^\text{in}(d_k^\text{in}-1)}{(\overline{d}n)^2} \biggr)\mathbb{E}(w^2),
\end{equation}
\begin{equation}
\begin{aligned}
\text{Var}(A_{ik}A_{jk}-A_{ki}A_{kj}) &\approx \mathbb{E}(A_{ik}^2A_{jk}^2) + \mathbb{E}(A_{ki}^2A_{kj}^2)\\
&\approx \biggl(  \frac{d_{i}^\text{in}d_{j}^\text{in}d_k^\text{out}(d_k^\text{out}-1)}{(\overline{d}n)^2} + \frac{d_{i}^\text{out}d_{j}^\text{out}d_k^\text{in}(d_k^\text{in}-1)}{(\overline{d}n)^2}  \biggr)(\mathbb{E}(w^2))^2,
\end{aligned}
\end{equation}
which lead to the weighted version of Eq.~\eqref{eqn:sumEXij}:
\begin{equation}\label{eqn:sumEijweighted}
    \mathop{\sum\sum}_{i \neq j} \mathbb{E}(Y_{ij}^2) \approx n[ \langle ( d^\text{in})^2 \rangle +\langle ( d^\text{out})^2 \rangle - 2\overline{d}](\mathbb{E}(w^2))^2.
\end{equation}
To approximate the first term on the right side of Eq.~\eqref{eqn:DF_deriv_weighted}, we note that
\begin{equation}\label{eqn:EAiksq2}
\begin{aligned}
\mathbb{E}(A_{ik}^2)
\approx \frac{d_{i}^\text{in}d_k^\text{out}}{\overline{d}n}\mathbb{E}(w^2), \ \ \ \ \ 
\mathbb{E}(A_{ki}^2)
\approx \frac{d_{i}^\text{out}d_k^\text{in}}{\overline{d}n}\mathbb{E}(w^2)
\end{aligned}
\end{equation}
and that there are exactly $d_{i}^\text{in}$ and $d_{i}^\text{out}$ nonzero terms in $\sum_{k\neq i}A_{ik}^2$ and $\sum_{k\neq i}A_{ki}^2$, respectively.  Thus, given fixed values of $d_{i}^\text{in}$ and $d_{i}^\text{out}$, the expectation and variance of $Y_{ii} = \sum_{k\neq i}(A_{ik}^2-A_{ki}^2)$ can be estimated as
\begin{equation}\label{eqn:EXii}
    \mathbb{E}(Y_{ii}) \approx (d_{i}^\text{in}-d_{i}^\text{out})\mathbb{E}(w^2),
\end{equation}
\begin{equation}\label{eqn:VarXii}
    \text{Var}(Y_{ii}) \approx (d_{i}^\text{in}+d_{i}^\text{out})\text{Var}(w^2),
\end{equation}
and hence we have
\begin{equation}\label{eqn:sumEiiweighted}
\mathbb{E}(Y_{ii}^2) = \mathbb{E}(Y_{ii})^2 + \text{Var}(Y_{ii})
\approx (d_{i}^\text{in}-d_{i}^\text{out})^2(\mathbb{E}(w^2))^2+(d_{i}^\text{in}+d_{i}^\text{out})\text{Var}(w^2).
\end{equation}
Combining Eqs.~\eqref{eqn:DF_deriv_weighted}, \eqref{eqn:sumEijweighted}, and \eqref{eqn:sumEiiweighted} and approximating $\lVert D \rVert_\text{F}^2$ by its conditional expectation $\mathbb{E}\bigl(\lVert D \rVert_\text{F}^2\bigr)$ yield Eq.~\eqref{eqn:D_F_theory_weighted}.

The above results can be further extended to allow for any given assignment of self-links.  Assume that $\hat{A} = A + \text{diag}(\beta)$ with diagonal elements $\beta_i$, $i=1,2,\cdots,n$. Then $Y_{ij} = \sum_{k\neq i,j}(A_{ik}A_{jk}-A_{ki}A_{kj}) - (A_{ij}-A_{ji})(\beta_i - \beta_j)$ and $Y_{ii}$ remains unchanged. Note that
\begin{equation}
\begin{aligned}\label{eqn:EX_ij2}
    \mathbb{E}({Y}_{ij}^2) \approx \text{Var}({Y}_{ij}) &= \text{Var}\Bigl(\sum_{k\neq i,j}(A_{ik}A_{jk}-A_{ki}A_{kj})\Bigr) + \text{Var}\bigl((A_{ij}-A_{ji})(\beta_i-\beta_j)\bigr) \\
    &\approx \biggl[ \frac{d_{i}^\text{in}d_{j}^\text{in} \bigl( \langle ( d^\text{out})^2 \rangle - \overline{d} \,\bigr) }{\overline{d}^2} + \frac{d_{i}^\text{out}d_{j}^\text{out} \bigl( \langle ( d^\text{in})^2 \rangle - \overline{d} \,\bigr) }{\overline{d}^2} \biggr] \cdot \frac{\mathbb{E}^2(w^2)}{n}\\
    & \ \ \ \ \ + (\beta_i - \beta_j)^2 \cdot \frac{d_{i}^\text{in}d_{j}^\text{out}+d_{j}^\text{in}d_{i}^\text{out}}{\bar{d}}\cdot \frac{\mathbb{E}(w^2)}{n}.
\end{aligned}
\end{equation}
Therefore, we have
\begin{equation}\label{eqn:DA_weighted}
\begin{aligned}
\norm{D(\hat{A})}_\text{F}^2 &\approx 2n \bigl[ \langle (d^\text{in} - d^\text{out})^2 \rangle + \langle d^\text{in} d^\text{out} \rangle  - \overline{d} \,\bigr]\mathbb{E}^2(w^2)+2n\bar{d}\,\text{Var}(w^2)\\
&\quad + 2n\mathbb{E}(w^2)[ 
\langle \beta^2 d^\text{in}  \rangle + 
\langle \beta^2  d^\text{out} \rangle -
\frac{2}{\bar{d}}\langle \beta d^\text{in}  \rangle \langle \beta d^\text{out} \rangle
] \\
&= \norm{D(A)}_\text{F}^2 + 2n\mathbb{E}(w^2)[ 
\langle \beta^2 d^\text{in}  \rangle + 
\langle \beta^2  d^\text{out} \rangle -
\frac{2}{\bar{d}}\langle \beta d^\text{in}  \rangle \langle \beta d^\text{out} \rangle
].
\end{aligned}
\end{equation}

\phantomsection
\addcontentsline{toc}{subsection}{Extension of the nonnormality approximation to Laplacian-coupled networks}
\subsection*{Extension of the nonnormality approximation to Laplacian-coupled networks}
Here, we consider the nonnormality of the Laplacian matrices of networks generated by the weighted GCLV model under the same conditions as in the derivation of Eq.~\eqref{eqn:D_F_theory_weighted}. The Laplacian matrix of a network with adjacency matrix $A$ is defined by $L := K - A$, where $K$ denotes the diagonal matrix with $K_{ii} = \sum_{k} A_{ik}$. A straightforward calculation yields
\begin{equation}
 \norm{D(L)}_\text{F}^2=\norm{L^TL-LL^T}_\text{F}^2 = \sum_{i=1}^n \mathbb{E}(\tilde{Y}_{ii}^2) + \mathop{\sum\sum}_{i \neq j} \mathbb{E}(\tilde{Y}_{ij}^2),
\end{equation}
where $\tilde{Y}_{ij} = \sum_{k\neq i,j}(A_{ik}A_{jk}-A_{ki}A_{kj})+(A_{ij}-A_{ji})(d_i^\text{in}-d_j^\text{in})$ and, in particular, $\tilde{Y}_{ii} = \sum_{k\neq i}(A_{ik}^2-A_{ki}^2)$. Following Eq.~\eqref{eqn:sumEiiweighted}, we have 
\begin{equation}
    \mathbb{E}(\tilde{Y}_{ii}^2) \approx (d_{i}^\text{in}-d_{i}^\text{out})^2(\mathbb{E}(w^2))^2+(d_{i}^\text{in}+d_{i}^\text{out})\text{Var}(w^2).
\end{equation}
In addition, following Eqs.~\eqref{eqn:EXij} and \eqref{eqn:sumEijweighted}, we obtain
\begin{equation}
\begin{aligned}
    \mathbb{E}(\tilde{Y}_{ij}^2) \approx \text{Var}(\tilde{Y}_{ij}) &= \text{Var}\Bigl(\sum_{k\neq i,j}(A_{ik}A_{jk}-A_{ki}A_{kj})\Bigr) + \text{Var}\bigl((A_{ij}-A_{ji})(d_i^\text{in}-d_j^\text{in})\bigr) \\
    &\approx \frac{\mathbb{E}^2(w^2)}{n} \biggl[ \frac{d_{i}^\text{in}d_{j}^\text{in} \bigl( \langle ( d^\text{out})^2 \rangle - \overline{d} \,\bigr) }{\overline{d}^2} + \frac{d_{i}^\text{out}d_{j}^\text{out} \bigl( \langle ( d^\text{in})^2 \rangle - \overline{d} \,\bigr) }{\overline{d}^2} \biggr]\\
    & \ \ \ \ \ + (d_i^\text{in} - d_j^\text{in})^2 \cdot \frac{d_{i}^\text{in}d_{j}^\text{out}+d_{j}^\text{in}d_{i}^\text{out}}{n \bar{d}}\cdot \mathbb{E}(w^2).
\end{aligned}
\end{equation}
Therefore, we have
\begin{equation}\label{eqn:D_F_theory_laplacian}
\begin{aligned}
\norm{D(L)}_\text{F}^2 &\approx 2n \bigl[ \langle (d^\text{in} - d^\text{out})^2 \rangle + \langle d^\text{in} d^\text{out} \rangle  - \overline{d} \,\bigr]\mathbb{E}^2(w^2)+2n\bar{d}\,\text{Var}(w^2)\\
&\quad + 2n\mathbb{E}(w^2)[ 
\langle ( d^\text{in} )^3 \rangle + 
\langle ( d^\text{in} )^2  d^\text{out} \rangle -
\frac{2}{\bar{d}}\langle ( d^\text{in} )^2 \rangle \langle d^\text{in} d^\text{out} \rangle
] \\
&= \norm{D(A)}_\text{F}^2 + 2n\mathbb{E}(w^2)[ 
\langle ( d^\text{in} )^3 \rangle + 
\langle ( d^\text{in} )^2  d^\text{out} \rangle -
\frac{2}{\bar{d}} \langle ( d^\text{in} )^2 \rangle\langle d^\text{in} d^\text{out} \rangle].
\end{aligned}
\end{equation}

\phantomsection
\addcontentsline{toc}{subsection}{Reactivity approximation for networks with dominant largest eigenvalue}
\subsection*{Reactivity approximation for networks with dominant largest eigenvalue}
Here, we derive Eq.~\eqref{eqn:lambda_delta_theory} assuming $\lambda_1(H) \approx \lambda_1(H_1)$ and the non-degeneracy of $\lambda_1(A)$.
We normalize the left and right eigenvectors associated with $\lambda_1(A)$, so that we have $v_1^T v_1 = 1$ and $u_1^T v_1 = v_1^T u_1 = 1$, and thus the angle $\theta_1$ between the left and right eigenvectors is given by $\cos\theta_1 = 1/\sqrt{u_1^T u_1}$.
Noting that the symmetric part of $A_1$ can be written as $H_1 = \frac{1}{2}\lambda_1(A)(u_1 v_1^T + v_1 u_1^T)$ and approximating the eigenvector associated with its largest eigenvalue by a linear combination $\alpha u_1 + \beta v_1$, we have
\begin{equation}\label{eqn:lambda1H}
\begin{aligned}
\lambda_1(H) &\approx \lambda_1(H_1)
= \max_{x\in \mathbb{R}^n, \, x \neq 0} \frac{x^T H_1 x}{x^T x} \\
&= \frac{1}{2}\,\lambda_1(A) \max_{\alpha^2 + \beta^2 \neq 0} \frac{(\alpha u_1 + \beta v_1)^T(u_1 v_1^T+v_1 u_1^T)(\alpha u_1 + \beta v_1)}{(\alpha u_1 + \beta v_1)^T (\alpha u_1 + \beta v_1)} \\
&= \lambda_1(A) \max_{y \in \mathbb{R}^2, \, y \neq 0} \frac{y^T \Phi y}{y^T \Psi y},
\end{aligned}
\end{equation}
where we defined
\begin{align}
\Phi &:= \Phi(\theta_1) =
\begin{bmatrix}
1 & \frac{1}{2}u_1^T u_1 + \frac{1}{2}\\
\frac{1}{2} u_1^T u_1 + \frac{1}{2} & u_1^T u_1
\end{bmatrix}
=
\begin{bmatrix}
1 & \frac{1}{2\cos^2\theta_1}+\frac{1}{2}\\
\frac{1}{2\cos^2\theta_1}+\frac{1}{2} & \frac{1}{\cos^2\theta_1}
\end{bmatrix}, \\
\Psi &:= \Psi(\theta_1) =
\begin{bmatrix}
1 & 1\\
1 & u_1^T u_1
\end{bmatrix}
=
\begin{bmatrix}
1 & 1\\
1 & \frac{1}{\cos^2\theta_1}
\end{bmatrix}.
\end{align}
The last maximum in Eq.~\eqref{eqn:lambda1H} can be computed as the largest generalized eigenvalue $\mu$ of the matrix pencil $(\Phi(\theta_1),\Psi(\theta_1))$ and satisfies the equation
\begin{equation}
\det\bigl(\Phi(\theta_1) - \mu \Psi(\theta_1)\bigr)=0.
\end{equation}
This equation can be explicitly solved to yield $\mu = \frac{1+\cos\theta_1}{2\cos\theta_1}$.
Substituting this into Eq.~\eqref{eqn:lambda1H} and using the definition of $\lambda_{\Delta}$, we obtain Eq.~\eqref{eqn:lambda_delta_theory}:
\begin{equation}
\lambda_{\Delta} 
= \lambda_1(H) - \lambda_1(A)
\approx \lambda_1(A) \cdot \frac{1+\cos\theta_1}{2\cos\theta_1} - \lambda_1(A)
= \lambda_1(A) \cdot \frac{1-\cos\theta_1}{2\cos\theta_1}.
\end{equation}

\phantomsection
\addcontentsline{toc}{subsection}{Random networks used in Fig.~\ref{fig:theory_validation}}
\subsection*{Random networks used in Fig.~\ref{fig:theory_validation}}
To generate these networks, we used the GCLV model with a given correlated identical distributions of the expected in-degree $\tilde{d}_i^{\text{in}}$ and the expected out-degree $\tilde{d}_i^{\text{out}}$ for each node $i$.
To realize such a joint distribution, we first drew $\tilde{d}_i^{\text{in}}$ and $\tilde{d}_i^{\text{out}}$ (independently for each node $i$) from a common distribution, which
was either \textit{i}) the gamma distribution, with probability density function $p(x) \sim (x - d_\text{min})^{a-1}e^{-b(x - d_\text{min})}$, $x>d_\text{min}$, where $d_\text{min}$ is the minimum degree, $a>0$ is the shape parameter, and $b>0$ is the rate parameter;
or \textit{ii}) a discrete power-law distribution with the probability mass function $p(x) \sim x^{-\beta}$ for integers $x = d_\text{min}, d_\text{min}+1, \ldots, d_\text{max}$, where $\beta>0$ is the scaling exponent, $d_\text{min}$ is the minimum degree, and $d_\text{max} = \sqrt{cn d_\text{min}}$ is the maximum degree imposed to ensure that the model assumption $\rho_{ij} \le c$ is satisfied (and here we set $c=0.99$).
For the gamma distribution, the mean degree $d$ and the parameter $1/b$ were drawn randomly from the intervals $[10,50]$ and $[2,10]$, respectively, and the parameter $a$ was then set to be $a = b(d - d_\text{min})$ (to ensure that the mean degree equals $d$).
For the power-law distribution, the scaling exponent $\beta$ was drawn randomly from the interval $[2,4]$. For both distributions, we used the minimum degree $d_\text{min}=10$.

After generating $\tilde{d}_i^{\text{in}}$ and $\tilde{d}_i^{\text{out}}$, correlation was added between them using
a parameter $\rho$ randomly chosen from the interval $[-1,1]$ to increase the range of $\lVert D \rVert_\text{F}$ and $\lambda_\Delta(A)$ observed.
The correlation was created by sorting $\tilde{d}_i^{\text{in}}$ and $\tilde{d}_i^{\text{out}}$ for a randomly chosen subset of nodes (with mean fraction $\lvert\rho\rvert$) in the same order for $\tilde{d}_i^{\text{in}}$ and $\tilde{d}_i^{\text{out}}$ if $\rho>0$ (leading to positive correlation) and in the opposite order if $\rho<0$ (leading to negative correlation).

For the weighted networks, the random weights were drawn either from the exponential distribution with mean $\mu$ and variance $\mu^2$ (for $\mu = 0.7$, $1.0$, or $1.3$) or from the positive domain of the Gaussian distribution with mean one and variance $\sigma^2$ (for $\sigma = 0.1$ or $0.2$).  For this figure, we allowed self-links in the GCLV model (with random weights drawn from the same distribution as the other links).

\phantomsection
\addcontentsline{toc}{subsection}{Randomization of real networks in Fig.~\ref{fig_scaling_reactivity_stability}A}
\subsection*{Randomization of real networks in Fig.~\ref{fig_scaling_reactivity_stability}A}
For a given network from the data set, we generated its randomization using the GCLV model.
The in- and out-degrees of node $i$ in the network were used as the expected degrees $\tilde{d}_i^{\text{in}}$ and $\tilde{d}_i^{\text{out}}$, respectively, to generate a random network topology.
Note that, for nodes $i$ and $j$ with the right side of Eq.~\eqref{eqn:def_rho} exceeding one, we set the connection probability $\rho_{ij} = 1$.
The link weights of the network were re-sampled (with replacement) to generate random weights.

\phantomsection
\addcontentsline{toc}{subsection}{Determination of the shaded region in Fig.~\ref{fig_scaling_reactivity_stability}C}
\subsection*{Determination of the shaded region in Fig.~\ref{fig_scaling_reactivity_stability}C}
First, for each $\beta$, we computed the averages of $\lambda_1(H)$ and $\lambda_1(A)$ over $10$ realizations of the GCLV model for $15$ values of $n$, equally spaced on the logarithmic scale between $n=10^2$ and $n=10^5$.
We then performed a least-squares fit of these averages as functions of $n$ on the logarithmic scale for $n \ge 300$ and used the resulting slopes as the scaling exponents for $\lambda_1(H)$ and $\lambda_1(A)$ for the given $\beta$ (shown in the background by blue and green dots, respectively).
The boundary curves were obtained by fitting these scaling exponents with seventh-order polynomials.
The shaded area between these two curves thus represents the parameter region in which the scaling exponent $\ell$ for $\alpha(n)$ is smaller than that for $\lambda_1(H)$ but larger than that for $\lambda_1(A)$, implying that, as $n$ increases, $\lambda_1(H)-\alpha(n)$ increases while $\lambda_1(A)-\alpha(n)$ decreases.

\phantomsection

\phantomsection
\addcontentsline{toc}{section}{Acknowledgments}
\bigskip\noindent{\bf\large Acknowledgments}

\noindent
\textbf{Funding:} This research was supported by ARO Grant No.~W911NF-19-1-0383. D.E.\ also acknowledges support from TÜBİTAK Grant No.~119F125.
\textbf{Author contributions:} All authors contributed to the design of the research.  C.D., T.N., and D.E.\ processed the network data and performed the simulations.  C.D., T.N., and A.E.M.\ led the modeling, analyzed the results, and wrote the paper.  All authors approved the final manuscript.
\textbf{Competing interests:} The authors declare no competing interests.
\textbf{Data and materials availability:} All data needed to evaluate the conclusions in the paper are present in the paper, in the Supplementary Materials, and/or at \url{https://doi.org/10.5281/zenodo.5964372}.

\bigskip\noindent{\bf\large List of supplementary materials}

\noindent
Supplementary Text\\
Figs.~S1 to S5\\
Table S1

\clearpage

\clearpage
\baselineskip18pt

\setcounter{page}{1}

\stepcounter{myequation}
\renewcommand{\theequation}{S\arabic{equation}}

\begin{center}

\vspace{36pt}{\Large Supplementary Materials for}\\[12pt]

{\bf\large Network structural origin of instabilities in large complex systems}

\vspace{6pt}Chao Duan, Takashi Nishikawa$^*$, Deniz Eroglu, and Adilson E. Motter

\vspace{6pt}$^*$Corresponding author. Email: tnishik21@gmail.com

\end{center}

\vspace{5mm}

\noindent{\bf This PDF file includes:}

\vspace{6pt}Supplementary Text

\vspace{-3pt}Figs.~S1 to S5

\vspace{-3pt}Table S1

\pagebreak

\noindent{\Large\bf Contents}\\[6pt]
{\bf Supplementary Text}\\[6pt]
\rule{12pt}{0pt}{S1\ \ Nonnormality and reactivity of Jacobian vs. adjacency matrices} \hfill 3\\[6pt]
\rule{12pt}{0pt}{S2\ \ Proof of nonnormality and reactivity for almost all large networks} \hfill 5\\
\rule{24pt}{0pt}S2.1\ \ Convergence properties of connection probabilities in the GCLV model \hfill 6\\
\rule{24pt}{0pt}S2.2\ \ Proof of nonnormality \hfill 7\\
\rule{24pt}{0pt}S2.3\ \ Proof of reactivity \hfill 14\\
\rule{36pt}{0pt}S2.3.1\ \ Sufficient condition for reactivity \hfill 15\\
\rule{36pt}{0pt}S2.3.2\ \ Proof that condition (C1) is satisfied for almost all large networks \hfill 16\\
\rule{24pt}{0pt}S2.4\ \ Extension to weighted networks with self-links and Laplacian-coupled networks \hfill 25\\[6pt]
\rule{12pt}{0pt}{S3\ \ Computational details for fig.~S4 and table S1} \hfill 25\\[6pt]
{\bf Supplementary Figures}\\[6pt]
\rule{12pt}{0pt}{Fig.~S1\ \ Version of Fig.~1 indicating network types} \hfill 27\\
\rule{12pt}{0pt}{Fig.~S2\ \ Topological and spectral imbalances in representative networks} \hfill 28\\
\rule{12pt}{0pt}{Fig.~S3\ \ Nonnormality of adjacency vs.\ Laplacian matrices} \hfill 29\\
\rule{12pt}{0pt}{Fig.~S4\ \ Nonnormality and reactivity of typical random networks} \hfill 30\\
\rule{12pt}{0pt}{Fig.~S5\ \ Validating the approximations underlying Eqs.~(5) and (6)} \hfill 31\\[6pt]
{\bf Supplementary Table}\\[6pt]
\rule{12pt}{0pt}{Table S1\ \ Probability that $A$ is normal or non-reactive} \hfill 32\\

\pagebreak

\titleformat{\section}{\normalfont\large\bfseries}{\thesection.}{0.5em}{}
\renewcommand{\thesection}{S\arabic{section}}
\titleformat{\subsection}{\normalfont\normalsize\bfseries}{\thesubsection.}{0.5em}{}
\renewcommand{\thesubsection}{S\arabic{section}.\arabic{subsection}}

\vspace{7mm}
\phantomsection
\addcontentsline{toc}{section}{Supplementary Text}
\noindent{\bf\Large Supplementary Text}

\bigskip\noindent
In Sec.~\ref{sec-si-contri-Jacobian} below, we establish a general relation between the Jacobian and adjacency matrices for a broad class of nonlinear network systems with multi-dimensional node dynamics.
In Sec.~\ref{sec-si-proofs}, we provide mathematical proofs of almost-sure nonnormality and reactivity for random weighted directed networks generated by the GCLV model.
In Sec.~\ref{sec-si-details-figS4-TableS1}, we detail computational procedures used to obtain the results in fig.~\ref{fig:prob-norm-nonreact} and table~\ref{table:prob-norm-nonreact}.

\section[S1. Nonnormality and reactivity of Jacobian vs. adjacency matrices]{Nonnormality and reactivity of Jacobian vs.\ adjacency matrices}
\label{sec-si-contri-Jacobian}

Here, we consider a class of network systems that are more general than Eq.~\eqref{eqn:main-system}:
\begin{equation}
\dot{{x}}_i = {f}({{x}}_i) + \sum_{j=1}^n A_{ij}{h}({x}_j), \quad i=1,\ldots,n,
\end{equation}
where vector $x_i \in \mathbb{R}^m$ represents the dynamical state of node $i$, function $f(\cdot) \in \mathbb{R}^m$ describes the node dynamics, function $h(\cdot) \in \mathbb{R}^m$ captures how the node is coupled to the rest of the network, and $A$ is the possibly weighted $n \times n$ adjacency matrix of the network (but note that the arguments below would remain unchanged if $A$ is replaced with the Laplacian matrix $L$).  We assume that the system has an equilibrium $x^*=({x^*_1}^T, {x^*_2}^T, \ldots, {x^*_n}^T)^T \in \mathbb{R}^N$, $N:=mn$, and seek to analyze its stability. We linearize the system at $x^*$ and assume that $D_x f(x_1^*) = D_x f(x_2^*) = \cdots = D_x f(x_n^*) =: F$. This assumption is valid under either of the following conditions: 1)~$f(\cdot)$ is a linear function (which is the case, e.g., for power grids and many mechanical networks); 2)~$x^*$ is a synchronous state, i.e., $x_1^*=x_2^*=\cdots x_n^*$, or can be transformed into a synchronous state by a suitable change of coordinates (as often done, e.g., for biological clocks). We also assume that $D_x h(x) = \zeta(x) H$ for some scalar-valued function $\zeta(\cdot)$ and constant matrix $H$. Under these assumptions, the $N \times N$ Jacobian matrix $M$ of the network system takes the form
\begin{equation}\label{eqn:def_M}
M = {I_n}\otimes {F} + A^*\otimes {H},
\end{equation}
where $\otimes$ denotes the Kronecker product, $A^*$ is defined by $A^*_{ij} = \zeta(x^*_j) {A}_{ij}$, and we recall that $I_n$ is the identity matrix of size $n$.
Equation~\eqref{eqn:main-system} corresponds to a special case of this formulation with $m=1$, $F=-\alpha$, and $\zeta(x)=H=1$ (and hence $M = -\alpha I_n + A$).

Letting $A^* = V \Lambda V^{-1}$ be the eigen-decomposition of $A^*$, we have
\begin{equation}
M = (V \otimes {I}_m)[{I_{n}}\otimes {F}+{\Lambda}\otimes {H}](V^{-1} \otimes {I}_m).
\end{equation}
It follows from this that, for each eigenvalue $\lambda$ of $A^*$, the eigenvalues of the matrix
\begin{equation}
C(\lambda) := F+\lambda H
\end{equation}
are also eigenvalues of the Jacobian matrix $M$. Writing the Jordan decomposition of $C(\lambda)$ as $C(\lambda) = Y(\lambda) \Sigma (\lambda) Y(\lambda)^{-1}$, we can express the Jordan decomposition of the Jacobian matrix $M$ in terms of $Y(\lambda)$ and $\Sigma(\lambda)$ as
\begin{equation}
M = R \Phi R^{-1},
\end{equation}
where
\begin{equation}
\begin{aligned}
R &= [v_1 \otimes Y(\lambda_1), v_2 \otimes Y(\lambda_2), \cdots, v_n \otimes Y(\lambda_n)], \\
\Phi &= \begin{bmatrix} \Sigma(\lambda_1) & O & \cdots & O \\ O & \Sigma(\lambda_2) & \ddots & \vdots \\ \vdots & \ddots & \ddots & O \\ O & \cdots & O &\Sigma(\lambda_n) \end{bmatrix},
\end{aligned}
\end{equation}
and $v_i$ denotes the right eigenvector associated with the $i$th eigenvalue $\lambda_i$ of $A^*$.
We note that the Jacobian matrix $M$ is normal if and only if all of its eigenvectors are orthogonal to each other (i.e., $R^\dagger R=I_{N}$, where $R^\dagger$ denotes the conjugate transpose of $R$).
Since the $(i,j)$ block of $R^\dagger R$ can be written as
\begin{equation}
(R^\dagger R)_{ij}  = (v_i^\dagger v_j)\cdot (Y(\lambda_i)^\dagger Y(\lambda_j)),
\end{equation}
a necessary and sufficient condition for $M$ to be nonnormal can be written as
\begin{equation}\label{nonnormaldecopose}
  \norm{R^\dagger R-I_{N}}_\text{F}^2   = \underbrace{\sum_{i} \norm{ Y(\lambda_i)^\dagger Y(\lambda_i)-I_{m}}_\text{F}^2 }_\text{node-induced nonnormality}+\underbrace{2 \mathop{\sum\sum}_{i<j} (v_i^\dagger v_j)^2 \cdot \norm{Y(\lambda_i)^\dagger Y(\lambda_j)}_\text{F}^2}_\text{network-induced nonnormality} > 0,
\end{equation}
where the first sum represents the contribution coming from the nonnormality of $C(\lambda)$ while the second represents the contribution from the nonnormality of $A^*$.
Note that $\abs{v_i^\dagger v_j}$ measures the non-orthogonality of the eigenvectors $v_i$ and $v_j$ and that we generically have $\norm{ Y(\lambda_i)^\dagger Y(\lambda_j)}_\text{F}^2>0$ if $\lambda_i \neq \lambda_j$.
Consequently, if $A^*$ for a given network is nonnormal, i.e., $v_i^\dagger v_j \neq 0$ for some $i$ and $j$, then generically the corresponding Jacobian matrix $M$ must also be nonnormal.
While the converse does not generally hold (i.e., the nonnormality of $M$ does not imply that of $A^*$), the decomposition in Eq.~\eqref{nonnormaldecopose} clearly shows that the nonnormality of $M$ must come from the node dynamics if it does not come from the network structure (i.e., if $A^*$ is normal). 

An analogous decomposition holds true also for reactivity.
To see this, we first note that having distinct left and right eigenvectors associated with the rightmost eigenvalue is a sufficient condition for reactivity (which is proved in materials and methods). Let $y_k(\lambda)$ denote the $k$th column of the matrix $Y(\lambda)$, i.e., the $k$th eigenvector of $C(\lambda)$. Let $\sigma_1 (\lambda_1)$ denote the rightmost eigenvalue of $M$, which is also an eigenvalue of $C(\lambda_1)$.  Then, $v_1 \otimes y_1(\lambda_1)$ is the eigenvector of $M$ associated with $\sigma_1 (\lambda_1)$, where $y_1(\lambda_1)$ is the eigenvector of $C(\lambda_1)$ associated with $\sigma_1(\lambda_1)$. Since having distinct left and right eigenvectors associated with $\sigma_1 (\lambda_1)$ is equivalent to the non-orthogonality between $v_1 \otimes y_1(\lambda_1)$ and some other right eigenvector of $M$, a sufficient condition for $M$ to be reactive can be expressed as
\begin{equation}\label{reactivedecopose}
\underbrace{\sum_{k=1}^m (y_1(\lambda_1)^\dagger y_k(\lambda_1))^2}_\text{node-induced} + \underbrace{ \sum_{j=2}^{n} (v_1^\dagger v_j)^2 \cdot \norm{ y_1(\lambda_1)^\dagger Y(\lambda_j)}_\text{F}^2}_\text{network-induced} >0.
\end{equation}
Similarly to the case of nonnormality, this condition shows that $A^*$ being reactive (i.e.,  $\abs{v_1^\dagger v_j} > 0$ for some $j$) generically implies $M$ being reactive.
(In the derivation of Eqs.~\eqref{nonnormaldecopose} and \eqref{reactivedecopose}, we implicitly assumed that $A^*$ is diagonalizable for simplicity, but this assumption can be lifted using the Jordan transformation to derive similar decomposition.)

Thus, if $A^*$ is nonnormal and reactive with high probability in a class of network systems, then the Jacobian matrix $M$ is nonnormal and reactive with high probability as well.
For the weighted GCLV model, the almost-sure nonnormality and reactivity of $A$ proved in Sec.~\ref{sec-si-weighted} below suggest that $\tilde{A}$ (with the modified link weight distribution) and thus the Jacobian matrix $M$ in Eq.~\eqref{eqn:def_M} would also be nonnormal and reactive almost surely in the limit of large network sizes.

\section[S2. Proof of nonnormality and reactivity for almost all large networks]{Proof of nonnormality and reactivity for almost all large networks}
\label{sec-si-proofs}
Here, we prove that, for the GCLV model with a given joint distribution of in- and out-degrees and a distribution of link weights, the probability that the network's adjacency matrix $A$ is nonnormal and the probability that $A$ is reactive both converge to one in the limit of large network size, $n \to \infty$.
The proofs are valid regardless of whether we allow for self-links or not, as they play no role in the arguments.
We first prove key convergence properties of the connection probabilities $\rho_{ij}$ in the model (Sec.~\ref{sec-si-conv}).
We then present proofs for the almost-sure nonnormality (Sec.~\ref{sec-si-nonnormal}) and the almost-sure reactivity (Sec.~\ref{sec-si-reactive}) for the case of unweighted GCLV model (i.e., assuming $A_{ij}=0$ or $1$).
Finally, we show how the proofs can be extended to weighted networks and also to Laplacian-coupled networks (Sec.~\ref{sec-si-weighted}).

\subsection[S2.1. Convergence properties of connection probabilities in the GCLV model]{Convergence properties of connection probabilities in the GCLV model}
\label{sec-si-conv}
Here, we first show that both $\underset{1\leq i\leq n}{\text{max}}\rho_{ij}$ for any fixed $j$ and $\underset{1\leq j\leq n}{\text{max}}\rho_{ij}$ for any fixed $i$ converge to zero in probability as $n \to \infty$.  This property is essential for the proofs in the sections below.  To prove this property for $\underset{1\leq i\leq n}{\text{max}}\rho_{ij}$, we first note that $\rho_{ij} \ge 0$, which implies that the convergence in probability is equivalent to
\begin{equation}\label{eqn:as_conv}
\underset{n\to\infty}{\lim}
\mathbb{P}\Bigl( \underset{1\leq i\leq n}{\text{max}}\rho_{ij} > \varepsilon \Bigr) = 0 \quad\text{for all $\varepsilon>0$}.
\end{equation}
Since $\underset{1\leq i\leq n}{\text{max}}\rho_{ij} = \tilde{d}_{\text{max}}^{\text{in}} \tilde{d}_j^{\text{out}}/\sum_{k=1}^{n}\tilde{d}_k^{\text{in}}$ by the definition of $\rho_{ij}$ in Eq.~\eqref{eqn:def_rho}, and thus $\tilde{d}_{\text{max}}^{\text{in}} = \underset{1\leq i\leq n}{\text{max}}\rho_{ij} \sum_{k=1}^{n} \tilde{d}_k^{\text{in}}/\tilde{d}_j^{\text{out}}$, it follows from Eq.~\eqref{eqn:profkn} that
\begin{equation}\label{eqn:gev_conv}
    \begin{aligned}
    \underset{n\to\infty}{\lim}
\mathbb{P}\Bigl( \underset{1\leq i\leq n}{\text{max}}\rho_{ij} > \varepsilon \Bigr)& = \underset{n\to\infty}{\lim} \mathbb{P}\biggl( {\tilde{d}_{\text{max}}^{\text{in}} } > \frac{\varepsilon \sum_{k=1}^{n} \tilde{d}_k^{\text{in}}}{\tilde{d}_j^{\text{out}}} \biggr)\\
&= \underset{n\to\infty}{\lim} \mathbb{P} \biggl( \frac{\tilde{d}_{\text{max}}^{\text{in}} - b_n}{a_n} > \frac{\varepsilon \sum_{k=1}^{n} \tilde{d}_k^{\text{in}} / \tilde{d}_j^{\text{out}} - b_n}{a_n} \biggr) \\
& =\underset{n\to\infty}{\lim} \biggl[ 1 -  G_{\gamma}\Bigl(\frac{\varepsilon \sum_{k=1}^{n} \tilde{d}_k^{\text{in}} / \tilde{d}_j^{\text{out}} - b_n}{a_n} \Bigr) \biggr] \\
&= \underset{n\to\infty}{\lim} \biggl[ 1 -  G_{\gamma}\Bigl(\frac{\frac{\varepsilon}{n} \sum_{k=1}^{n} \tilde{d}_k^{\text{in}} / \tilde{d}_j^{\text{out}} - b_n/n}{a_n/n} \Bigr) \biggr] =0
    \end{aligned}
\end{equation}
for any $\varepsilon>0$, where the last equality is due to the assumption $\underset{n\to\infty}{\lim}a_n/n = \underset{n\to\infty}{\lim}b_n/n = 0$, and further due to strong law of large numbers, $\frac{\varepsilon}{n} \sum_k \tilde{d}_k^{\text{in}} / \tilde{d}_j^{\text{out}} \rightarrow \varepsilon \tilde{d}/\tilde{d}_j^{\text{out}} >0$ as $n \rightarrow \infty$ (recalling the definition $\tilde{d}=\mathbb{E}(\tilde{d}_k^{\text{in}})$).
This proves that $\underset{1\leq i\leq n}{\text{max}}\rho_{ij}$ converges to $0$ in probability as $n \rightarrow \infty$. The convergence of $\underset{1\leq j\leq n}{\text{max}}\rho_{ij}$ in probability can be proved following the same argument, with index $i$ replaced by $j$ and the ``in'' superscript replaced by ``out'' in appropriate places.

In addition, $\rho_{ij}$ satisfies a similar but slightly different property: $\underset{1\leq i\leq n}{\text{max}}\rho_{in}$ converges to zero in probability as $n \to \infty$.
To see this, we first note that, for any $a>0$, we have
\begin{equation}
        \begin{aligned}
  \Bigl\{ \underset{1\leq i\leq n}{\text{max}}\rho_{in} > \varepsilon \Bigr\} = \Bigl\{
  \underset{1\leq i\leq n}{\text{max}}\rho_{in} > \varepsilon \text{ and } \tilde{d}_n^{\text{out}}\leq a
  \Bigr\} \cup   \Bigl\{ 
   \underset{1\leq i\leq n}{\text{max}}\rho_{in} > \varepsilon \text{ and } \tilde{d}_n^{\text{out}}> a
  \Bigr\},
\end{aligned}
\end{equation}
and thus
\begin{equation}
\begin{aligned}
\mathbb{P}\Bigl( \underset{1\leq i\leq n}{\text{max}}\rho_{in} > \varepsilon \Bigr)
&= \mathbb{P}\Bigl( \underset{1\leq i\leq n}{\text{max}}\rho_{in} > \varepsilon \text{ and } \tilde{d}_n^{\text{out}}\leq a \Bigr)
+ \mathbb{P}\Bigl( \underset{1\leq i\leq n}{\text{max}}\rho_{in} > \varepsilon \text{ and } \tilde{d}_n^{\text{out}}> a \Bigr)\\
&\leq \mathbb{P}\Bigl( \underset{1\leq i\leq n}{\text{max}}\rho_{in} > \varepsilon \text{ and } \tilde{d}_n^{\text{out}}
\leq a \Bigr) + \mathbb{P}\bigl( \tilde{d}_n^{\text{out}} > a \bigr)\\
&\leq \mathbb{P}\Bigl( {\tilde{d}_{\text{max}}^{\text{in}} } > \frac{\varepsilon \sum_{k=1}^{n} \tilde{d}_k^{\text{in}}}{a} \Bigr) + \mathbb{P}\bigl( \tilde{d}_n^{\text{out}} > a \bigr)\\
&\leq \mathbb{P}\Bigl( {\tilde{d}_{\text{max}}^{\text{in}} } > \frac{\varepsilon \sum_{k=1}^{n} \tilde{d}_k^{\text{in}}}{a} \Bigr)+ \frac{\tilde{d}}{a},
  \end{aligned}
\end{equation}
where the last inequality is due to Markov's inequality. Following the same procedure as in Eq.~\eqref{eqn:gev_conv}, we have
\begin{equation}
\begin{aligned}
\underset{n\to\infty}{\lim} \mathbb{P}\Bigl( \underset{1\leq i\leq n}{\text{max}}\rho_{in} > \varepsilon \Bigr) 
& \leq \underset{n\to\infty}{\lim} \biggl[ 1 -  G_{\gamma}\biggl(\frac{\frac{\varepsilon}{n} \sum_{k=1}^{n} \tilde{d}_k^{\text{in}} / a - b_n/n}{a_n/n} \biggr) \biggr]+ \frac{\tilde{d}}{a} 
= \frac{\tilde{d}}{a} .
\end{aligned}
\end{equation}
Since $a>0$ can be chosen arbitrarily, we conclude that
\begin{equation}\label{eqn:rhoincon}
\underset{n\to\infty}{\lim} \mathbb{P}\Bigl( \underset{1\leq i\leq n}{\text{max}}\rho_{in} > \varepsilon \Bigr)=0,
\end{equation}
i.e., $ \underset{1\leq i\leq n}{\text{max}}\rho_{in}$ converges to $0$ as $n\rightarrow \infty$.
Again, the convergence of $\underset{1\leq j\leq n}{\text{max}}\rho_{nj}$ in probability can be proved by following the same argument with index $i$ replaced by $j$ and the ``in'' superscript replaced by ``out'' in appropriate places.
Combining the results above, we see that $\varepsilon_i(n) := \underset{1\leq j\leq n}{\max}(\rho_{ij}+\rho_{ji}) \rightarrow 0$ for any fixed $i$ and that $\varepsilon_n(n) \rightarrow 0$ as $n\to\infty$.
If the support of the distribution of expected degrees is bounded (i.e., constrained to the finite interval $[0,d_\text{max}]$), the convergence $\rho_{ij} \to 0$ occurs with probability one and is uniform over all $1\leq i,j\leq n$, since we have $\sum_{k=1}^n \tilde{d}_k^\text{in}/n \to \tilde{d} > 0$ almost surely and thus have $\underset{1\leq i,j\leq n}{\text{max}}\rho_{ij} \leq \frac{1}{n} d_\text{max}^2/(\sum_{k=1}^n \tilde{d}_k^\text{in}/n) \to 0$ (independently of $i$ and $j$).

\subsection[S2.2. Proof of nonnormality]{Proof of nonnormality}
\label{sec-si-nonnormal}

We now show that $A$ is nonnormal with probability approaching one as $n\to\infty$ under the assumption that $A_{ij}\in\{0,1\}$ (the proof will be extended to general weighted $A_{ij}$ in Sec.~\ref{sec-si-weighted}).
Since a sufficient condition for $A$ to be nonnormal is that there exists $1\leq i\leq n$ such that $d_{i}^{\text{in}} \neq d_{i}^{\text{out}}$ 
(i.e., at least one term is strictly positive in the first summation in Eq.~\eqref{eqn:D_F_def}),
we have
\begin{equation}\label{eqn:prob-nonnormal}
\begin{split}
\mathbb{P}(\text{$A$ is nonnormal})
&\ge \mathbb{P}(\text{$d_{i}^{\text{in}} \neq d_{i}^{\text{out}}$ for some node $i$}) = \mathbb{E}(Q(n)),
\end{split}
\end{equation}
where we defined
\begin{equation}\label{eqn:Qn}
Q(n) := \mathbb{P}\bigl(\text{$d_{i}^{\text{in}} \neq d_{i}^{\text{out}}$ for some node $i$} \,\big\vert\, \tilde{d}_1^{\text{in}}, \ldots, \tilde{d}_n^{\text{in}}, \tilde{d}_1^{\text{out}}, \ldots, \tilde{d}_n^{\text{out}} \bigr),
\end{equation}
i.e., the conditional probability that the sufficient condition is satisfied given a realization of the (random) expected degrees $\tilde{d}_1^{\text{in}}, \ldots, \tilde{d}_n^{\text{in}}, \tilde{d}_1^{\text{out}}, \ldots, \tilde{d}_n^{\text{out}}$.
We note that $Q(n)$ itself is a random variable because the expected degrees are random in the GCLV model.
Below, we will show that $Q(n) \to 1$ in probability as $n\to\infty$, implying that its expected value $\mathbb{E}(Q(n)) \to 1$, and thus that $\mathbb{P}(\text{$A$ is nonnormal}) \to 1$, in view of Eq.~\eqref{eqn:prob-nonnormal}.

To estimate $Q(n)$, we first seek to estimate an analogous conditional probability for a given node $i$:
\begin{equation}
Q_i(n) := \mathbb{P}(d_{i}^{\text{in}} \neq d_{i}^{\text{out}} \,\big\vert\, \tilde{d}_1^{\text{in}}, \ldots, \tilde{d}_n^{\text{in}}, \tilde{d}_1^{\text{out}}, \ldots, \tilde{d}_n^{\text{out}}).
\end{equation}
We note that, for a given realization of the expected degrees, the difference between the (actual) in- and out-degrees of node $i$ is a sum of independent random variables: $d_{i}^{\text{in}} - d_{i}^{\text{out}}=\sum_j Y_{ij}$, where $Y_{ij}:=A_{ij}-A_{ji}$ has mean $\mu_{ij}:=\rho_{ij}-\rho_{ji}$, 
variance $\sigma^2_{ij}:=\rho_{ij}+\rho_{ji}-\rho_{ij}^2-\rho_{ji}^2$, and finite third moment $h_{ij}:=\mathbb{E}(\abs{Y_{ij}-\mu_{ij}}^3)$ for $i \neq j$.  Since $Y_{ii}=0$, its mean $\mu_{ii}$, variance $\sigma_{ii}$, and third moment $h_{ii}$ are all equal to zero.
Since $Y_{ij} \in \{-1,0,1\}$, we have
\begin{equation}\label{eqn:proof-nonnormal-1}
\begin{split}
\frac{h_{ij}}{\sigma^2_{ij}}
&= \frac{\mathbb{E}(\abs{Y_{ij}-\mu_{ij}}^3)}{\mathbb{E}(\abs{Y_{ij}-\mu_{ij}}^2)}
\leq \max_{x\in \{-1,0,1\}} \frac{\abs{x-\mu_{ij}}^3}{\abs{x-\mu_{ij}}^2}
= \max_{x\in \{-1,0,1\}} \abs{x-\rho_{ij}+\rho_{ji}}
\leq 1+\rho_{ij}+\rho_{ji}
\end{split}
\end{equation}
for $i \neq j$, where the first inequality follows from the general inequality $\sum_i a_i / \sum_i b_i \le \max_i (a_i/b_i)$.
We note that the standard deviation $s_i$ of $\sum_j Y_{ij}$ is given by $s_i^2 := \sum_j \sigma^2_{ij}$.
According to the Berry--Esseen Theorem \cite{Esseen:1942}, the distribution of the standardized sum $\sum_j (Y_{ij}-\mu_{ij})/s_i$ converges to the standard normal distribution with the approximation error bounded as
\begin{equation}\label{eqn:proof-nonnormal-2}
\sup_{x\in \mathbb{R}}\,\abs{F_{i}(x)-\Phi(x)}
\leq C_0 \sum_{j=1}^n \mathbb{E} \Bigl(\, \Bigl\lvert\frac{Y_{ij}-\mu_{ij}}{s_i} \Bigr\rvert^3 \,\Bigr)
= \frac{C_0}{ {s_{i}^3}} \sum_{j=1}^n h_{ij}
\leq \frac{C_0}{s_{i}}+\frac{C_0}{ {s_{i}^3}} \sum_{j=1}^n(\rho_{ij}+\rho_{ji})^2,
\end{equation}
where we denote the CDF of the random variable $\sum_j (Y_{ij}-\mu_{ij})/s_i$ by $F_{i}(x)$ and the CDF of standard normal distribution by $\Phi(x)$, and $C_0$ is a constant, which is known \cite{shevtsova2010improvement} to satisfy $0.40 \le C_0 \le 0.56$.
The last inequality in Eq.~\eqref{eqn:proof-nonnormal-2} follows from Eq.~\eqref{eqn:proof-nonnormal-1} and the definition of $\sigma_{ij}$.
Observing that
\begin{equation}
\begin{aligned}
Q_i(n) &\geq 1 - \mathbb{P}(-\xi < d_{i}^{\text{in}} - d_{i}^{\text{out}}\leq \xi) \\
&= 1 - \mathbb{P}({\textstyle\sum_j Y_{ij}} \leq \xi) + \mathbb{P}({\textstyle\sum_j Y_{ij}} \leq -\xi) \\
&\geq 1- \Phi\Bigl(\frac{\xi - \sum_j \mu_{ij}}{s_{i}}\Bigr) + \Phi\Bigl(\frac{-\xi-\sum_j \mu_{ij}}{s_{i}}\Bigr) - \frac{2C_0}{s_{i}}-\frac{2C_0}{ {s_{i}^3}} \sum_{j=1}^n(\rho_{ij}+\rho_{ji})^2,
\end{aligned}
\end{equation}
for any $\xi>0$ and taking the limit $\xi \rightarrow 0$, we obtain a lower bound for $Q_i(n)$:
\begin{equation}\label{eqn:Pdinout}
Q_i(n) \geq P_i(n) := 1 - \frac{2C_0}{s_{i}}-\frac{2C_0}{ {s_{i}^3}} \sum_{j=1}^n(\rho_{ij}+\rho_{ji})^2.
\end{equation}

We now estimate $Q(n)$ using a recursive argument involving $Q_i(n)$ and $P_i(n)$.
For a given $1 \le i \le n$, let $f_i(n)$ denote the conditional probability that the subnetwork induced by nodes $1,\ldots,i$ satisfies the sufficient condition for nonnormality, i.e., there exists $1 \leq j \leq i$ for which the in- and out-degrees of node $j$ defined within the subnetwork are distinct (given a realization of all the expected degrees).
For $i=n$, we have $f_n(n)=Q(n)$.
For $i=1$, we have $f_1(n)=0$, since the adjacency matrix of a single isolated node is always normal.
For $i=2$, a direct calculation yields $f_2(n)=\rho_{12}(1-\rho_{21})+\rho_{21}(1-\rho_{12})$, noting that the adjacency matrix elements $A_{12}$ and $A_{21}$ are the only random variables involved.
For the general case, by considering the addition of node $i$ to the subnetwork induced by nodes $1,\ldots,i-1$ (and the links between node $i$ and the subnetwork in both directions), we have the following inequality for any $i \ge 2$:
\begin{equation}
f_i(n) \geq f_{i-1}(n) \cdot \underset{1 \leq j< i}{\min}\bigl((1-\rho_{ji})(1-\rho_{ij})+\rho_{ji}\rho_{ij}\bigr)  +(1-f_{i-1}(n)) Q_i(n),
\end{equation}
which implies
\begin{equation}\label{eqn:fincre}
f_i(n) \geq f_{i-1}(n) \cdot (1- \tilde{\varepsilon}_i(n)-Q_i(n)) + Q_i(n),
\end{equation}
where we define $\tilde{\varepsilon}_i(n) := \underset{1 \le j < i}{\max}(\rho_{ji}+\rho_{ij})$.
Noting that the r.h.s.\ of Eq.~\eqref{eqn:fincre} is monotonically increasing in 
$Q_i(n)$ and using $Q_i(n) \geq P_i(n)$ from Eq.~\eqref{eqn:Pdinout}, this leads to
\begin{equation}\label{eqn:recurvelb}
\begin{aligned}
f_i(n) &\geq f_{i-1}(n)\cdot(1 - \tilde{\varepsilon}_i(n)-P_i(n))+ P_i(n) \\
&\geq f_{i-1}(n)\cdot(1-P_i(n))+ P_i(n)- \tilde{\varepsilon}_i(n),
\end{aligned}
\end{equation}
which is equivalent to 
\begin{equation}\label{eqn:recurvelb2}
1-f_i(n) \leq (1-f_{i-1}(n)) \cdot ( 1 - P_i(n)) + \tilde{\varepsilon}_i(n).
\end{equation}
Recursively applying this inequality, we obtain
\begin{equation}\label{eqn:diffto1bd}
1-Q(n) = 1 - f_n(n) \leq \tilde{\varepsilon}_n(n) + \prod_{i=2}^n (1-P_i(n)) + \sum_{j=3}^{n}\tilde{\varepsilon}_{j-1}(n) \prod_{i=j}^n (1-P_i(n)).
\end{equation}
Since the derivation of the bound in Eq.~\eqref{eqn:diffto1bd} does not depend on a particular ordering of the nodes.
By reversing the order of the nodes, i.e., interchanging node $i$ with node $n-i+1$, we can rewrite this relation as
\begin{equation}\label{eqn:diffto1bd2}
1 - Q(n) \leq \tilde{\varepsilon}_n(n) + \prod_{i=1}^{n-1} (1-P_i(n)) +  \sum_{j=1}^{n-2}\tilde{\varepsilon}_{j+1}(n) \prod_{i=1}^{j} (1-P_i(n)).
\end{equation}
We would thus prove our claim if we show that the r.h.s.\ converges to zero in probability (recalling that all these terms are random since $\rho_{ij}$ and $s_i$ are).

To help estimate the r.h.s.\ of Eq.~\eqref{eqn:diffto1bd2}, we consider the double sequence $g_{jn}$ defined by
\begin{equation}\label{eqn:Pin}
g_{jn} = \begin{cases}
{\displaystyle\frac{1}{j}\sum_{i=1}^j} P_i(n), &j \leq n,\\
{\displaystyle\frac{1}{n}\sum_{i=1}^n} P_i(n), &j>n.
\end{cases}
\end{equation}
According to the Moore-Osgood Theorem \cite{Zakon:2011}, if $\underset{j\to\infty}{\lim} g_{jn}$ exists for every $n$ and $\underset{n\to\infty}{\lim} g_{jn}$ exists for every $j$, with the convergence in the latter limit uniform in $j$, then $\underset{j,n\to\infty}{\lim} g_{jn}$ exists (regardless of how $j$ and $n$ are taken to $\infty$) and can be calculated as iterated limits, i.e.,
\begin{equation}
\lim_{j,n \rightarrow \infty}g_{jn}
= \lim_{j \rightarrow \infty}\lim_{n \rightarrow \infty} g_{jn}
= \lim_{n \rightarrow \infty}\lim_{j \rightarrow \infty} g_{jn}.
\end{equation}
We now show that the required conditions are satisfied, so that the theorem can be applied.
First, we see that $\underset{j\to\infty}{\lim} g_{jn} = \frac{1}{n}\sum_{i=1}^n P_i(n)$ for every fixed $n$, since $g_{jn}$ itself equals $\frac{1}{n}\sum_{i=1}^n P_i(n)$ and does not vary with $j$ for any $j>n$ by the definition of $g_{jn}$.
Next, we compute $\underset{n\to\infty}{\lim} g_{jn}$ for a given $j$.
From the definition of $g_{jn}$ for $j \leq n$ in Eq.~\eqref{eqn:Pin} and the definition of $P_i(n)$ in Eq.~\eqref{eqn:Pdinout}, we have
\begin{equation}\label{eqn:gjnlimit2}
\begin{split}
\lim_{n\to\infty} g_{jn} 
&= \frac{1}{j}\sum_{i=1}^j \lim_{n\to\infty} P_i(n) \\
&= 1 - \frac{2C_0}{j} \sum_{i=1}^j \lim_{n\to\infty} \frac{1}{s_i} -  \frac{2C_0}{j} \sum_{i=1}^j \lim_{n\to\infty} \frac{1}{s_i^3} \sum_{k=1}^n (\rho_{ik} + \rho_{ki})^2.\\
\end{split}
\end{equation}
To compute the limit of $1/s_i$, we first consider $s_i^2$ and observed that
\begin{equation}\label{vardiff}
\begin{split}
s_{i}^2
&= \sum_{k=1}^n \sigma^2_{ik} = \sum_{k=1}^n (\rho_{ik}+\rho_{ki}-\rho_{ik}^2-\rho_{ki}^2) \\
&= \frac{\frac{1}{n}\sum_k \tilde{d}_k^{\text{out}}}{\frac{1}{n}\sum_k \tilde{d}_k^{\text{in}}} \tilde{d}_i^{\text{in}}
+ \tilde{d}_i^{\text{out}}
- \frac{(\tilde{d}_i^{\text{in}})^2 \cdot \frac{1}{n} \sum_k (\tilde{d}_k^{\text{out}})^2}{n \cdot (\frac{1}{n}\sum_k \tilde{d}_k^{\text{in}})^2}
- \frac{(\tilde{d}_i^{\text{out}})^2 \cdot \frac{1}{n} \sum_k (\tilde{d}_k^{\text{in}})^2}{n \cdot (\frac{1}{n}\sum_k \tilde{d}_k^{\text{in}})^2}.
\end{split}
\end{equation}
We note that $\frac{1}{n} \sum_k \tilde{d}_k^{\text{in}} \to \tilde{d}$ and $\frac{1}{n} \sum_k \tilde{d}_k^{\text{out}} \to \tilde{d}$ as $n\to\infty$ by the strong law of large numbers, implying that the first term in Eq.~\eqref{vardiff} converges to $\tilde{d}_i^{\text{in}}$.
For the last two terms, the strong law of large numbers can be used again to see that $\frac{1}{n} \sum_k (\tilde{d}_k^{\text{in}})^2 \to \mathbb{E}\bigl((\tilde{d}_*^{\text{in}})^2\bigr)$ and $\frac{1}{n} \sum_k (\tilde{d}_k^{\text{out}})^2 \to \mathbb{E}\bigl((\tilde{d}_*^{\text{out}})^2\bigr)$, where $\tilde{d}_*^{\text{in}}$ and $\tilde{d}_*^{\text{out}}$ denote random variables drawn from the joint distribution of $\tilde{d}_i^{\text{in}}$ and $\tilde{d}_i^{\text{out}}$, which is independent of $i$.
With a factor of $n$ in the denominators, we see that both of these two terms converge to zero.
We thus have 
$s_{i} \to (\tilde{d}_i^{\text{in}}+\tilde{d}_i^{\text{out}})^{1/2}$
almost surely.
For the last term in Eq.~\eqref{eqn:gjnlimit2}, we note that
\begin{equation}\label{eqn:rhoik2}
\begin{split}
\sum_{k=1}^n (\rho_{ik} + \rho_{ki})^2
&= \sum_{k=1}^n (\rho_{ik}^2 + \rho_{ki}^2 + 2\rho_{ik}\rho_{ki}) \\
&= \frac{(\tilde{d}_i^{\text{in}})^2 \cdot \frac{1}{n} \sum_k (\tilde{d}_k^{\text{out}})^2
+ (\tilde{d}_i^{\text{out}})^2 \cdot \frac{1}{n} \sum_k (\tilde{d}_k^{\text{in}})^2
+ 2 \tilde{d}_i^{\text{in}} \tilde{d}_i^{\text{out}} \cdot \frac{1}{n} \sum_k \tilde{d}_k^{\text{in}} \tilde{d}_k^{\text{out}}}{n \cdot (\frac{1}{n}\sum_k \tilde{d}_k^{\text{in}})^2}
\end{split}
\end{equation}
converges to zero as $n\to\infty$ almost surely, since $\frac{1}{n} \sum_k \tilde{d}_k^{\text{in}} \tilde{d}_k^{\text{out}} \to \mathbb{E}\bigl( \tilde{d}_*^{\text{in}} \tilde{d}_*^{\text{out}} \bigr)$ by the strong law of large numbers.
Thus, Eq.~\eqref{eqn:gjnlimit2} becomes
\begin{equation}\label{eqn:gjnlimit3}
\lim_{n\to\infty} g_{jn} 
= 1 - \frac{2C_0}{j} \sum_{i=1}^j \frac{1}{\sqrt{\tilde{d}_i^{\text{in}}+\tilde{d}_i^{\text{out}}}}\\
\end{equation}

To see that this convergence is uniform in $j$, we note that the random variables
\begin{equation}
\frac{\frac{1}{n}\sum_k \tilde{d}_k^{\text{out}}}{\frac{1}{n}\sum_k \tilde{d}_k^{\text{in}}}, \quad
\frac{\frac{1}{n} \sum_k (\tilde{d}_k^{\text{in}})^2}{n\cdot(\frac{1}{n}\sum_k \tilde{d}_k^{\text{in}})^2}, \quad \frac{\frac{1}{n} \sum_k (\tilde{d}_k^{\text{out}})^2}{n\cdot(\frac{1}{n}\sum_k \tilde{d}_k^{\text{in}})^2}, \quad
\frac{\frac{1}{n} \sum_k \tilde{d}_k^{\text{in}} \tilde{d}_k^{\text{out}}}{n\cdot(\frac{1}{n}\sum_k \tilde{d}_k^{\text{in}})^2}
\end{equation}
appearing in Eqs.~\eqref{vardiff} and \eqref{eqn:rhoik2} are all independent of $j$, and so is the convergence to their respective limits ($1$ for the first one and $0$ for the other three).
Thus, there exists a function $\varepsilon(n)$ satisfying $\underset{n\to\infty}{\lim} \varepsilon(n) = 0$ and
\begin{equation}\label{eqn:tmp1}
\begin{split}
&\biggl\lvert \frac{\frac{1}{n} \sum_k \tilde{d}_k^{\text{out}}}{\frac{1}{n} \sum_k \tilde{d}_k^{\text{in}}} - 1 \biggr\rvert \leq \varepsilon(n),\quad\hspace{6.75pt}
\Biggl\lvert \frac{\frac{1}{n} \sum_k (\tilde{d}_k^{\text{in}})^2}{n\cdot\bigl(\frac{1}{n}\sum_k \tilde{d}_k^{\text{in}}\bigr)^2} \Biggr\rvert \leq \varepsilon(n), \\
&\Biggl\lvert \frac{\frac{1}{n} \sum_k (\tilde{d}_k^{\text{out}})^2}{n\cdot\bigl(\frac{1}{n}\sum_k \tilde{d}_k^{\text{in}}\bigr)^2} \Biggr\rvert \leq \varepsilon(n),\quad
\Biggl\lvert \frac{\frac{1}{n} \sum_k \tilde{d}_k^{\text{in}} \tilde{d}_k^{\text{out}}}{n\cdot\bigl(\frac{1}{n}\sum_k \tilde{d}_k^{\text{in}}\bigr)^2} \Biggr\rvert \leq \varepsilon(n).
\end{split}
\end{equation}
This, combined with Eqs.~\eqref{vardiff} and \eqref{eqn:rhoik2}, leads to the estimates
\begin{equation}
\abs{s_i^2 - (\tilde{d}_i^\text{in}+\tilde{d}_i^\text{out})}
\leq \bigl( \tilde{d}_i^\text{in} + (\tilde{d}_i^\text{in})^2 + (\tilde{d}_i^\text{out})^2 \bigr) \cdot \varepsilon(n)
\end{equation}
and
\begin{equation}\label{eqn:rhoik22}
\sum_{k=1}^n (\rho_{ik} + \rho_{ki})^2
\leq \bigl( (\tilde{d}_i^\text{in})^2 + (\tilde{d}_i^\text{out})^2 + 2 \tilde{d}_i^\text{in} \tilde{d}_i^\text{out} \bigr) \cdot \varepsilon(n).
\end{equation}
We also have a constant lower bound for $s_i^2$:
\begin{equation}\label{eqn:si-lower-bound}
\begin{split}
s_i^2 &= \sum_{k=1}^n \sigma^2_{ik}
= \sum_{k=1}^n \bigl( (1-\rho_{ik})\rho_{ik} + (1-\rho_{ki})\rho_{ki} \bigr) \\
&\geq (1-c) \sum_{k=1}^n (\rho_{ik} + \rho_{ki}) 
\geq (1-c) \sum_{k=1}^n \rho_{ki} = (1-c) \tilde{d}_i^\text{out} \geq 1-c,
\end{split}
\end{equation}
where we recall from the definition of the GCLV model that $c<1$ is a constant and that we have $0 \leq \rho_{ij} \leq c$ and $\tilde{d}_i^\text{out} \geq 1$.
For the convergence of $1/s_i$, we have the estimate 
\begin{equation}\label{eqn:si1conv}
\begin{aligned}
\Biggl\lvert \frac{1}{s_i} - \frac{1}{\sqrt{\tilde{d}_i^{\text{in}} + \tilde{d}_i^{\text{out}}}} \Biggr\rvert
&= \frac{\abs{s^2_i - ({\tilde{d}_i^{\text{in}} + \tilde{d}_i^{\text{out}}})}}{s_i \sqrt{\tilde{d}_i^{\text{in}}+\tilde{d}_i^{\text{out}}} \Bigl(s_i + \sqrt{\tilde{d}_i^{\text{in}}+\tilde{d}_i^{\text{out}}}\,\Bigr)} 
\leq \frac{\bigl( \tilde{d}_i^\text{in} + (\tilde{d}_i^\text{in})^2 + (\tilde{d}_i^\text{out})^2 \bigr)\cdot\varepsilon(n)}{2\sqrt{1-c}},
\end{aligned}
\end{equation}
since $s_i \ge \sqrt{1-c} > 0$ from Eq.~\eqref{eqn:si-lower-bound} and $\tilde{d}_i^{\text{in}}, \tilde{d}_i^{\text{out}} \ge 1$ from the model definition.
Combining Eqs.~\eqref{eqn:Pdinout}, \eqref{eqn:rhoik22}, \eqref{eqn:si-lower-bound}, and \eqref{eqn:si1conv} for $j \le n$, we have
\begingroup
\allowdisplaybreaks
\begin{align}\label{eqn:gjn-uni-est}
\Biggl\lvert g_{jn} &- \Biggl( 1 - \frac{2C_0}{j} \sum_{i=1}^j \frac{1}{\sqrt{\tilde{d}_i^{\text{in}}+\tilde{d}_i^{\text{out}}}} \Biggr) \Biggr\rvert \nonumber\\
&= \Biggl\lvert \frac{1}{j}\sum_{i=1}^j P_i(n) - 1 + \frac{2C_0}{j} \sum_{i=1}^j \frac{1}{\sqrt{\tilde{d}_i^{\text{in}}+\tilde{d}_i^{\text{out}}}} \Biggr\rvert \nonumber\\
&= \Biggl\lvert \frac{1}{j}\sum_{i=1}^j \Bigl( 1 - \frac{2C_0}{s_{i}}-\frac{2C_0}{{s_{i}^3}} \sum_{k=1}^n(\rho_{ik}+\rho_{ki})^2 \Bigr) - 1 + \frac{2C_0}{j} \sum_{i=1}^j \frac{1}{\sqrt{\tilde{d}_i^{\text{in}}+\tilde{d}_i^{\text{out}}}} \Biggr\rvert \nonumber\\
&= \Biggl\lvert \frac{2C_0}{j}\sum_{i=1}^j \biggl( -\frac{1}{s_i} + \frac{1}{\sqrt{\tilde{d}_i^{\text{in}}+\tilde{d}_i^{\text{out}}}} - \frac{1}{s_{i}^3} \sum_{k=1}^n(\rho_{ik}+\rho_{ki})^2 \biggr) \Biggr\rvert \nonumber\\
&\leq \frac{2C_0}{j} \sum_{i=1}^j \Biggl( \Biggl\lvert \frac{1}{s_i}-\frac{1}{\sqrt{\tilde{d}_i^{\text{in}}+\tilde{d}_i^{\text{out}}}} \Biggr\rvert + \frac{1}{{s_{i}^3}} \sum_{k=1}^n(\rho_{ik}+\rho_{ki})^2 \Biggr) \nonumber\\
&\leq 2C_0 \varepsilon(n) \cdot \frac{1}{j} \sum_{i=1}^j \biggl( \frac{\tilde{d}_i^\text{in} + (\tilde{d}_i^\text{in})^2 + (\tilde{d}_i^\text{out})^2}{2\sqrt{1-c}} + \frac{(\tilde{d}_i^\text{in})^2 + (\tilde{d}_i^\text{out})^2 + 2 \tilde{d}_i^\text{in} \tilde{d}_i^\text{out}}{{(1-c)^{3/2}}} \biggr) \nonumber\\
& \leq 2C_0 \varepsilon(n) M,
\end{align}
\endgroup
where the average over $j$ in the second to the last line above is bounded by a finite constant $M$ because the averages of the individual terms $\tilde{d}_i^\text{in}$, $(\tilde{d}_i^\text{in})^2$, $(\tilde{d}_i^\text{out})^2$, and $\tilde{d}_i^\text{in} \tilde{d}_i^\text{out}$ all converge to finite values as $j\to\infty$ due to the strong law of large numbers.
Since $2C_0 \varepsilon(n) M$ converges to zero as $n\to\infty$ with a rate that does not depend on $j$ (since $\varepsilon(n) \to 0$), the convergence in Eq.~\eqref{eqn:gjnlimit3} is indeed uniform in $j$. 
We can now apply the Moore-Osgood Theorem to conclude that the limit of the double sequence $g_{jn}$ exists and can be computed as an iterated limit:
\begin{equation}\label{eqn:gdoublelimit}
\begin{split}
\lim_{j,n \rightarrow \infty} g_{jn}
&= \lim_{j\rightarrow \infty}\lim_{n\rightarrow \infty} g_{jn}
= 1 - 2C_0 \cdot \lim_{j\to\infty} \frac{1}{j} \sum_{i=1}^j \frac{1}{\sqrt{\tilde{d}_i^{\text{in}} + \tilde{d}_i^{\text{out}}}} \\
&= 1 - 2C_0 \cdot \mathbb{E}\Biggl( \frac{1}{\sqrt{\tilde{d}_*^{\text{in}} + \tilde{d}_*^{\text{out}}}} \Biggr) 
\ge 1 - \frac{2 \cdot 0.56}{\sqrt{1+1}} \ge 0.2,
\end{split}
\end{equation}
where we also used $C_0 \le 0.56$ and $\tilde{d}_i^{\text{in}},\tilde{d}_i^{\text{out}}\geq 1$.
Thus, the double sequence $g_{jn}$ converges to the same value regardless of how $j$ and $n$ are taken to $\infty$, and the limit is bounded away from zero.

We now return to Eq.~\eqref{eqn:diffto1bd2} and show the convergence of $1-Q(n)$ to zero by estimating the terms on the r.h.s.\ one by one.
For the first term, we have $\tilde{\varepsilon}_n(n) = \underset{1 \le j \le n-1}{\max}(\rho_{jn}+\rho_{nj}) \le \underset{1 \le j \le n}{\max}(\rho_{jn}+\rho_{nj}) = \varepsilon_n(n)$, and hence the term converges to zero in probability as $n \rightarrow \infty$, since we proved $\varepsilon_n(n) \to 0$ in the previous section. 

For the second term, we have
\begin{equation}\label{eqn:diffto1bd5}
\prod_{i=1}^{n-1} (1-P_i(n)) \leq \exp\Bigl( -\sum_{i=1}^{n-1} P_i(n) \Bigr)= \exp \bigl( -(n-1) \cdot g_{n-1,n} \bigr),
\end{equation}
where we used the fact that $1-x\leq e^{-x}$ for any $0\leq x \leq 1$.
According to Eq.~\eqref{eqn:gdoublelimit}, the factor $g_{n-1,n}$ above converges to a strictly positive value while the factor $n-1$ diverges, implying that $\exp \bigl( -(n-1) \cdot g_{n-1,n} \bigr)$, and hence the second term on the r.h.s.\ of Eq.~\eqref{eqn:diffto1bd2}, converges to zero as $n\to\infty$.

For the third term, 
we choose a positive integer $N<n-2$ and split the sum to obtain
\begin{equation}\label{eqn:diffto1bd6}
\begin{aligned}
\sum_{j=1}^{n-2} \tilde{\varepsilon}_{j+1}(n) \prod_{i=1}^j (1-P_i(n))
&\leq \sum_{j=1}^{n-2}\tilde{\varepsilon}_{j+1}(n) \exp \biggl( -\sum_{i=1}^j P_i(n) \biggr) \\
&< \sum_{j=1}^{N} \tilde{\varepsilon}_{j+1}(n) \exp \biggl( -\sum_{i=1}^j P_i(n) \biggr) + 2\sum_{j=N+1}^{n-2} \exp(-j \cdot g_{jn}),
\end{aligned}
\end{equation}
where we used the fact that $\rho_{ij}<1$ and thus $\tilde{\varepsilon}_i(n) = \underset{1 \le j < i}{\max}(\rho_{ji}+\rho_{ij}) < 2$ for any $i$.
Taking the limit $n\to\infty$ on the r.h.s., we see that the first term converges to zero in probability, since each $\tilde{\varepsilon}_{j+1}(n)$ converges to zero, the argument of the exponential function is bounded (as it converges; see Eq.~\eqref{eqn:gjnlimit3}), and the number of terms in the sum is finite.
For the second term, we see that
\begin{equation}\label{eqn:gjn-sum}
\sum_{j=N+1}^{n-2} \exp(-j \cdot g_{jn})
\le \sum_{j=N+1}^{n-2} e^{-0.1j}
\le \sum_{j=N+1}^{\infty} e^{-0.1j}
= \frac{e^{-0.1(N+1)}}{1 - e^{-0.1}},
\end{equation}
where we used Eq.~\eqref{eqn:gjn-uni-est} and $\tilde{d}_i^{\text{in}},\tilde{d}_i^{\text{out}}\geq 1$ to estimate $g_{jn}$ as
\begin{equation}
g_{jn} \ge 1 - \frac{2C_0}{j} \sum_{i=1}^j \frac{1}{\sqrt{\tilde{d}_i^{\text{in}}+\tilde{d}_i^{\text{out}}}} - 2C_0 \varepsilon(n) M
\ge 0.2 - 2C_0 \varepsilon(n) M \ge 0.1
\end{equation}
for sufficiently large $n$ (and sufficiently small $\varepsilon(n)$).
Taking the limit $n\to\infty$ in Eq.~\eqref{eqn:gjn-sum} and noting that the r.h.s.\ can be made arbitrarily small by choosing sufficiently large $N$, we see that the last sum in Eq.~\eqref{eqn:diffto1bd6} converge to zero.
We thus conclude that the third term of the r.h.s.\ of Eq.~\eqref{eqn:diffto1bd2} also converges to zero in probability. 

Putting everything together, we have proved that $Q(n)$, and thus its expected value $\mathbb{E}(Q(n))$, converges to one in probability, i.e., the probability that $A$ is nonnormal converges to one as the network size $n$ approaches infinity.

\subsection[S2.3. Proof of reactivity]{Proof of reactivity}
\label{sec-si-reactive}

Now we show that $A$ is reactive with probability converging to one as $n\to\infty$, assuming that $A_{ij}\in\{0,1\}$ (which will be relaxed to allow for weighted $A_{ij}$ in Sec.~\ref{sec-si-weighted}).
We first present a sufficient condition for reactivity and then show that the probability of satisfying this condition approaches one with increasing $n$.

\subsubsection[S2.3.1. Sufficient condition for reactivity]{Sufficient condition for reactivity}
\label{sec-si-reactive-suff-cond}
Consider a square matrix $X = (X_{ij})$ with $X_{ij} \ge 0$ for $i \neq j$ and denote by $G$ the network in which link $j \to i$ exists if and only if $X_{ij} \neq 0$.  We will prove that the following is a sufficient condition for $X$ to be reactive:
\begin{quote}
\textbf{(C1)} The network $G$ has a strongly connected component $\tilde{G}$ containing a node with a nonzero eigenvector in-centrality, and the adjacency matrix $\tilde{A}$ of $\tilde{G}$ has at least one column that strictly dominates the corresponding row.
\end{quote}
Here, given column/row vectors $a$ and $b$, we say that $a$ dominates $b$ if their $i$th components $a_i$ and $b_i$ satisfy $a_i\geq b_i$ for all $i$.
If at least one of these inequalities is strict, then we say $a$ strictly dominates $b$.

Thus, for a network satisfying this condition, we have $v_{1i} \neq 0$ for some $i\in\tilde{G}$ (recalling that the eigenvector in-centrality $v_{1i}$ is the $i$th component of the right eigenvector associated with $\lambda_1(A)$) and $\tilde{A}_{kj} \geq \tilde{A}_{jk}, \forall k$ for some $j$ (with at least one of these inequalities being strict).
For an unweighted directed network $G$, the latter part of the condition regarding $\tilde{A}$
can be interpreted in terms of network topology: there is at least one node whose in-neighbors form a proper subset of the out-neighbors within the strongly connected component containing that node.

To establish the sufficiency of condition (C1) for the reactivity of $A$, suppose that (C1) is satisfied.
Then, $A$ can be transformed by an appropriate node permutation into the block form~\cite{bullo2018lectures}
\begin{equation}\label{Ablockform}
A = \begin{bmatrix}
C & B & O\\ 
O & \tilde{A} & D \\
O & O & E
\end{bmatrix},
\end{equation}
where $O$ denotes the matrix of all zeros (of an appropriate size), and the middle block $\tilde{A}$ is the adjacency matrix of the strongly connected component $\tilde{G}$ in (C1).
Such a permutation can be constructed by re-indexing the nodes in the following order: those nodes $i$ outside $\tilde{G}$ to which there is a directed path from a node in $\tilde{G}$, the nodes in $\tilde{G}$, and all the remaining nodes (where the ordering within each group can be arbitrary).

To prove that $A$ must be reactive, we now assume that $A$ is not reactive and show that this assumption leads to a contradiction.
We first partition the right eigenvector associated with $\lambda_1(A)$ as $v_1 = [v_{c}^T, \tilde{v}_1^T, v_{e}^T]^T$ according to the block structure in Eq.~\eqref{Ablockform}, where (C1) guarantees that $\tilde{v}_1^T$ is nonzero.
By explicitly writing the eigenvalue relation $Av_1 = \lambda_1(A) v_1$ for the second and third rows of the block form in Eq.~\eqref{Ablockform}, we see that $\lambda_1(A)$ is also an eigenvalue of the submatrix 
$A' := \begin{bmatrix}
\tilde{A} & D \\
O & E
\end{bmatrix}$
with eigenvector $[\tilde{v}_1^T, v_{e}^T]^T$.
Since the largest (Perron-Frobenius) eigenvalue $\lambda_1(A')$ of the submatrix $A'$ cannot exceed that of the entire matrix $A$, we must have $\lambda_1(A)=\lambda_1(A')$.
If $A'$ were reactive, then we would have $\lambda_1(A) = \lambda_1(A') < \lambda_1\bigl(\frac{A'+(A')^T}{2}\bigr) \leq \lambda_1\bigl(\frac{A+A^T}{2}\bigr)$, implying that $A$ is also reactive, contradicting the assumption we made above.
Hence, $A'$ must be non-reactive, and thus the right eigenvector 
$[\tilde{v}_1^T, v_{e}^T]^T$
is also a left eigenvector of $A'$ corresponding to $\lambda_1(A')$.
This further implies 
$\tilde{v}_1^T \tilde{A} = \lambda_1(A') \tilde{v}_1^T$, 
and thus $\lambda_1(A')$ is an eigenvalue of $\tilde{A}$ with left eigenvector 
$\tilde{v}_1$.
With the arguments we used above for $A'$ now applied to $\tilde{A}$, we see that $\lambda_1(A')=\lambda_1(\tilde{A})$ (the largest eigenvalue of $\tilde{A}$) and that $\tilde{A}$ must be non-reactive.
This implies that 
$\tilde{v}_1$
is not only the left eigenvector but also the right eigenvector corresponding to $\lambda_1(\tilde{A})$, and thus 
\begin{equation}\label{eqn:left-right-v2}
\tilde{A}^T \tilde{v}_1 = \lambda_1(\tilde{A}) \tilde{v}_1 = \tilde{A} \tilde{v}_1.
\end{equation}
Because $\tilde{G}$ is strongly connected, $\tilde{A}$ is irreducible, and the components of 
$\tilde{v}_1$,
which we denote by 
$\tilde{v}_{1i}$,
are all strictly positive by the Perron-Frobenius Theorem.
Since (C1) is satisfied, there exists an index $j$ for which the $j$th column of $\tilde{A}$ (and hence the $j$th row of $\tilde{A}^T$) strictly dominates the $j$th row of $\tilde{A}$.
By the positivity of 
$\tilde{v}_{1i}$,
this implies 
$(\tilde{A}^T \tilde{v}_1)_j = \sum_i (\tilde{A}^T)_{ji} \tilde{v}_{1i} > \sum_i \tilde{A}_{ji} \tilde{v}_{1i} = (\tilde{A} \tilde{v}_1)_j$,
contradicting Eq.~\eqref{eqn:left-right-v2}.
Therefore, $A$ must be reactive.

\subsubsection[S2.3.2. Proof that condition (C1) is satisfied for almost all large networks]{Proof that condition (C1) is satisfied for almost all large networks}
\label{subsec:proof_reactivity}

The GCLV model is a special case of the more general model discussed in Ref.~\citen{Cao:2020} for which the two functions $\kappa$ and $\varphi_n$ defining the model are given by
\begin{equation}
\kappa(\mathbf{x}_i,\mathbf{x}_j)=\frac{\tilde{d}_i^\text{in}\tilde{d}_j^\text{out}}{\tilde{d}}
\end{equation}
and
\begin{equation}
\varphi_n(\mathbf{x}_i,\mathbf{x}_j) = \frac{n\tilde{d}}{\sum_{k=1}^n \tilde{d}_k^\text{in}} \biggl( 1 \wedge \frac{c\sum_{k=1}^n \tilde{d}_k^\text{in}}{\tilde{d}_i^\text{in}\tilde{d}_j^\text{out}} \biggr)-1,
\end{equation}
where we used the notations $x \wedge y := \min\{x, y\}$ and $\mathbf{x}_i := (\tilde{d}_i^\text{in},\tilde{d}_i^\text{out})$ for each node $i$ and recall that $\tilde{d} = \mathbb{E}(\tilde{d}_i^\text{in}) = \mathbb{E}(\tilde{d}_i^\text{out})$ for any $i$.
By Proposition 3.13 in Ref.~\citen{Cao:2020} and our model assumption $\tilde{d}>1$, there exists a constant $\tau>0$ such that the largest strongly connected component $\mathcal{C}_n$ has approximately $\tau n$ nodes in the limit of large network size $n$.
In other words, if we denote the size of that component by $\tilde{n}=\abs{\mathcal{C}_n}$, we have $\tilde{n} / n \rightarrow \tau$ as $n\rightarrow \infty$ (a defining property of a giant strongly connected component of the network).
In addition, according to Ref.~\citen{Neri:2020}, the largest eigenvalue $\lambda_1(A)$ is also the largest eigenvalue of the adjacency matrix $\tilde{A}$ of $\mathcal{C}_n$ in the limit $n\to\infty$.
From this, it follows that the probability of satisfying the in-centrality part of condition (C1) (that there is a node $i$ in $\tilde{G}=\mathcal{C}_n$ for which $v_{1i} \neq 0$) approaches one as $n\to\infty$. 
Therefore, the probability that $A$ satisfies (C1) is asymptotically bounded from below by the probability that there is a column of $A$ that strictly dominates the corresponding row of $A$ within $\mathcal{C}_n$.

We can now estimate the probability that $A$ is reactive as
\begin{equation}\label{eqn:A-reactive}
\mathbb{P}(\text{$A$ is reactive}) \ge \mathbb{E}\bigl( p(n) \bigr),
\end{equation}
where $p(n)$ denotes the conditional probability that there exists a node $i$ in the network for which the $i$th column of $A$ strictly dominates the $i$th row of $A$ within $\mathcal{C}_n$ given realizations of $\{ \tilde{d}^\text{in}_i \}_{i=1}^n$ and $\{ \tilde{d}^\text{out}_i \}_{i=1}^n$.
We thus seek to establish a lower bound for $p(n)$ and use it to show that $p(n)$ approaches one in the large network limit.
For that purpose, we first consider an analogous probability for a given node $i$.
Specifically, we define $p_i(n)$ to be the conditional probability that the $i$th column strictly dominates the $i$th row in $\tilde{A}(n)$ (again, given a realization of $\{ \tilde{d}^\text{in}_i \}_{i=1}^n$ and $\{ \tilde{d}^\text{out}_i \}_{i=1}^n$).
This probability can be computed as
\begin{equation}\label{eqn:pi}
\begin{split}
p_i(n) &= \mathbb{P}(A_{ji} \ge A_{ij}, \forall j \in \mathcal{C}_n) - \mathbb{P}(A_{ji} = A_{ij}, \forall j \in \mathcal{C}_n)\\
&= \underset{j\in \mathcal{C}_n,\ j\neq i}{\prod}\left( 1- r_{ij}(n) \right) -  \underset{j\in \mathcal{C}_n,\ j\neq i}{\prod} \left( 1-s_{ij}(n) \right),
\end{split}
\end{equation}
where $r_{ij}(n)$ is the probability that $A_{ji} < A_{ij}$ for given $i,j \in \mathcal{C}_n$, which can be expressed using the definition of the model as
\begin{equation}\label{eqn:defrij}
r_{ij}(n) = (1-\rho_{ji})\rho_{ij} 
= \frac{\tilde{d}_i^{\text{in}}\tilde{d}_j^{\text{out}}}{\sum_{k=1}^{n}\tilde{d}_k^{\text{in}}}- \frac{\tilde{d}_j^{\text{in}}\tilde{d}_i^{\text{out}}}{\sum_{k=1}^{n}\tilde{d}_k^{\text{in}}}\cdot \frac{\tilde{d}_i^{\text{in}}\tilde{d}_j^{\text{out}}}{\sum_{k=1}^{n}\tilde{d}_k^{\text{in}}},
\end{equation}
and $s_{ij}(n)$ is the probability that $A_{ji}\neq A_{ij}$, which can be written as
\begin{equation}\label{eqn:defsij}
\begin{split}
s_{ij}(n) &= (1-\rho_{ji})\rho_{ij} + (1-\rho_{ij})\rho_{ji} \\
&= \frac{\tilde{d}_i^{\text{in}}\tilde{d}_j^{\text{out}}}{\sum_{k=1}^{n}\tilde{d}_k^{\text{in}}} + \frac{\tilde{d}_j^{\text{in}}\tilde{d}_i^{\text{out}}}{\sum_{k=1}^{n}\tilde{d}_k^{\text{in}}}-2\cdot\frac{\tilde{d}_j^{\text{in}}\tilde{d}_i^{\text{out}}}{\sum_{k=1}^{n}\tilde{d}_k^{\text{in}}}\cdot \frac{\tilde{d}_i^{\text{in}}\tilde{d}_j^{\text{out}}}{\sum_{k=1}^{n}\tilde{d}_k^{\text{in}}}.
\end{split}
\end{equation}
Note that $p_i(n)$ in Eq.~\eqref{eqn:pi} is strictly positive for any given $i$ and $n$, since $0<\rho_{ij}<1$ and thus $s_{ij}(n) > r_{ij}(n)$ for all $i$ and $j$. 

To derive a lower bound for $p_i(n)$ in Eq.~\eqref{eqn:pi}, we will use several elementary inequalities.
We first note that the derivative of the function $-x-\text{ln}(1-x) $ is $\frac{x}{1-x}$, which is monotonically increasing on the interval $[0,1)$.
Applying the Mean Value Theorem to this function, we obtain the inequality $-x-\text{ln}(1-x) \leq \frac{x^2}{1-x}$.
We further note that the inequalities $x(1-y)+y(1-x)\leq c$ and $x(1-y)\leq c$ can be shown to hold true for any $0\leq x,y\leq c$ under the model assumption $c \ge 1/2$.
From this, together with $\rho_{ij}\leq c$ as well as Eqs.~(\ref{eqn:defrij}) and (\ref{eqn:defsij}), we see that $r_{ij}(n)\leq c$ and $s_{ij}(n)\leq c$.
Further noting that $\abs{e^{-x}-e^{-y}}\leq \abs{x-y}$ for any $x,y>0$, we estimate the second term in Eq.~\eqref{eqn:pi} using an exponential function:
\begin{equation}\label{eqn:sijconbd2}
\begin{aligned}
\Biggl\lvert \exp\Bigl(-&\sum_{j\in \mathcal{C}_n,\ j\neq i} s_{ij}(n) \Bigr) - \prod_{j\in \mathcal{C}_n,\ j\neq i} \bigl( 1-s_{ij}(n) \bigr) \Biggr\rvert \\
&\leq \Biggl\lvert -\sum_{j\in \mathcal{C}_n,\ j\neq i}  s_{ij}(n) - \ln\Bigl( {\prod_{j\in \mathcal{C}_n,\ j\neq i}} \bigl( 1-s_{ij}(n) \bigr) \Bigr) \Biggr\rvert \\
&\leq \sum_{j\in \mathcal{C}_n,\ j\neq i} \frac{s_{ij}^2(n)}{1-s_{ij}(n)}\\
&\leq \frac{1}{1-c} \sum_{j\in \mathcal{C}_n,\ j\neq i} \left( \rho_{ij} + \rho_{ji} - 2\rho_{ij}\rho_{ji} \right)^2\\
&\leq  \frac{1}{1-c} \sum_{1\leq j\leq n,\ j\neq i} \left( \rho_{ij} + \rho_{ji} \right)^2 \\
& \leq  \frac{\bigl( (\tilde{d}_i^{\text{in}})^2 + (\tilde{d}_i^\text{out})^2 + 2\tilde{d}_i^\text{in}\tilde{d}_i^\text{out} \bigr) \cdot \varepsilon(n)}{1-c},
\end{aligned}
\end{equation}
where the third and last line follow from the inequality $-x-\text{ln}(1-x) \leq \frac{x^2}{1-x}$ and Eq.~\eqref{eqn:rhoik22}, respectively.
We also have a similar estimate for $r_{ij}(n)$:
\begin{equation}\label{eqn:rijconbd2}
\begin{aligned}
\Biggl\lvert \exp\Bigl( - &\sum_{j\in \mathcal{C}_n,\ j\neq i} r_{ij}(n) \Bigr) - \prod_{j\in \mathcal{C}_n,\ j\neq i} \bigl( 1-r_{ij}(n) \bigr) \Biggr\rvert \\ 
&\leq \Biggl\lvert - \sum_{j\in \mathcal{C}_n,\ j\neq i}  r_{ij}(n) - \ln\Bigl(\, {\prod_{j\in \mathcal{C}_n,\ j\neq i}} \bigl( 1-r_{ij}(n) \bigr) \Bigr) \Biggr\rvert \\
&\leq \sum_{j\in \mathcal{C}_n,\ j\neq i} \frac{r_{ij}^2(n)}{1-r_{ij}(n)}\\
&\leq \frac{1}{1-c} \sum_{j\in \mathcal{C}_n,\ j\neq i} \bigl( (1-\rho_{ji})\rho_{ij} \bigr)^2\\
&\leq \frac{1}{1-c} \sum_{1\leq j\leq n,\ j\neq i} \rho_{ij}^2 \\
& \leq \frac{(\tilde{d}_i^{\text{in}})^2 \cdot \varepsilon(n)}{1-c},
\end{aligned}
\end{equation}
where the last inequality follows from the lower left inequality in Eq.~\eqref{eqn:tmp1}.
Combining Eqs.~(\ref{eqn:pi}), (\ref{eqn:sijconbd2}), and (\ref{eqn:rijconbd2}), we obtain a lower bound for $p_i(n)$:
\begin{multline}\label{eqn:lowerbdpi}
p_i(n) \geq \exp\Bigl(-\sum_{j\in \mathcal{C}_n,\ j\neq i} r_{ij}(n) \Bigr) - \exp\Bigl(-\sum_{j\in \mathcal{C}_n,\ j\neq i} s_{ij}(n) \Bigr) \\
- \frac{\bigl( 2(\tilde{d}_i^{\text{in}})^2 + (\tilde{d}_i^\text{out})^2 + \tilde{d}_i^\text{in}\tilde{d}_i^\text{out} \bigr) \cdot \varepsilon(n)}{1-c}.
\end{multline}
To further estimate the r.h.s., we note that
\begin{equation}
\begin{split}
\sum_{j\in \mathcal{C}_n,\ j\neq i} r_{ij}(n)
&= \sum_{j\in \mathcal{C}_n,\ j\neq i} \left( \frac{\tilde{d}_i^{\text{in}}\tilde{d}_j^{\text{out}}}{\sum_{k=1}^{n}\tilde{d}_k^{\text{in}}}- \frac{\tilde{d}_j^{\text{in}}\tilde{d}_i^{\text{out}}}{\sum_{k=1}^{n}\tilde{d}_k^{\text{in}}}\cdot \frac{\tilde{d}_i^{\text{in}}\tilde{d}_j^{\text{out}}}{\sum_{k=1}^{n}\tilde{d}_k^{\text{in}}} \right)\\
&=\tilde{d}_i^{\text{in}} \frac{\sum_{j\in \mathcal{C}_n,\ j\neq i}\tilde{d}_j^{\text{out}}}{\sum_{k=1}^{n}\tilde{d}_k^{\text{in}}} -\tilde{d}_i^{\text{out}}\tilde{d}_i^{\text{in}}
\frac{\sum_{j\in \mathcal{C}_n,\ j\neq i}\tilde{d}_j^{\text{in}}\tilde{d}_j^{\text{out}}}{(\sum_{k=1}^{n}\tilde{d}_k^{\text{in}})^2}
\end{split}
\end{equation}
and
\begin{equation}
\begin{split}
\sum_{j\in \mathcal{C}_n,\ j\neq i} s_{ij}(n) &=  \sum_{j\in \mathcal{C}_n,\ j\neq i} \left( \frac{\tilde{d}_i^{\text{in}}\tilde{d}_j^{\text{out}}}{\sum_{k=1}^{n}\tilde{d}_k^{\text{in}}} + \frac{\tilde{d}_j^{\text{in}}\tilde{d}_i^{\text{out}}}{\sum_{k=1}^{n}\tilde{d}_k^{\text{in}}}-2\frac{\tilde{d}_j^{\text{in}}\tilde{d}_i^{\text{out}}}{\sum_{k=1}^{n}\tilde{d}_k^{\text{in}}}\cdot \frac{\tilde{d}_i^{\text{in}}\tilde{d}_j^{\text{out}}}{\sum_{k=1}^{n}\tilde{d}_k^{\text{in}}} \right) \\
&=\tilde{d}_i^{\text{in}} \frac{\sum_{j\in \mathcal{C}_n,\ j\neq i}\tilde{d}_j^{\text{out}}}{\sum_{k=1}^{n}\tilde{d}_k^{\text{in}}}+
\tilde{d}_i^{\text{out}} \frac{\sum_{j\in \mathcal{C}_n,\ j\neq i}\tilde{d}_j^{\text{in}}}{\sum_{k=1}^{n}\tilde{d}_k^{\text{in}}}
-2\tilde{d}_i^{\text{out}}\tilde{d}_i^{\text{in}}
\frac{\sum_{j\in \mathcal{C}_n,\ j\neq i}\tilde{d}_j^{\text{in}}\tilde{d}_j^{\text{out}}}{(\sum_{k=1}^{n}\tilde{d}_k^{\text{in}})^2},
\end{split}
\end{equation}
from which we see that these sums satisfy
\begin{equation}
\sum_{j\in \mathcal{C}_n,\ j\neq i}r_{ij}(n) <\sum_{j\in \mathcal{C}_n,\ j\neq i} s_{ij}(n)\leq \tilde{d}_i^\text{in}(1+\varepsilon(n))+\tilde{d}_i^\text{out},
\end{equation}
where $\varepsilon(n)$ is defined in Eq.~\eqref{eqn:tmp1}.
Since $\varepsilon(n)$ converges to zero as $n\rightarrow \infty$, it is also bounded, implying that there is a constant $\eta >1$ for which $1+\varepsilon(n)<\eta$ for all $n$.
Hence, we have 
\begin{equation}
\sum_{j\in \mathcal{C}_n,\ j\neq i}r_{ij}(n) <\sum_{j\in \mathcal{C}_n,\ j\neq i} s_{ij}(n)<\eta\tilde{d}_i^\text{in}+\tilde{d}_i^\text{out}.
\end{equation}
Now, noting that $e^{-x}$ is a monotonically decreasing function, and its derivative, $-e^{-x}$, is monotonically increasing, we obtain the following estimate for the first two terms of Eq.~\eqref{eqn:lowerbdpi}:
\begin{equation}\label{eqn:lowerbddiffexp}
\begin{aligned}
&\exp\Bigl( -\sum_{j\in \mathcal{C}_n,\ j\neq i} r_{ij}(n) \Bigr) - \exp\Bigl( -\sum_{j\in \mathcal{C}_n,\ j\neq i} s_{ij}(n) \Bigr) \\
&\geq \exp\bigl(-(\eta \tilde{d}_i^\text{in}+\tilde{d}_i^\text{out}) \bigr) \cdot \Bigl( \sum_{j\in \mathcal{C}_n,\ j\neq i}  s_{ij}(n)-\sum_{j\in \mathcal{C}_n,\ j\neq i}  r_{ij}(n) \Bigr)\\
&= \exp\bigl( -(\eta \tilde{d}_i^\text{in}+\tilde{d}_i^\text{out}) \bigr) \cdot \biggl( \tilde{d}_i^{\text{out}} \frac{\sum_{j\in \mathcal{C}_n,\ j\neq i}\tilde{d}_j^{\text{in}}}{\sum_{k=1}^{n}\tilde{d}_k^{\text{in}}} -\tilde{d}_i^{\text{out}}\tilde{d}_i^{\text{in}} \frac{\sum_{j\in \mathcal{C}_n,\ j\neq i}\tilde{d}_j^{\text{in}}\tilde{d}_j^{\text{out}}}{\bigl( \sum_{k=1}^{n}\tilde{d}_k^{\text{in}} \bigr)^2} \biggr) \\
&\geq \tilde{d}_i^\text{out} \exp\bigl(-(\eta \tilde{d}_i^\text{in}+\tilde{d}_i^\text{out}) \bigr) \biggl( \frac{\sum_{j\in \mathcal{C}_n,\ j\neq i}\tilde{d}_j^{\text{in}}}{\sum_{k=1}^{n}\tilde{d}_k^{\text{in}}} -\tilde{d}_i^{\text{in}} \varepsilon(n) \biggr) \\
&\geq \tilde{d}_i^\text{out} \exp\bigl(-(\eta \tilde{d}_i^\text{in}+\tilde{d}_i^\text{out})  \bigr)
    \biggl(
    \frac{\sum_{1\leq j\leq \tilde{n},\ j\neq i}\tilde{d}_{[n-j+1]}^{\text{in}}}{\sum_{k=1}^{n}\tilde{d}_k^{\text{in}}}
    -\tilde{d}_i^{\text{in}}
    \varepsilon(n)
    \biggr) \\
    & \geq \tilde{d}_i^\text{out} \exp\bigl(-(\eta \tilde{d}_i^\text{in}+\tilde{d}_i^\text{out})  \bigr)
    \biggl(
    \frac{\sum_{1\leq j \leq \tilde{n},\ j\neq i}\min \{\tilde{d}_{[n-j+1]}^{\text{in}},\tilde{d}\}}{\sum_{k=1}^{n}\tilde{d}_k^{\text{in}}}
    -\tilde{d}_i^{\text{in}}
    \varepsilon(n)
    \biggr)\\
    &= \tilde{d}_i^\text{out} \exp\bigl(-(\eta \tilde{d}_i^\text{in}+\tilde{d}_i^\text{out})  \bigr)
    \biggl(
    \frac{\frac{\tilde{n} - 1}{\tilde{n}}\tilde{d}-\frac{1}{\tilde{n}}\sum_{1\leq j \leq \tilde{n},\ j\neq i}\bigl(\tilde{d}-\min \{\tilde{d}_{[n-j+1]}^{\text{in}},\tilde{d}\}\bigr)}{\frac{1}{\tilde{n}}\sum_{k=1}^{n}\tilde{d}_k^{\text{in}}}
    -\tilde{d}_i^{\text{in}}
    \varepsilon(n)
    \biggr)\\
    & = \tilde{d}_i^\text{out} \exp\bigl(-(\eta \tilde{d}_i^\text{in}+\tilde{d}_i^\text{out})  \bigr)
    \biggl(
    \frac{\frac{\tilde{n} - 1}{\tilde{n}}\tilde{d}-\frac{1}{\tilde{n}}\sum_{1\leq j \leq \tilde{n},\ j\neq i}
    \tilde{h}_{[j]}^\text{in}
    }{\frac{1}{\tilde{n}}\sum_{k=1}^{n}\tilde{d}_k^{\text{in}}}
    -\tilde{d}_i^{\text{in}}
    \varepsilon(n)
    \biggr),
    \end{aligned}
\end{equation}
where $\{\tilde{d}_{[j]}^{\text{in}}\}_{j=1}^n$ denotes the order statistics of $\{\tilde{d}_{j}^{\text{in}}\}_{j=1}^n$, i.e., the re-indexed version of $\{\tilde{d}_{j}^{\text{in}}\}_{j=1}^n$ in which $\tilde{d}_{[1]}^{\text{in}}\geq \tilde{d}_{[2]}^{\text{in}}\geq \cdots \geq \tilde{d}_{[n]}^{\text{in}}$, and we define $\tilde{h}^\text{in}_{[j]}:=\tilde{d}-\min \{\tilde{d}^{\text{in}}_{[n-j+1]},\tilde{d}\}$.

To further estimate the term $\frac{1}{\tilde{n}}\sum_{1\leq j \leq \tilde{n},\ j\neq i} \tilde{h}_{[j]}^\text{in}$ in Eq.~\eqref{eqn:lowerbddiffexp}, we consider the so-called conditional value at risk \cite{Wang:2010}, given by
\begin{equation}
c_{\tau} := \mathbb{E}\bigl( \tilde{h}^\text{in} \,\big\vert\, \tilde{h}^\text{in} \geq v_{\tau} \bigr)
\end{equation}
in the case of the random variable $\tilde{h}^\text{in} := \tilde{d}-\min \{\tilde{d}^{\text{in}},\tilde{d}\}$, where $v_{\tau} := \sup\{\xi : \mathbb{P}(\tilde{h}^\text{in}> \xi) \geq \tau\}$ is called the value at risk and $\tilde{d}^{\text{in}}$ is a random variable following the same distribution as $\tilde{d}_i^{\text{in}}$ (for any $i$).
We note that both $c_{\tau}$ and $v_{\tau}$ are constants determined by the parameter $\tau$ and the distribution of $\tilde{d}^{\text{in}}$. 
A finite-sample estimator for $c_{\tau}$ is given by
\begin{equation}
\hat{c}_{\tau} := \frac{1}{\tau n}\sum_{1\leq j \leq \lfloor \tau n \rfloor} \tilde{h}_{[j]}^\text{in}
\end{equation}
and can be used to approximate $\frac{1}{\tilde{n}}\sum_{1\leq j \leq \tilde{n},\ j\neq i} \tilde{h}_{[j]}^\text{in}$ as
\begin{equation}
\begin{split}
\biggl\lvert \frac{1}{\tilde{n}} &\sum_{1\leq j \leq \tilde{n},\ j\neq i} \tilde{h}_{[j]}^\text{in} - \hat{c}_{\tau} \biggr\rvert\\
&\leq \biggl\lvert \frac{1}{\tilde{n}}\sum_{1\leq j \leq \tilde{n},\ j\neq i} \tilde{h}_{[j]}^\text{in}-\frac{1}{\tilde{n}}\sum_{1\leq j \leq \lfloor \tau n \rfloor} \tilde{h}_{[j]}^\text{in} \biggr\rvert + \biggl\lvert \frac{1}{\tilde{n}}\sum_{1\leq j \leq \lfloor \tau n \rfloor} \tilde{h}_{[j]}^\text{in}-\frac{1}{\tau n}\sum_{1\leq j \leq \lfloor \tau n \rfloor} \tilde{h}_{[j]}^\text{in} \biggr\rvert \\
&\leq \delta (n) := \Bigl( 2 \Bigl\lvert 1-\frac{ \tau n }{\tilde{n}} \Bigr\rvert + \frac{1}{\tilde{n}} \Bigr)\,\tilde{d},
\end{split}
\end{equation}
where we used $\tilde{h}^\text{in}_{[j]} \le \tilde{d}$ and we have $\underset{n\to\infty}{\lim} \delta(n) = 0$ (since $\underset{n\to\infty}{\lim} \tilde{n}/n = \tau$).
Using this in Eq.~\eqref{eqn:lowerbddiffexp}, we obtain
\begin{multline}\label{eqn:lowerbddiffexp2}
\exp\Bigl( -\sum_{j\in \mathcal{C}_n,\ j\neq i}  r_{ij}(n) \Bigr) - \exp\Bigl( -\sum_{j\in \mathcal{C}_n,\ j\neq i} s_{ij}(n) \Bigr) \\
\geq \tilde{d}_i^\text{out} \exp\bigl(-(\eta \tilde{d}_i^\text{in}+\tilde{d}_i^\text{out}) \bigr)
    \biggl(
    \frac{\frac{\tilde{n} - 1}{\tilde{n}}\tilde{d} - \hat{c}_{\tau} - \delta (n)
    }{\frac{1}{\tilde{n}}\sum_{k=1}^{n}\tilde{d}_k^{\text{in}}}
    -\tilde{d}_i^{\text{in}}
    \varepsilon(n)
    \biggr).
\end{multline}
Using the concentration bounds proved in Ref.~\citen{Wang:2010} and noting the fact that $x \geq \epsilon$ if and only if $\text{max}\{x,0\} \geq \epsilon$ for any given $\epsilon>0$, we have
\begin{equation}\label{eqn:CVaRbdd}
\mathbb{P} \bigl( ( \hat{c}_{\tau} - {c}_{\tau} )^+ > \epsilon \bigr) \leq 3\exp\biggl( -\frac{n\tau\epsilon^2}{11\tilde{d}^2} \biggr)
\end{equation}
for any $\epsilon>0$, where we use the notation $(x)^+ = \text{max}\{x,0\}$.
Applying Borel-Cantelli lemma (Proposition 2.6 in Ref.~\citen{Cinlar:2011}) and noting that $\sum_{n=1}^{\infty} \exp\bigl( -\frac{n\tau\epsilon^2}{11\tilde{d}^2} \bigr) <\infty$, Eq.~\eqref{eqn:CVaRbdd} implies
\begin{equation}\label{eqn:CVaRbddlim}
\chi(n) := \bigl( \hat{c}_{\tau} - {c}_{\tau} \bigr)^+ \longrightarrow 0
\end{equation}
as $n\to\infty$ almost surely.
Combining
\begin{equation}\label{eqn:lowerbdcvar}
\hat{c}_{\tau} \leq \max \{ \hat{c}_{\tau}, {c}_{\tau} \}
 = {c}_{\tau} + \bigl( \hat{c}_{\tau} - {c}_{\tau} \bigr)^+ 
= {c}_{\tau} + \chi (n).
\end{equation}
with Eqs. (\ref{eqn:lowerbdpi}), (\ref{eqn:lowerbddiffexp2}), and (\ref{eqn:CVaRbddlim}), we see that
\begin{equation}\label{eqn:defhatpin}
\begin{split}
p_i(n) \geq \hat{p}_i(n) &:= \tilde{d}_i^\text{out} \exp\bigl(-(\eta \tilde{d}_i^\text{in}+\tilde{d}_i^\text{out}) \bigr) \biggl( \frac{\frac{\tilde{n} - 1}{\tilde{n}}\tilde{d} - {c}_{\tau} - \chi (n) - \delta(n) }{\frac{n}{\tilde{n}}\cdot\frac{1}{n}\sum_{k=1}^{n}\tilde{d}_k^{\text{in}}} - \tilde{d}_i^{\text{in}} \varepsilon(n) \biggr) \\
&\quad\quad\quad - \frac{\bigl( 2(\tilde{d}_i^{\text{in}})^2 + (\tilde{d}_i^\text{out})^2 + \tilde{d}_i^\text{in}\tilde{d}_i^\text{out} \bigr) \cdot \varepsilon(n)}{1-c} \\
&\longrightarrow \tilde{d}_i^\text{out} \exp\bigl(-(\eta\tilde{d}_i^\text{in}+\tilde{d}_i^\text{out}) \bigr) \cdot\frac{\tau(\tilde{d} - {c}_{\tau})}{\tilde{d}}
\end{split}
\end{equation}
as $n\to\infty$ almost surely.
Without loss of generality, we may modify the function $\varepsilon(n)$ in Eq.~\eqref{eqn:tmp1} to additionally satisfy
\begin{equation}
\biggl\lvert \frac{\frac{\tilde{n} - 1}{\tilde{n}}\tilde{d} - {c}_{\tau} -\chi(n)-\delta(n)}{\frac{n}{\tilde{n}} \cdot \frac{1}{n}\sum_{k=1}^{n}\tilde{d}_k^{\text{in}}}-\frac{\tau(\tilde{d} - {c}_{\tau})}{\tilde{d}} \biggr\rvert  \leq \varepsilon(n),
\end{equation}
and thus the deviation from the limit in Eq.~\eqref{eqn:defhatpin} can be estimated as
\begin{multline}\label{eqn:hatpconverge}
\biggl\lvert \hat{p}_i(n) - \tilde{d}_i^\text{out} \exp\bigl(-(\eta\tilde{d}_i^\text{in}+\tilde{d}_i^\text{out})  \bigr) \cdot \frac{\tau(\tilde{d} - {c}_{\tau})}{\tilde{d}} \biggr\rvert \\
\leq (\tilde{d}_i^\text{out}+ \tilde{d}_i^\text{in}\tilde{d}_i^\text{out}) \exp\bigl(-(\eta\tilde{d}_i^\text{in}+\tilde{d}_i^\text{out})  \bigr) \cdot \varepsilon(n)+ \frac{\bigl( 2(\tilde{d}_i^{\text{in}})^2 + (\tilde{d}_i^\text{out})^2 + \tilde{d}_i^\text{in}\tilde{d}_i^\text{out} \bigr) \cdot \varepsilon(n)}{1-c}.
\end{multline}

We now seek to estimate $p(n)$ by a recursive argument involving $p_i(n)$ and $\hat{p}_i(n)$.
Let $\tilde{A}_k(n)$ denote the principal submatrix of $A$ obtained by keeping only those rows $i$ and columns $i$ for which $i \le k$ and $i \in \mathcal{C}_n$.
We further let $q_k(n)$ denote the probability that there is at least one column strictly dominating the corresponding row in $\tilde{A}_k(n)$.
We thus have $q_n(n) = p(n)$.
Then, we have the following recursive inequality:
\begin{equation}\label{eqn:recurqkn}
q_k(n) \geq q_{k-1}(n) \cdot \underset{1\leq j \leq k-1}{\text{min}} \left(1-r_{kj}(n)\right)   + (1-q_{k-1}(n))p_k(n).
\end{equation}
Let $\epsilon_k (n) := \underset{1\leq j \leq k-1}{\text{max}} \rho_{kj} \geq  \underset{1\leq j \leq k-1}{\text{max}} r_{kj}(n)$.
Using this notation, Eq.~\eqref{eqn:recurqkn} can be rewritten as
\begin{equation}\label{eqn:recurqkn2}
1-q_k(n) \leq (1-q_{k-1}(n))\cdot(1- p_k(n))+\epsilon_k(n).
\end{equation}
Since the r.h.s.\ of Eq.~\eqref{eqn:recurqkn2} is monotonically decreasing in $p_k(n)$, the inequality remains true if $p_k(n)$ is replaced with its lower bound $\hat{p}_k(n)$.
We thus have
\begin{equation}\label{eqn:recurqkn3}
1-q_k(n) \leq (1-q_{k-1}(n))\cdot(1- \hat{p}_k(n))+\epsilon_k(n).
\end{equation}
Recursive application of this inequality yields
\begin{equation}
1 - q_{n}(n) \leq \epsilon_n(n) + \prod_{j=2}^n (1-\hat{p}_j(n)) +  \sum_{j=3}^{n}\epsilon_{j-1}(n) \prod_{i=j}^n (1-\hat{p}_i(n)).
\end{equation}
Applying the same reordering of nodes we used for Eq.~\eqref{eqn:diffto1bd} to obtain Eq.~\eqref{eqn:diffto1bd2}, we obtain
\begin{equation}\label{eqn:diffto1bd3}
1 - q_{n}(n) \leq \epsilon_n(n) + \prod_{j=1}^{n-1} (1-\hat{p}_j(n)) + \sum_{j=1}^{n-2}\epsilon_{j+1}(n) \prod_{i=1}^{j} (1-\hat{p}_i(n)).
\end{equation}

To help estimate the r.h.s.\ of Eq.~\eqref{eqn:diffto1bd3}, we consider the double sequence defined by
\begin{equation}
m_{jn} = \begin{cases}
\displaystyle \frac{1}{j}\sum_{i=1}^j \hat{p}_i(n), \ j\leq n,\\
\displaystyle \frac{1}{n}\sum_{i=1}^n \hat{p}_i(n),\ j>n.
\end{cases}
\end{equation}
As was done in Sec.~\ref{sec-si-nonnormal} for a similar double sequence (see Eq.~\eqref{eqn:Pin}), we will use the Moore-Osgood theorem \cite{Zakon:2011} to calculate the limit of this double sequence.
We first note that $\underset{j\to\infty}{\lim} m_{jn} = m_{nn}$ for each fixed $n$ by definition and that
\begin{equation}\label{eqn:conergemjn}
\lim_{n\to\infty} m_{jn}
= \frac{1}{j}\sum_{i=1}^j \lim_{n\to\infty} \hat{p}_i(n)
= \frac{1}{j}\sum_{i=1}^j \tilde{d}_i^\text{out} \exp\bigl(-(\eta\tilde{d}_i^\text{in}+\tilde{d}_i^\text{out}) \bigr) \cdot\frac{\tau(\tilde{d} - {c}_{\tau})}{\tilde{d}}
\end{equation}
for each fixed $j$.
To see that this convergence is uniform over all $j$, we note that Eq.~\eqref{eqn:hatpconverge} implies
\begin{equation}\label{eqn:mjnconbdd}
\begin{aligned}
\biggl\lvert \frac{1}{j}&\sum_{i=1}^j \hat{p}_i(n) - \frac{1}{j}\sum_{i=1}^j \tilde{d}_i^\text{out} \exp\bigl(-(\eta\tilde{d}_i^\text{in}+\tilde{d}_i^\text{out}) \bigr) \cdot\frac{\tau(\tilde{d} - {c}_{\tau})}{\tilde{d}} \biggr\rvert\\
& \leq \frac{1}{j}\sum_{i=1}^j \biggl\lvert (\tilde{d}_i^\text{out}+ \tilde{d}_i^\text{in}\tilde{d}_i^\text{out}) \exp\bigl(-(\eta\tilde{d}_i^\text{in}+\tilde{d}_i^\text{out})  \bigr) + \frac{2(\tilde{d}_i^{\text{in}})^2 + (\tilde{d}_i^\text{out})^2 + \tilde{d}_i^\text{in}\tilde{d}_i^\text{out}}{1-c} \biggr\rvert \cdot \varepsilon(n)\\
& \leq M' \varepsilon(n),
\end{aligned}
\end{equation}
where the average over $j$ on the second line is bounded by a finite constant $M'$ because the averages of the individual terms all converge to finite values as $j\to\infty$ due to the strong law of large numbers.
Since $M' \varepsilon(n)$ converges to zero as $n\to\infty$ with a rate independent of $j$, the convergence in Eq.~\eqref{eqn:conergemjn} is indeed uniform in $j$.
Applying the Moore-Osgood theorem and using the strong law of large numbers again, we conclude that the limit $m_\infty$ of the double sequence $m_{jn}$ exists and can be computed using the following iterated limit:
\begin{equation}\label{eqn:mdoublelimit}
\begin{split}
\lim_{j,n \rightarrow \infty}m_{jn}
&= \lim_{j\rightarrow \infty}\lim_{n\rightarrow \infty} m_{jn} \\
&= m_\infty := \mathbb{E}\bigl( \tilde{d}_i^\text{out} \exp\bigl(-(\eta\tilde{d}_i^\text{in}+\tilde{d}_i^\text{out})  \bigr) \bigr) \cdot \frac{\tau(\tilde{d} - {c}_{\tau})}{\tilde{d}} > 0.
\end{split}
\end{equation}
We note that the limit $m_\infty$ is strictly positive because $\tilde{d}_i^\text{out} \ge 1$, $\tau>0$, and $\tilde{d} - {c}_{\tau} \ge 1$ (which follows from $\tilde{d}^\text{in} \ge 1$ and the definition of ${c}_{\tau}$).

With the limit $m_\infty$ in hand, we now return to Eq.~\eqref{eqn:diffto1bd3} and show the convergence of $1 - q_{n}(n)$ to zero as $n\to\infty$ by estimating the terms on the r.h.s.\ one by one.
We first note that, from Eqs.~(\ref{eqn:gev_conv}) and (\ref{eqn:rhoincon}), both $\epsilon_k (n) \in [0,1]$ and $\epsilon_n (n) \in [0,1]$ converge to zero in probability as $n \rightarrow \infty$.
Thus, the first term in Eq.~\eqref{eqn:diffto1bd3} converges to zero.
For the second term, we have
\begin{equation}\label{eqn:secterm}
\prod_{j=1}^{n-1} (1-\hat{p}_j(n)) \leq \exp\Bigl(-\sum_{j=1}^{n-1} \hat{p}_j(n) \Bigr)
= \exp \bigl( -(n-1)\cdot m_{n-1,n} \bigr).
\end{equation}
According to Eq.~\eqref{eqn:mdoublelimit}, the factor $m_{n-1,n}$ above converges to a strictly positive value while the factor $n-1$ diverges, implying that $\exp \bigl( -(n-1) \cdot m_{n-1,n} \bigr)$, and hence the second term on the r.h.s.\ of Eq.~\eqref{eqn:diffto1bd3}, converges to zero as $n\to\infty$.

For the third term on the r.h.s.\ of Eq.~\eqref{eqn:diffto1bd3}, we first note that, for any $0<N<n-2$, we have
\begin{equation}\label{eqn:bnbound}
\begin{aligned}
\sum_{j=1}^{n-2} \epsilon_{j+1}(n) &\prod_{i=1}^{j} (1-\hat{p}_i(n)) 
\leq \sum_{j=1}^{n-2} \epsilon_{j+1}(n) \exp \Bigl( -\sum_{i=1}^j \hat{p}_i(n) \Bigr) \\
&= \sum_{j=1}^{N} \epsilon_{j+1}(n) \exp \Bigl( -\sum_{i=1}^j \hat{p}_i(n) \Bigr)+ \sum_{j=N+1}^{n-2} \epsilon_{j+1}(n) \exp \Bigl( -\sum_{i=1}^j \hat{p}_i(n) \Bigr)\\
& \leq \sum_{j=1}^{N} \epsilon_{j+1}(n) \exp \Bigl( -\sum_{i=1}^j \hat{p}_i(n) \Bigr)+ \sum_{j=N+1}^{n-2} \exp \Bigl( -\sum_{i=1}^j \hat{p}_i(n) \Bigr)\\
& \leq \sum_{j=1}^{N} \epsilon_{j+1}(n) \exp \Bigl( -\sum_{i=1}^j \hat{p}_i(n) \Bigr) + \sum_{j=N+1}^{n-2} \exp \bigl( -j\cdot m_{jn} \bigr).
\end{aligned}
\end{equation}
The first sum on the last line converges to zero in probability as $n\rightarrow \infty$, since each $\epsilon_{j+1}(n)$ converges to zero in probability, and the sum has only a finite number of terms.
For the second sum, we observe that
\begin{equation}\label{eqn:reacthird}
\begin{aligned}
\sum_{j=N+1}^{n-2} \exp\bigl( -j\cdot m_{jn} \bigr)
\leq \sum_{j=N+1}^{\infty} \exp \bigl(-j \cdot m_\infty/2 \bigr) 
= \frac{\exp\bigl( -(N+1) \cdot m_\infty/2 \bigr)}{1-\exp\bigl( -m_\infty/2 \bigr)},
\end{aligned}
\end{equation}
where we used Eq.~\eqref{eqn:mjnconbdd} to estimate $m_{jn}$ for $j,n > N$ with sufficiently large $N$ as
\begin{equation}
m_{jn} \geq \frac{1}{j}\sum_{i=1}^j \tilde{d}_i^\text{out} \exp\bigl(-(\eta\tilde{d}_i^\text{in}+\tilde{d}_i^\text{out}) \bigr) \cdot\frac{\tau(\tilde{d} - {c}_{\tau})}{\tilde{d}} - M' \epsilon(n) \geq \frac{m_\infty}{2}.
\end{equation}
Noting that the r.h.s. of Eq.~\eqref{eqn:reacthird} can be made arbitrarily close to zero by sufficiently increasing $N$, we conclude that the third term on the r.h.s.\ of Eq.~\eqref{eqn:diffto1bd3} also converges to zero in probability. 

Therefore, all three terms on the r.h.s.\ of Eq.~\eqref{eqn:diffto1bd3} converge to zero in probability as $n \rightarrow \infty$, implying $\underset{n\to\infty}{\lim} p(n) = \underset{n\to\infty}{\lim} q_n(n) = 1$.
In view of Eq.~\eqref{eqn:A-reactive}, this proves that the probability that $A$ is reactive converges to one in the limit of large networks.

\subsection[S2.4. Extension to weighted networks with self-links and Laplacian-coupled networks]{Extension to weighted networks with self-links and Laplacian-coupled networks}
\label{sec-si-weighted}

We now show that a weighted adjacency matrix $A$ is nonnormal and reactive with probability tending to one as $n\to\infty$ for the weighted GCLV model, in which the link weights are independently and identically distributed. 
We also allow arbitrary nonzero diagonal elements in the matrix $A$, which does not affect the column-row strict dominance condition and thus condition~(C1).
For these weighted networks, the probability $r_{ij}(n)$ that $A_{ji} < A_{ij}$ and the probability $s_{ij}(n)$ that $A_{ji}\neq A_{ij}$ have upper and lower bounds, respectively:
\begin{equation}\label{eqn:rijsijbound}
\begin{aligned}
r_{ij}(n) &\leq (1-\rho_{ji})\rho_{ij} +\frac{1}{2}\rho_{ij}\rho_{ji}=\rho_{ij}-\frac{1}{2}\rho_{ij}\rho_{ji}, \\
s_{ij}(n) &\geq (1-\rho_{ji})\rho_{ij} + (1-\rho_{ij})\rho_{ji} = \rho_{ij}+\rho_{ji}-2  \rho_{ij}\rho_{ji}.
\end{aligned}
\end{equation}
Using Eq.~\eqref{eqn:rijsijbound} in Eq.~\eqref{eqn:pi} yields a lower bound for the probability $p_i(n)$ that the $i$th column strictly dominates the $i$th row for $i \in \mathcal{C}_n$. The rest of the proof in Sec.~\ref{subsec:proof_reactivity} would then remain valid if we simply redefine $r_{ij}(n) := \rho_{ij}-\frac{1}{2}\rho_{ij}\rho_{ji}$ and $s_{ij}(n) := \rho_{ij}+\rho_{ji}-2  \rho_{ij}\rho_{ji}$, since the difference it creates in the coefficient of the high-order term $\rho_{ij}\rho_{ji}$ in Eq.~\eqref{eqn:rijconbd2} does not affect this and the subsequent inequalities. This shows that almost all random weighted directed networks are reactive and thus also nonnormal.

In the case of Laplacian-coupled networks, the adjacency matrix $A$ in Eq.~\eqref{eqn:main-system} is replaced by $-L = -K + A$, the negative of the Laplacian matrix.  Since condition (C1) is satisfied for $X=A$ with probability approaching one as $n\to\infty$ and the addition of diagonal elements from the term $-K$ does not affect condition~(C1), we conclude that the probability that $-L$ satisfies condition~(C1), and thus the probability that $-L$ is reactive approaches one in the limit of large networks.  Since reactivity implies nonnormality, we also conclude that $-L$ is also nonnormal with probability approaching one as $n\to\infty$.

\section[S3. Computational details for fig.~\ref{fig:prob-norm-nonreact} and table~\ref{table:prob-norm-nonreact}]{Computational details for fig.~\ref{fig:prob-norm-nonreact} and table~\ref{table:prob-norm-nonreact}}
\label{sec-si-details-figS4-TableS1}
For a given random network model and for each network size $n$, the probabilities were estimated from $10^6$ network realizations.  The unweighted networks were generated using four different network topology models.  For the first two, we used the GCLV model with two different in- and out-degree distributions: the gamma distribution $p(x) \sim x^{a-1}e^{-bx}$ with parameters $a=2$ and $b=1/2$ and the Dirac delta distribution concentrated at $d=2$ (which is equivalent to the Erd\H{o}s--R\'{e}nyi (ER) model with connection probability $p=d/n$ and fixed $d$).
In both cases, the in- and out-degrees were uncorrelated.  The remaining two models are the ER model with a fixed connection probability $p$ and the random $d$-regular networks.  For the ER model, we used $p=0.8$.  For the $d$-regular networks, we used $d = 3$ and generated realizations using the configuration model \cite{Newman:2001}.
When using these four models in fig.~\ref{fig:prob-norm-nonreact} and table~\ref{table:prob-norm-nonreact}, we prohibit self-links (i.e., we set $A_{ii} = 0$ for all $i$), which play a limited role in the condition for nonnormality based on Eq.~\eqref{eqn:D_F_def} and the condition $\theta_1 > 0$ for reactivity and thus are not expected to significantly affect the probability estimates.
In addition, Eq.~\eqref{eqn:EX_ij2} indicates that neglecting self-links (when they are present) would not overestimate the probability that $A$ is nonnormal (it would, in fact, underestimate it if the $A_{ii}$ are not all identical).
The estimated probabilities plotted in fig.~\ref{fig:prob-norm-nonreact}, A to C are also shown in table~\ref{table:prob-norm-nonreact} under ``unweighted networks.''  For weighted networks in table~\ref{table:prob-norm-nonreact}, each realization was generated by first creating the network topology using one of the models described above and then assigning to each link a random weight drawn from the (discrete) Poisson distribution with mean $9$.
The numerical results are presented using a threshold of $10^{-8}$ for both nonnormality $\norm{D}_\text{F}$ and reactivity $\lambda_{\Delta}(A)$.  In the absence of any threshold, it follows from the expected impact of link weights on the imbalances underlying nonnormality and reactivity that networks with continuously distributed random weights are nonnormal and reactive with probability one.

\clearpage
\newcounter{sfigure}
\renewcommand{\figurename}{Fig.}
\renewcommand{\thefigure}{S\arabic{sfigure}}

\vspace{5mm}

\addtocounter{sfigure}{1}
\begin{figure}
\phantomsection
\addcontentsline{toc}{section}{Fig.~\ref{fig:real_networks_SI}: Version of Fig.~\ref{fig:real_networks} indicating network types}
\begin{center}
\vspace{-6mm}
\includegraphics[width=6in]{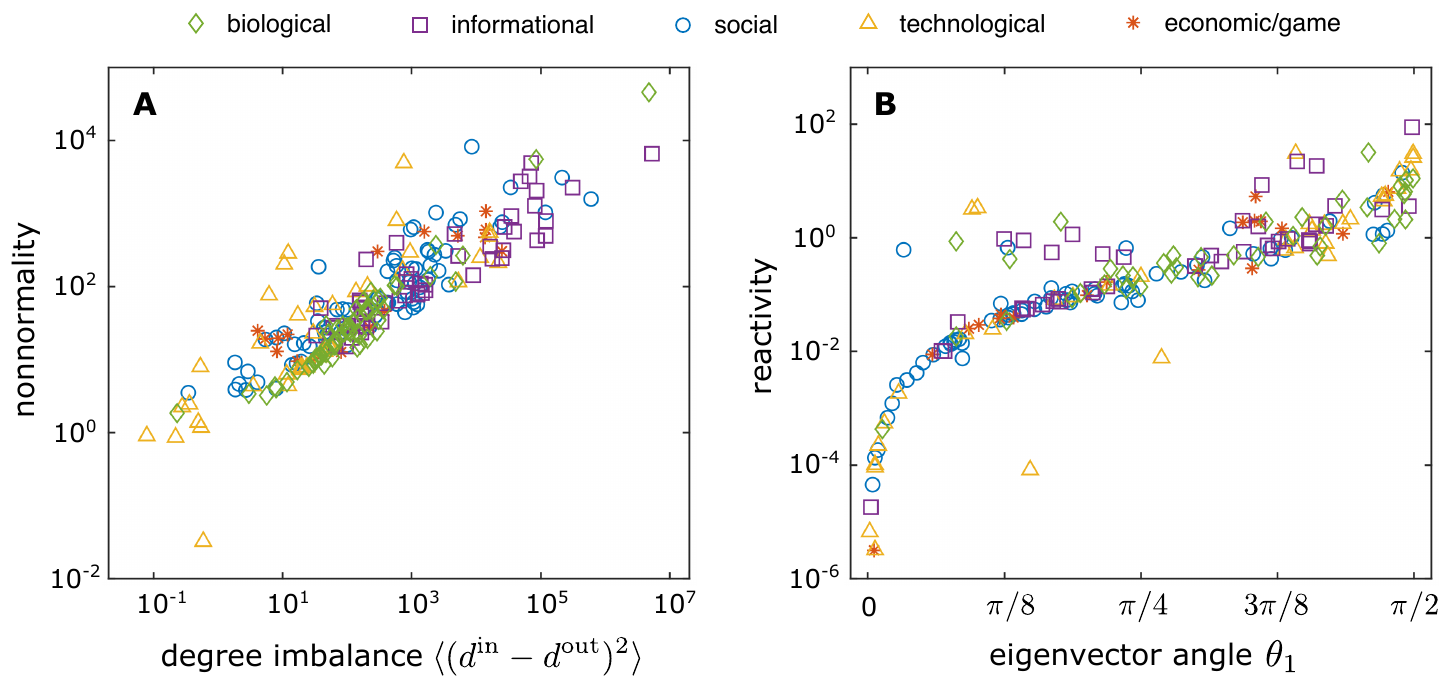}
\caption{\textbf{Version of Fig.~\ref{fig:real_networks} indicating network types.}
Both the marker symbols and colors encode the five types of networks as labeled in the data set.  We observe that the distributions of nonnormality and reactivity are comparable for (and with large overlaps between) different types of networks.}
\label{fig:real_networks_SI}
\end{center}
\end{figure}

\addtocounter{sfigure}{1}
\begin{figure}
\phantomsection
\addcontentsline{toc}{section}{Fig. \ref{fig:real_networks_examples}: Topological and spectral imbalances in representative networks}
\begin{center}
\vspace{-6mm}
\includegraphics[width=6in]{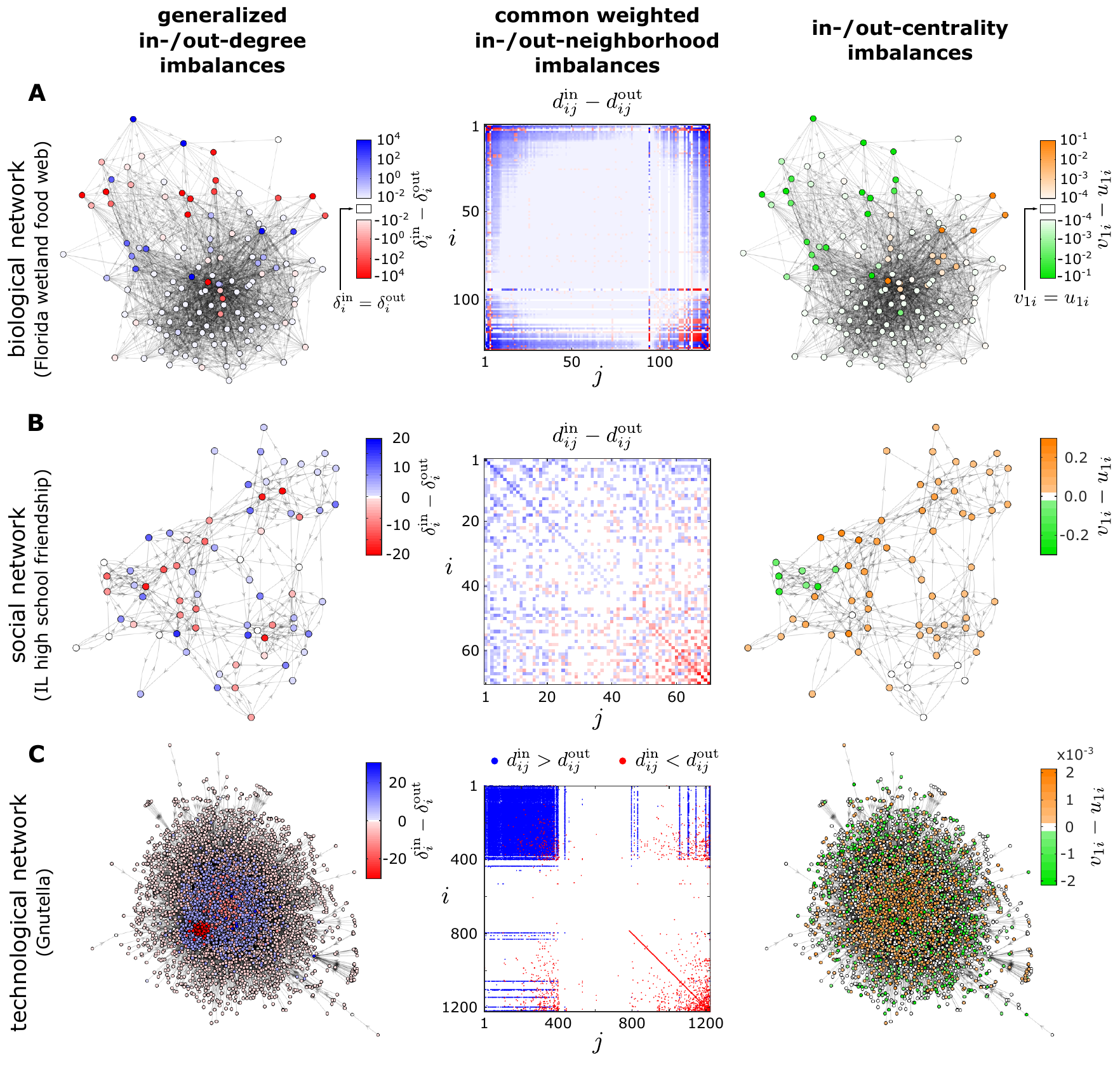}
\caption{\textbf{Topological and spectral imbalances in representative networks.}
Here, we show the examples of biological, social, and technological networks indicated in Fig.~\ref{fig:real_networks}.
(\textbf{A})~Network of feeding relations between species or groups of species in the cypress wetlands of South Florida during the dry season.
(\textbf{B})~Network of friendship relations between boys in an Illinois high school in 1957--1958.
(\textbf{C})~Network of connections between Gnutella host computers in 2002.
In the first column, the color of node $i$ indicates the generalized degree imbalance $\delta_i^\text{in} - \delta_i^\text{out}$ (positive and negative imbalances in shades of blue and red, respectively), showing that the imbalances are heterogeneously distributed across the network.
The second column shows the matrix whose $(i,j)$ component is $d_{ij}^\text{in} - d_{ij}^\text{out}$ if $i \neq j$ and $\delta_i^\text{in} - \delta_i^\text{out}$ if $i=j$.
Each matrix components is color coded in (A) and (B), while only the sign of the component is shown in (C) (to highlight the small but numerous imbalances).
The node indices are ordered so that $\delta_i^\text{in} - \delta_i^\text{out}$ is non-increasing with respect to $i$.
In the third column, the color of node $i$ represents the imbalance $v_{1i} - u_{1i}$ between the node's in- and out-centralities (the corresponding components of left and right eigenvectors normalized so that $\norm{v_1}=\norm{u_1}=1$), 
which is distributed heterogeneously across the network.
}
\label{fig:real_networks_examples}
\end{center}
\end{figure}

\addtocounter{sfigure}{1}
\begin{figure}
\phantomsection
\addcontentsline{toc}{section}{Fig. \ref{fig:A_L_normality}: Nonnormality of adjacency vs. Laplacian matrices}
\begin{center}
\includegraphics[width=5.8in]{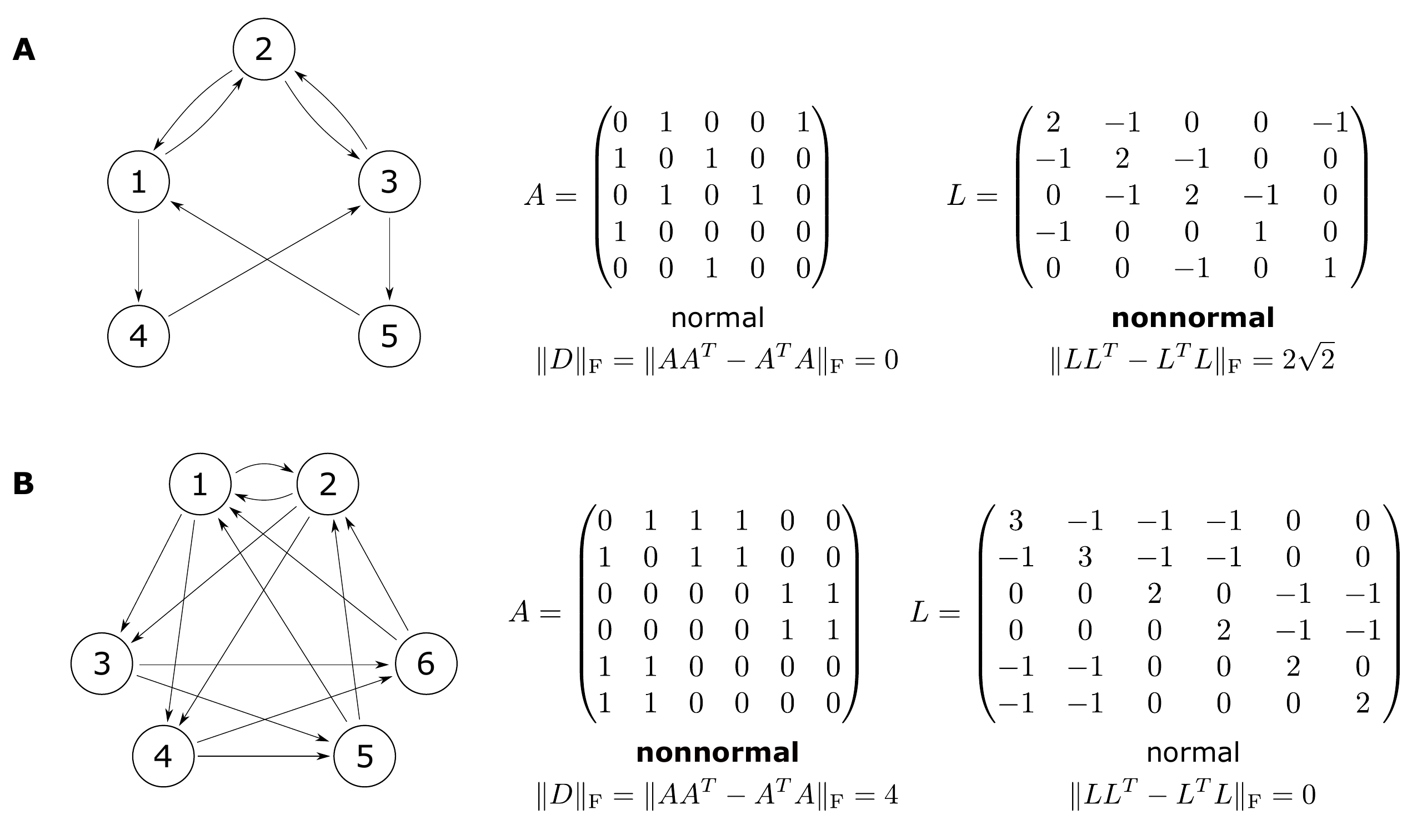}
\caption{
\textbf{Nonnormality of adjacency vs.\ Laplacian matrices.}
For a given network, nonnormality of the adjacency matrix $A$ does not necessarily imply nonnormality of the Laplacian matrix 
$L=K-A$, and vice versa, where we recall that $K$ denotes the diagonal matrix with $\sum_{k \neq i} A_{ik}$ on the diagonal.
Indeed, the Laplacian matrix $L$ is nonnormal if and only if $LL^T-L^TL = (AA^T-A^TA) + [K(A-A^T)-(A-A^T)K]$ is nonzero, which is different from the nonnormality condition for $A$, i.e., $AA^T-A^TA \neq 0$.
(\textbf{A})~Example (unweighted) network for which $A$ is normal but $L$ is nonnormal.
In this case, $AA^T-A^TA$ is zero, but $K(A-A^T)-(A-A^T)K$ is nonzero, rendering $L$ nonnormal.
(\textbf{B})~Example (unweighted) network for which $A$ is nonnormal but $L$ is normal.
In this case, $AA^T-A^TA$ is nonzero but is canceled exactly by $K(A-A^T)-(A-A^T)K$, rendering $L$ normal.
}
\label{fig:A_L_normality}
\end{center}
\end{figure}

\addtocounter{sfigure}{1}
\begin{figure}
\phantomsection
\addcontentsline{toc}{section}{Fig. \ref{fig:prob-norm-nonreact}: Nonnormality and reactivity of typical random networks}
\begin{center}
\includegraphics[width=6.3in]{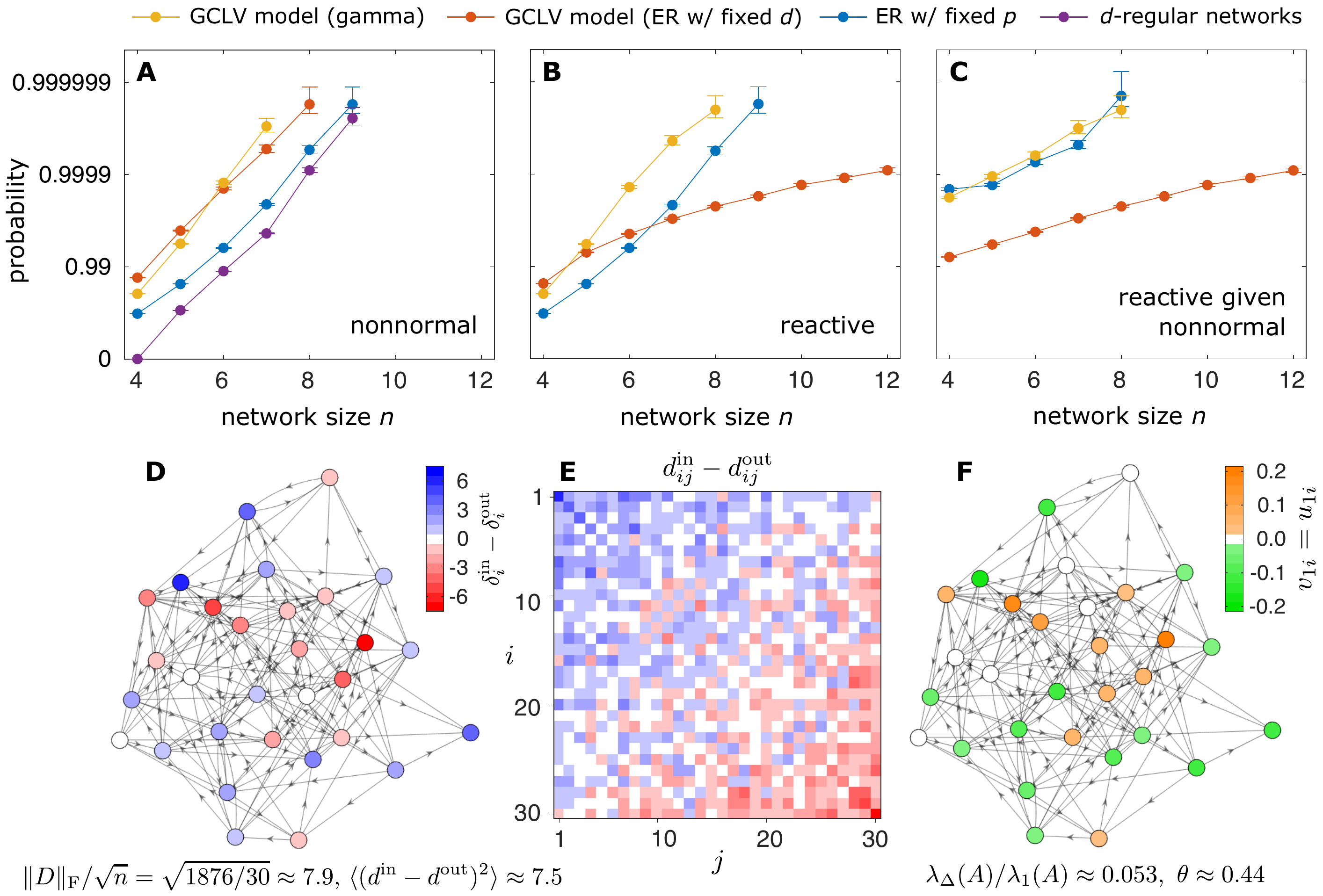}
\caption{\textbf{Nonnormality and reactivity of typical random networks.}  (\textbf{A} to \textbf{C})~Probability that $A$ is nonnormal (A), probability that $A$ is reactive (B), and conditional probability that $A$ is reactive given that it is nonnormal (C), plotted as functions of the network size $n$ for four different models of random unweighted networks (see table~\ref{table:prob-norm-nonreact} for the corresponding cases of weighted networks).  The error bars indicate the estimated standard deviation (too small to be visible in some cases).  The estimates were computed for each $4 \le n \le 12$ but not shown if they equal one, since the vertical axis is on logarithmic scale.
The estimates for $n>12$ are one within numerical precision.  The results suggest that, in each case, all three probabilities approach one at least exponentially as $n$ increases.  (\textbf{D} to \textbf{F})~Typical ER network with $30$ nodes and a connection probability of $p=0.2$, for which $A$ is both nonnormal and reactive.  Node-level imbalances are visualized as in fig.~\ref{fig:real_networks_examples}.
See supplementary text, Sec.~\ref{sec-si-details-figS4-TableS1} for computational details.
}
\label{fig:prob-norm-nonreact}
\end{center}
\end{figure}

\addtocounter{sfigure}{1}
\begin{figure}
\phantomsection
\addcontentsline{toc}{section}{Fig. \ref{fig_A1_approx}: Validating the approximations underlying Eqs. (5) and (6)}
\begin{center}
\includegraphics[width=5.7in]{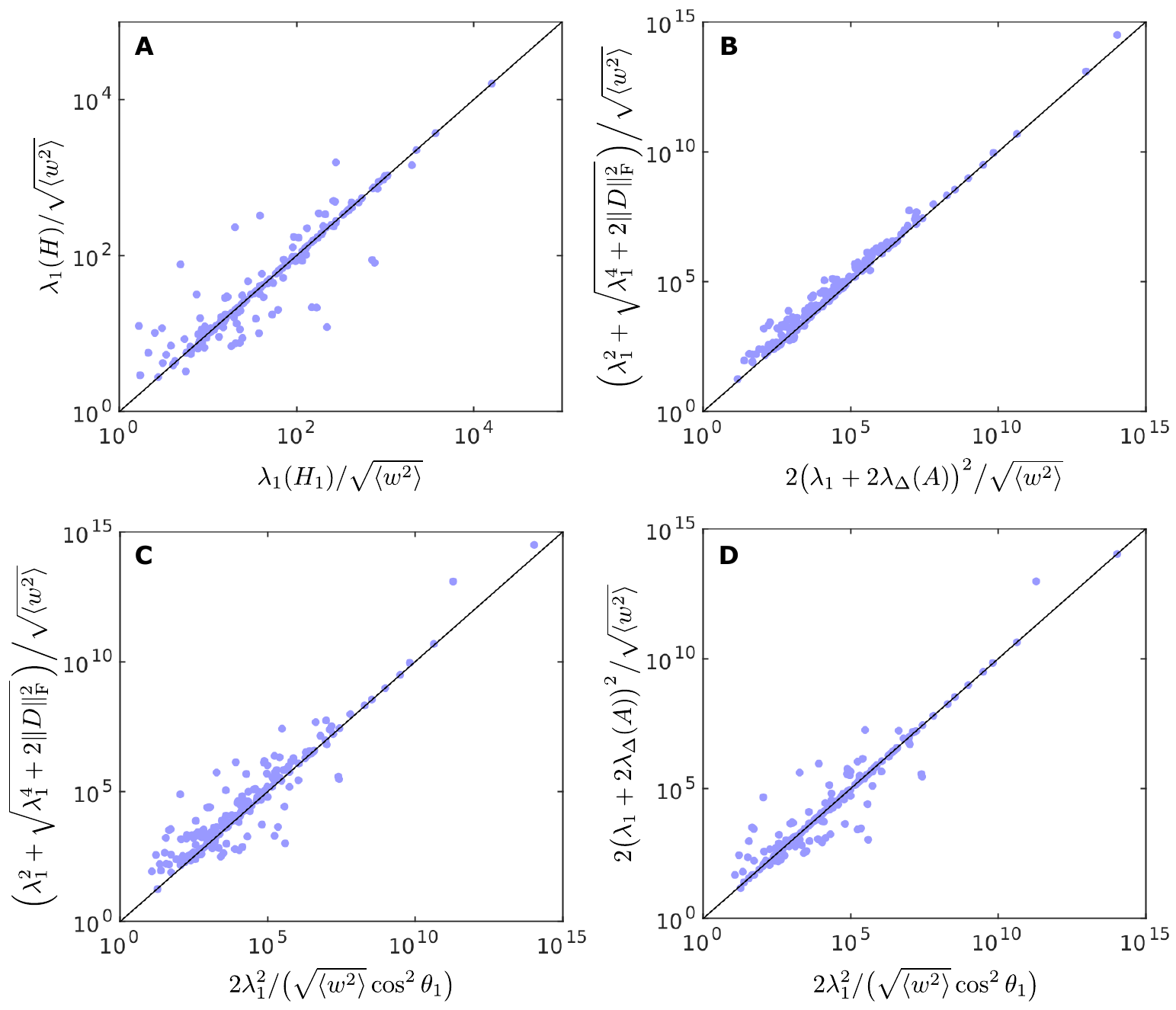}
\caption{
\textbf{Validating the approximations underlying Eqs.~\eqref{eqn:lambda_delta_theory} and \eqref{eqn:relation-dominant}.}
(\textbf{A})~Leading eigenvalues of $H$ and $H_1$ are plotted against each other for the real networks used in Fig.~\ref{fig:theory_validation}B, validating the approximation $\lambda_1(H) \approx \lambda_1(H_1)$ underlying Eq.~\eqref{eqn:lambda_delta_theory} in most cases.
(\textbf{B} to \textbf{D})~Comparing the three quantities in Eq.~\eqref{eqn:relation-dominant} with $A_1$ replaced by $A$ for the same set of networks as in (A).
In all panels, the quantities are normalized by$\sqrt{\langle w^2 \rangle}$.
We observe good agreement between these quantities for most of the networks.}
\label{fig_A1_approx}
\end{center}
\end{figure}

\clearpage
\newcounter{SItable}
\renewcommand{\tablename}{Table}
\renewcommand{\thetable}{S\arabic{SItable}}

\addtocounter{SItable}{1}
\begin{table}[]
\phantomsection
\addcontentsline{toc}{section}{Table \ref{table:prob-norm-nonreact}: Probability that A is normal and probability that $A$ is non-reactive}
\centering
\caption{Probability that $A$ is normal, probability that $A$ is non-reactive, and conditional probability that $A$ is non-reactive given that $A$ is nonnormal for both weighted and unweighted random networks.
See supplementary text, Sec.~\ref{sec-si-details-figS4-TableS1} for computational details.
}
\label{table:prob-norm-nonreact}
\begin{tabular}{lccccc}
\hline\hline
 & $n=4$ & $n=6$ & $n=8$ & $n=10$ & $n=12$ \\
\hline
\textbf{Unweighted networks:}\rule{0pt}{18pt} \\
$\mathbb{P}(\text{$A$ is normal})$\rule{0pt}{16pt} &&&&& \\
\rowcolor{Gray}
\hspace{5mm}GCLV model (gamma) & $0.04$ & $0.0002$ & $0$ & $0$ & $0$ \\
\hspace{5mm}GCLV model (ER w/ fixed $d$) & $0.02$ & $0.0003$ & $9\times10^{-6}$ & $0$ & $0$ \\
\rowcolor{Gray}
\hspace{5mm}ER w/ fixed $p$ & $0.1$ & $0.005$ & $5\times10^{-5}$ & $0$ & $0$ \\
\hspace{5mm}$d$-regular networks & $1$ & $0.09$ & $0.0002$ & $0$ & $0$ \\
$\mathbb{P}(\text{$A$ is non-reactive})$\rule{0pt}{16pt} &&&&& \\
\rowcolor{Gray}
\hspace{5mm}GCLV model (gamma) & $0.04$ & $0.0002$ & $3\times10^{-6}$ & $0$ & $0$ \\
\hspace{5mm}GCLV model (ER w/ fixed $d$) & $0.03$ & $0.002$ & $0.0003$ & $6\times10^{-5}$ & $2\times10^{-5}$ \\
\rowcolor{Gray}
\hspace{5mm}ER w/ fixed $p$ & $0.1$ & $0.005$ & $5\times10^{-5}$ & $0$ & $0$ \\
\hspace{5mm}$d$-regular networks & $1$ & $1$ & $1$ & $1$ & $1$ \\
\multicolumn{2}{l}{$\mathbb{P}(\text{$A$ is non-reactive} \,\vert\, \text{$A$ is nonnormal})$\rule{0pt}{16pt}} &&&& \\
\rowcolor{Gray}
\hspace{5mm}GCLV model (gamma) & $0.0002$ & $3\times10^{-5}$ & $3\times10^{-6}$ & $0$ & $0$ \\
\hspace{5mm}GCLV model (ER w/ fixed $d$) & $0.006$ & $0.001$ & $0.0003$ & $6\times10^{-5}$ & $2\times10^{-5}$ \\
\rowcolor{Gray}
\hspace{5mm}ER w/ fixed $p$ & $0.0002$ & $7\times10^{-5}$ & $2\times10^{-6}$ & $0$ & $0$ \\
\hspace{5mm}$d$-regular networks & -- & $1$ & $1$ & $1$ & $1$ \\
\textbf{Weighted networks:}\rule{0pt}{18pt} \\
$\mathbb{P}(\text{$A$ is normal})$\rule{0pt}{16pt} &&&&& \\
\rowcolor{Gray}
\hspace{5mm}GCLV model (gamma) & $0.0001$ & $10^{-6}$ & $0$ & $0$ & $0$ \\
\hspace{5mm}GCLV model (ER w/ fixed $d$) & $0.0004$ & $7\times10^{-6}$ & $0$ & $0$ & $0$ \\
\rowcolor{Gray}
\hspace{5mm}ER w/ fixed $p$ & $2\times10^{-6}$ & $0$ & $0$ & $0$ & $0$ \\
\hspace{5mm}$d$-regular networks & $0.003$ & $0.0001$ & $10^{-6}$ & $0$ & $0$ \\
$\mathbb{P}(\text{$A$ is non-reactive})$\rule{0pt}{16pt} &&&&& \\
\rowcolor{Gray}
\hspace{5mm}GCLV model (gamma) & $0.0001$ & $4\times10^{-6}$ & $0$ & $0$ & $0$ \\
\hspace{5mm}GCLV model (ER w/ fixed $d$) & $0.0007$ & $8\times10^{-5}$ & $2\times10^{-5}$ & $8\times10^{-6}$ & $4\times10^{-6}$ \\
\rowcolor{Gray}
\hspace{5mm}ER w/ fixed $p$ & $2\times10^{-6}$ & $0$ & $0$ & $0$ & $0$ \\
\hspace{5mm}$d$-regular networks & $0.03$ & $0.02$ & $0.009$ & $0.008$ & $0.005$ \\
\multicolumn{2}{l}{$\mathbb{P}(\text{$A$ is non-reactive} \,\vert\, \text{$A$ is nonnormal})$\rule{0pt}{16pt}} &&&& \\
\rowcolor{Gray}
\hspace{5mm}GCLV model (gamma) & $2\times10^{-5}$ & $3\times10^{-6}$ & $0$ & $0$ & $0$ \\
\hspace{5mm}GCLV model (ER w/ fixed $d$) & $0.0003$ & $8\times10^{-5}$ & $2\times10^{-5}$ & $8\times10^{-6}$ & $4\times10^{-6}$ \\
\rowcolor{Gray}
\hspace{5mm}ER w/ fixed $p$ & $0$ & $0$ & $0$ & $0$ & $0$ \\
\hspace{5mm}$d$-regular networks & $0.03$ & $0.02$ & $0.009$ & $0.008$ & $0.005$ \\
\hline\hline
\end{tabular}
\end{table}

\end{document}